\documentclass[12pt]{article}
\usepackage{graphicx}
\usepackage{epstopdf}
\usepackage{amsthm}
\usepackage{amsmath}
\usepackage{amssymb}
\usepackage{amstext}
\usepackage{amsfonts}
\usepackage{enumerate}
\usepackage[authoryear,nonamebreak,bibstyle]{natbib}
\usepackage{footnote}
\makesavenoteenv{tabular}
\makesavenoteenv{table}
\usepackage[onehalfspacing]{setspace}
\usepackage[shortlabels, inline]{enumitem}
\usepackage{multirow}
\usepackage[title]{appendix}
\usepackage{xcolor}
\usepackage{cancel}  
\usepackage{hyperref} 
\usepackage[normalem]{ulem}
\usepackage{tikz}
\usetikzlibrary{shapes,arrows}
\usepackage{tikz-cd}
\usepackage{sectsty}
\usepackage{caption, subcaption}
\usepackage[numbib]{tocbibind}
\usepackage[colorinlistoftodos,textsize=tiny]{todonotes}
\usepackage{endnotes}

\usepackage{tocloft}

\begin{document}
\baselineskip=.22in\parindent=30pt

\newtheorem{tm}{Theorem}
\newtheorem{dfn}{Definition}
\newtheorem{lma}{Lemma}
\newtheorem{assu}{Assumption}
\newtheorem{prop}{Proposition}
\newtheorem{cro}{Corollary}
\newtheorem*{theorem*}{Theorem}
\newtheorem{example}{Example}
\newtheorem{observation}{Observation}
\newcommand{\exm}{\begin{example}}
\newcommand{\exmm}{\end{example}}
\newcommand{\obs}{\begin{observation}}
\newcommand{\obss}{\end{observation}}
\newcommand{\cor}{\begin{cro}}
\newcommand{\corr}{\end{cro}}
\newtheorem{exa}{Example}
\newcommand{\ex}{\begin{exa}}
\newcommand{\exx}{\end{exa}}
\newtheorem{remak}{Remark}
\newcommand{\rmk}{\begin{remak}}
\newcommand{\rmkk}{\end{remak}}
\newcommand{\thm}{\begin{tm}}
\newcommand{\nt}{\noindent}
\newcommand{\thmm}{\end{tm}}
\newcommand{\lm}{\begin{lma}}
\newcommand{\lmm}{\end{lma}}
\newcommand{\ass}{\begin{assu}}
\newcommand{\asss}{\end{assu}}
\newcommand{\df}{\begin{dfn}  }
\newcommand{\dff}{\end{dfn}}
\newcommand{\prp}{\begin{prop}}
\newcommand{\prpp}{\end{prop}}
\newcommand{\bqu}{\sloppy \small \begin{quote}}
\newcommand{\equ}{\end{quote} \sloppy \large}
\newcommand\cites[1]{\citeauthor{#1}'s\ (\citeyear{#1})}

\renewcommand{\leq}{\leqslant}
\renewcommand{\geq}{\geqslant}

\newcommand{\eq}{\begin{equation}}
\newcommand{\eqq}{\end{equation}}
\newtheorem{claim}{\it Claim}
\newcommand{\cl}{\begin{claim}}
\newcommand{\cll}{\end{claim}}
\newcommand{\bit}{\begin{itemize}}
\newcommand{\eit}{\end{itemize}}
\newcommand{\ben}{\begin{enumerate}}
\newcommand{\een}{\end{enumerate}}
\newcommand{\bcen}{\begin{center}}
\newcommand{\ecen}{\end{center}}
\newcommand{\fn}{\footnote}
\newcommand{\ds}{\begin{description}}
\newcommand{\dss}{\end{description}}
\newcommand{\prf}{\begin{proof}}
\newcommand{\prff}{\end{proof}}
\newcommand{\cs}{\begin{cases}}
\newcommand{\css}{\end{cases}}
\newcommand{\ml}{\item}
\newcommand{\lb}{\label}
\newcommand{\ra}{\rightarrow}
\newcommand{\tra}{\twoheadrightarrow}
\newcommand*{\supp}{\operatornamewithlimits{sup}\limits}
\newcommand*{\inff}{\operatornamewithlimits{inf}\limits}
\newcommand{\nf}{\normalfont}
\renewcommand{\Re}{\mathbb{R}}
\newcommand*{\mmax}{\operatornamewithlimits{max}\limits}
\newcommand*{\mmin}{\operatornamewithlimits{min}\limits}
\newcommand*{\argmax}{\operatornamewithlimits{arg max}\limits}
\newcommand*{\argmin}{\operatornamewithlimits{arg min}\limits}
\newcommand{\uhr}{\!\! \upharpoonright  \!\! }

\newcommand{\CR}{\mathcal R}
\newcommand{\CC}{\mathcal C}
\newcommand{\CT}{\mathcal T}
\newcommand{\CS}{\mathcal S}
\newcommand{\CM}{\mathcal M}
\newcommand{\CL}{\mathcal L}
\newcommand{\CP}{\mathcal P}
\newcommand{\CN}{\mathcal N}

\newtheorem{innercustomthm}{Theorem}
\newenvironment{customthm}[1]
  {\renewcommand\theinnercustomthm{#1}\innercustomthm}
  {\endinnercustomthm}
\newtheorem{einnercustomthm}{Extended Theorem}
\newenvironment{ecustomthm}[1]
  {\renewcommand\theeinnercustomthm{#1}\einnercustomthm}
  {\endeinnercustomthm}
  
  \newtheorem{innercustomcor}{Corollary}
\newenvironment{customcor}[1]
  {\renewcommand\theinnercustomcor{#1}\innercustomcor}
  {\endinnercustomcor}
\newtheorem{einnercustomcor}{Extended Theorem}
\newenvironment{ecustomcor}[1]
  {\renewcommand\theeinnercustomcor{#1}\einnercustomcor}
  {\endeinnercustomcor}
    \newtheorem{innercustomlm}{Lemma}
\newenvironment{customlm}[1]
  {\renewcommand\theinnercustomlm{#1}\innercustomlm}
  {\endinnercustomlm}

\newcommand{\red}{\textcolor{red}}
\newcommand{\blue}{\textcolor{blue}}
\newcommand{\purple}{\textcolor{purple}}
\newcommand{\mred}[1]{\color{red}{#1}\color{black}}
\newcommand{\mblue}[1]{\color{blue}{#1}\color{black}}
\newcommand{\mpurple}[1]{\color{purple}{#1}\color{black}}

\newcommand{\norm}[1]{\left\lVert#1\right\rVert}

\subsubsectionfont{\normalfont\itshape}

\makeatletter
\newcommand{\customlabel}[2]{%
\protected@write \@auxout {}{\string \newlabel {#1}{{#2}{}}}}
\makeatother

 


\def\qed{\hfill\vrule height4pt width4pt
depth0pt}
\def\reff #1\par{\noindent\hangindent =\parindent
\hangafter =1 #1\par}
\def\title #1{\begin{center}
{\Large {\bf #1}}
\end{center}}
\def\author #1{\begin{center} {\large #1}
\end{center}}
\def\date #1{\centerline {\large #1}}
\def\place #1{\begin{center}{\large #1}
\end{center}}

\def\date #1{\centerline {\large #1}}
\def\place #1{\begin{center}{\large #1}\end{center}}
\def\intr #1{\stackrel {\circ}{#1}}
\def\R{{\rm I\kern-1.7pt R}}
 \def\N{{\rm I}\hskip-.13em{\rm N}}
 \newcommand{\cprod}{\Pi_{i=1}^\ell}
\let\Large=\large
\let\large=\normalsize


\begin{titlepage}

\def\thefootnote{\fnsymbol{footnote}}
\vspace*{0.1in}

\title{The Continuity Postulate  in  Economic Theory:  \\ \vskip 0.5em A Deconstruction and an Integration\fn{This work is presented as a celebration of Tajliing Koopmans' $110^{th}$-birthday: its authors constitute  the second and third generation of his student-followers, the first being represented by Beckman, Christ and Nerlove, the editors of the first volume of his collected papers.  The authors 
thank Max Amarante, Elena Antoniadou, Aurelien Baillon, Juan Dubra, H\"ulya Eraslan, Yorgos Gerasimou, Aniruddha Ghosh, Alfio Giarlotta, Carlos Herv{\'e}s-Beloso, Jeff Kline, David Levy, Paulo Monteiro, Efe Ok,  Eddie Schlee and Eric Schliesser for encouragement and stimulating conversation and correspondence. The authors also thank  seminar participants at the School of Economics at UTS and ANU, and conference participants at the 38th AETW held at the University of Adelaide, Australia.   Some of the results reported here were presented by the second author  at {\it Positivity X}, a conference held at the University of Pretoria July 8-12, 2019, and also  announced without proof at the 
NSF/NBER/CEME {\it Conference on Mathematical Economics: In Honor of Robert M.
Anderson} in Berkeley, October 25, 2019.  Khan should like to thank the respective organizers for their hospitality.  We thank the Handling Editor and   two anonymous referees for their careful and professional reading of the manuscript; acknowledgments to their specific questions and comments are detailed in the footnotes that follow.}} 

\vskip 1pt

\author{ Metin Uyanik\fn{School of Economics, University of Queensland, Brisbane, QLD 4072.  {\bf E-mail}
{m.uyanik@uq.edu.au}. We use the {\it certified random order} in order to list the authors; see Ray-Robson (2018, American Economic Review).}  \textcircled{r} 
M. Ali Khan\fn{Department of Economics, Johns Hopkins University, Baltimore, MD 21218. {\bf E-mail}
{akhan@jhu.edu}} 
}

\vskip 1.00em
\date{January 16, 2022}

\vskip 1.75em

\vskip 1.00em

\baselineskip=.18in

\noindent{\bf Abstract:}  This paper presents seven theorems and  nine propositions that can be read as  deconstructing  and integrating  the continuity postulate under the rubric of  pioneering work of Eilenberg, Wold, von Neumann-Morgenstern, Herstein-Milnor  and Debreu.  Its point of departure is the fact that  the adjective {\it continuous} applied to a function or a binary relation does not acknowledge the many meanings that can be given to the concept it names, and that under a variety of technical  mathematical structures, its many meanings can be whittled down to novel and unexpected equivalences that  have been missed  in the theory of choice. Specifically, it provides a systematic investigation of the two-way relation between {\it restricted} and {\it full} continuity of a function and a binary relation that, under  convex, monotonic and differentiable  structures, draws out the behavioral implications of the  postulate.  
  \hfill (135~words)
 
 \vskip 0.50in


\noindent {\it Journal of Economic Literature} Classification
Numbers: C00, C02, D01, D11, D21, D51, D81

\medskip

\noindent {\it 2020 Mathematics Subject} Classification Numbers: 26B05, 91B02, 91B08, 91B42, 06A06

\bigskip


\bigskip

\noindent {\it Running Title:} The Continuity Postulate

\end{titlepage}



\tableofcontents 

\thispagestyle{empty}

\vspace{15pt}

~~~~~~~ --------------------------------------------------------------- 

\vspace{15pt}

\pagebreak

\setcounter{page}{1}

\setcounter{footnote}{0}


\setlength{\abovedisplayskip}{-10pt}
\setlength{\belowdisplayskip}{-10pt}


\newif\ifall
\alltrue 

%
%
%
%
%
%
%


\bqu {\it Natura non facit saltus.}\fn{See Gottfried Wilhelm Leibniz {\it Nouveaux essais sur l'entendement humain,} 1704, p. 50, and  his 1702 essay {\it On the Principles of Continuity} in  \cite{we51}.  Also see    
 Alexander Baumgarten {\it Metaphysics: A Critical Translation with Kant's Elucidations,} translated and edited by Courtney D. Fugate and John Hymers, Bloomsbury, 2013, \lq\lq Preface of the Third Edition (1750)", p. 79, n. d: \lq\lq [Baumgarten] must also have in mind Leibniz's \lq natura non facit saltus [nature does not make leaps]' (NE IV, 16)." It is worth stating that sentence is also an epigraph of Marshall's {\it Principles.}  We engage this epigraph in the concluding section of this essay by emphasizing  the importance of {\it perception} to  {\it theorization}  as  the methodological backbone of this essay. This is 
   in response to a referee's comment that \lq\lq  the paper should convey some awareness of the approximations in use, and that it intends continuity to convey something at the perceptible level." We take this opportunity to thank  the referee for his/her generous and erudite comments.}   \hfill{Gottfried Leibniz~(1704)}  \equ 

\smallskip

\bqu {\it The postulational structure of the mathematical tool parallels that of the substantive theory to be constructed, and the two are studied and apprehended simultaneously. The welcome result is that \lq\lq mathematical" and \lq\lq literary" economics are moving closer. They meet on the ground of a common requirement of good hard thought from explicit basic postulates, rather than for manipulative skills.  If there is a difference, it is one of succinctness of expression rather than of content, concepts or objective.}\fn{See page 177 in \cite{ko57}; we skip some phrases. Koopmans offers as an illustration: \lq\lq [I]n some intuitive sense the \lq\lq distance" between Lerner's {\it The Economics of Control} and the mathematical formulations of the propositions of welfare economics ... is, I believe, not large;" see Koopmans for relevant references. We return to Koopmans in Footnote~\ref{fn:ko}, and in the text it footnotes.} \hfill{Tjalling Koopmans~(1957)}  \equ 

\section{Introduction} \label{sec: introduction}   

Tjalling Koopmans mentions the continuity postulate early on in the first of his 1957 magisterial essays  \lq\lq on the state of economic science."  After a footnote reference to a {\it continuous} function, an adjective he does not define but freely use,   and then  to a  {\it continuously representable} preference ordering that he does define, he masks the continuity postulate under the assumption  of {\it local non-satiation,}  referring to the latter   as \lq\lq a rather weak continuity property of preferences."  Subsequently, and in the interest of precision,  he sees the completeness postulate on preferences as \lq\lq  suggested by considerations of continuity and nonsaturation."\fn{See Footnote 5 on page 11 for the first reference; page 19 for the second; and Footnote 1 on page 20 for the third, all in \cite{ko57}.  For his free use of the notion of a continuous function and a continuously representable binary relation, see respectively Footnote 2 on page 33,  Footnote 3 on page 34, and page 55. It is of interest that the problem of representation of a binary relation by a function is introduced so early on the essay, right after the notion of profit maximization, and in the context of supply and consumption decisions. More to the point, Koopmans disregards the chronological order to introduce Wold before Hicks-Allen \citeyearpar{ha34ec}. } It is only in the second half of the essay, well after he has developed the basic theorems that the essay is to present,  that he uses the adjective {\it continuous} for a correspondence, but here again does not define the \lq\lq appropriate generalization of the concept of continuity.\fn{See page  57 in \cite{ko57} for the discussion of a \lq\lq class of theorems known as fixed point theorems, the appropriate concepts and theorems [for which] are found in the relatively young mathematical field of {\it topology} -- [which] crudely put, [is] the study of ways in which a set can \lq\lq hang together." Arrow-Hahn \citeyearpar{ah71} will refer to \lq\lq deeper methods," see Footnote~\ref{fn:ah} below.}  The point is that the notion of continuity is one of the signature notions in the {\it Essays,}\fn{A tracking of the word and its derivatives yields 84 instances, ranging its technical use in economic and statistical theory to its colloquial use in ordinary discourse, as for example, in the importance of the continuity of research support for the investigation of mathematical tools.}  and his reliance on it continues on  with more or less increasing force and emphasis throughout  his  work  subsequent to them.\fn{In his 1957 lectures at Stanford Summer School, (Koopmans-Bausch in \cite{ko70}, the notion is precisely defined in the context of a function, correspondence and a binary relation. In the {\it Essays} more generally, there are 21 uses of the word and its derivatives, but  in his 1964 paper \lq\lq On flexibility of future preference," originally dating to a 1950 abstract in which he tackles \lq\lq utility analysis of consumer choice," any reference to continuity is conspicuously missing. }

We shall have more to say about  Koopmans' fraught relationship to the continuity postulate: its  exploitation of  the interplay between its bearing on a function as opposed to that on a binary relation and/or a correspondence is an interplay that runs throughout his later {\it oeuvre.}  However, this is not a paper on Koopmans'  place  in the history of economic thought, and  we begin with him only because of the  impact that the 1957 {\it Essays,} and his  subsequent papers,  have had on the direction of the literature: it is this direction and its consequence for ongoing research  that is  of  primary concern of this paper's attempted  examination of the postulate.  In this connection,  it is worth pointing out that in his 1960 and   1964 work, Koopmans  assumes preferences to be parametrized by a continuous function, and investigates additional axioms under which such a function  takes on a specific form.  Later in  1972, in his two papers in honor of Jacob Marschak,  he considers  the representation of  a binary (preference) relation by a (utility)  function, the so-called representation problem,   and of necessity, introduces  continuous relations.  Since the  work reported here draws on the insights of the antecedent  mathematical literature on the continuity of functions,  specifically on the application of Rosenthal's 1955  theorem to  continuous relations (see below),  both aspects of his work are relevant for us.\fn{For Koopmans' work referred to here, see Footnote \ref{fn:koop} below. The reference to Rosenthal's theorem follows in Section \ref{sec:math} below.} 

Remaining with the {\it Essays,} but now shifting from the first to the second, Koopmans returns  in his masterful discussion of Herstein-Milnor \citeyearpar{hm53} and \cite{sa54} to the continuity postulate in the context of a \lq\lq single decision maker faced with uncertainty." In this discussion, he refers to the  \lq\lq  surprisingly far-reaching implications of simple postulates of consistency and sharpness of preferences concerning objects capable of a particular type of continuous variation," but again  masks the continuity property of preferences, herein {\it scalar continuity,}  in a postulate that exerts the existence of a  probability mixture equivalent to the one intermediate between three ranked    objects of choice, referred to as {\it prospects.}\fn{See Section 10 on pages 155 to 161 in \cite{ko57}.  The quote is taken from the last paragraph on page 159.  Also see Footnote 2 on page 155 for the relevant literature concerning the Herstein-Milnor paper.}   In his discussion of the representability problem,  he writes:

\bqu Herman Wold and Gerard Debreu have realized that the [assertion that] every complete ordering can be represented by an ordinal utility function needs proof, and have successively and with increasing generality formulated conditions, essentially of continuity, under which they prove that a complete ordering permits expression by a continuous utility function.\fn{See p. 19 in \cite{ko57}. In Footnote 2 on the same page the author cites the 1943 paper of Wold  and 1954 paper of Debreu.  The readers should note that the authors have rearranged this sentence hopefully without any loss or gain in meaning. Koopmans' emphasis on Wold  and Debreu is repeated in Arrow-Hahn \citeyearpar{ah71}.} \equ       
 
 \nt However, in this entire discussion there is no reference to \cite{ei41}, a  pioneering paper  only acknowledged as a precursor of Debreu fifteen years later in  \cite{ko72}.\fn{In his reproduction of Debreu's theorem [Proposition 1] on the sufficiency of connectedness of a choice set in a finite-dimensional Euclidean space,  Koopmans corrects his oversight by   observing that  \lq\lq Debreu credits a paper of \cite{ei41} as containing the mathematical essence of [his] Proposition 1."  Arrow-Hahn  \citeyearpar{ah71} cite Eilenberg, but in a dismissive way:  \lq\lq A very considerable generalization, based on a mathematical paper by \cite{ei41}, was achieved with the deeper methods by \cite{de54}."  We return to this in Section \ref{sec:math}. \label{fn:ah} }  More generally, it is  this omission of Eilenberg's  paper, as well as the   masking of the continuity postulate, both in the theory of competitive equilibrium and  in uncertainty theory as presented by  von Neumann-Morgenstern, that serves as  the point of departure for,  and motivational center of,  the results to follow.  There is no bridging between the two subjects, and they are confined to separate essays, a separation and a consequent obscuring that is seen here as a consequence of not beginning with the  primitive raw notion of continuity. In short, this essay  is an angled contribution to the theory,  its two arms are represented by the twin {\it omission} and {\it masking,}  each hitched to Koopmans' {\it Essays.}

After a section on mathematical underpinnings,  we discuss the omission in Section \ref{sec: results} below and turn to what we mean by the term \lq\lq masking."  This is really brought out 
     in Koopmans'  presentation of continuity as a non-satiation or an existence property.  The point is that  there is no single and unambiguous rendering of the continuity postulate: different notions of continuity have been used in economic and decision theory depending on the different needs and requirements of a particular result, as indeed they ought.   There is no standard continuity assumption even when differences in topologies are disregarded.  In terms of a further  unpackaging, 
  one can  discern  three types of continuity pertaining to a binary relation and labeled as   {\it graph} continuity,   {\it section} continuity, and  {\it scalar} continuity.    All  impose topological assumptions on the relation itself -- graph continuity on it  as a subset of the product space and   section continuity  on its sections --  but  while  the first two use the topological structure of the space on which the choice function or relation may be defined,  the third, in its reference to the {\it mixture-continuity,} and {\it Archimedean} properties,    works with an algebraic structure on the choice set, and   limits itself only to the topological structure of the unit interval.\fn{As we show in the sequel, {\it  Wold-continuity} and its various versions also need to be added to the mix, and they overlap with questions of {\it solvability;} see Figure 3 below.}   All  properties are widely used in economic theory,  but perhaps the last is more pervasive in decision theory, while the first two  constitute more of a staple of  Walrasian  general equilibrium and game-theoretic analyses.  The point that constantly guides this essay is that one has to go back to the most raw, primitive and  undefined, meaning of the term if the analytical and substantive connections between the subsequent conceptual proliferations are not to be missed.

With this preamble to the antecedent literature on 
 \lq\lq decision theory," we can turn to another dissonance and exclusion, that in the now-standard expositions of neoclassical theory of the consumer.    In seminal work on the theory of individual choice, \citet{wo43} and Wold-Jureen \citeyearpar{wj53},  also touch on the problem of the representation of a preference relation by a real-valued (utility) function in the context of a  finite-dimensional Euclidean space.  
However, unlike  \cite{ei41}, they work  with binary relations that are not necessarily anti-symmetric:, and  whose  indifference curves are therefore not singletons.   In their discussion of Wold's theorem,  Arrow-Hahn \citeyearpar[p. 106]{ah71} write: 
\bqu Since the introduction of indifference surfaces by Pareto
and Irving Fisher, it has been taken for granted that they could be
represented by a utility function. Wold seems to have been the first to
see the need of specifying assumptions under which the representation
by a continuous utility function exists.
\equ 
 In terms of bringing additional  mathematical (non-topological) registers to bear on the continuity postulate,  Wold's theorem for relations constitutes  another motivating result for this paper, and especially when coupled  with Rosenthal's 1955 theorem for functions.   Wold seems to be the first author to  formulate and use {\it scalar continuity} in economics in the framework of the consumer theory, and not only do Herstein and Milnor  not connect to him, but as stated above, Koopmans separates them in two different essays.  But even more than the  representation theorem itself, Wold's continuity assumption serves as a link between the results of Rosenthal and Herstein-Milnor, as well as articulating of a viewpoint that sees Debreu's generalization of Wold's work in the context of the criteria of  {\it inessentiality} in Kim-Richter \citeyearpar{kr86}, and  {\it hiddenness} and {\it redundancy} explicated in  Khan-Uyanik \citeyearpar{ku21et}.  
In the reference to Wold in his own study of the representation problem, 
  \cite{de54}  found his assumptions to be \lq\lq restrictive,"  and in particular,  singled out the Euclidean assumption, noting that to \lq\lq treat the problem in a more general frame involves no additional mathematical cost." In his comment on the antecedent literature, he  found only the \lq\lq particular case of the set of {\it prospects} as having received a \lq\lq rigorous and extensive investigation,"\fn{See paragraph 2 of \cite{de54} where he singles out von Neumann-Morgenstern \citeyearpar{vm47}, \cite{ma50} and Herstein-Milnor \citeyearpar{hm53}. We return to a reading of Debreu's work below.} and in his  subsequent treatment of the problem a decade later, he dropped any reference to Wold.

 In summary, adjectives change their meaning  depending on the noun they modify, and the   adjective in question here in this work is `continuous,' and is being used to modify different classes of functions and binary relations. In addition to the  usual variations involving topologies in use, `continuous' may also refer to   
(i) continuity defined on a choice space in terms of the topological structure of a possibly {\it distinct} parameter set, e.g. as in mixture continuity, or 
(ii) continuity restricted to  pertinent subsets  of the choice set, e.g. as in continuous on lines in a vector space or on paths in a path-connected space. Of particular import are the complementarities between different forms of continuity and the geometry of the individual functions or relations.\fn{ We thank a referee for language that helped us draft this  paragraph.  } 

With this introductory framing as the backdrop,  we are now in a position to spell out succinctly the substantive  contribution of the paper.  In a nutshell, we 
 introduce new  continuity concepts for relations by using the concepts developed for functions, and use them not only  to unify and simplify a range of results, but also  to bring to light novel equivalences hitherto unseen in the economic literature.  And perhaps most importantly, to see the neoclassical consumer theory as including decision theory now associated with the names of von Neumann, Morgenstern, Savage, Anscombe and Aumann.  To be sure, as brought out above, some of the the continuity concepts for functions are part of the historical record, linear continuity and Rosenthal's restricted continuity readily come to mind, 
  but the translation from functions to to relations has also yielded a new result for functions themselves; see, for example,  the very first theorem below.\fn{We are surprised that could not find any reference to the fact that  for
  a quasi-concave (or a quasi-convex) function, linear continuity is equivalent to its continuity!}  In Section \ref{sec:math}, we recall the theorems of Rosenthal and Eilenberg along with the Genocchi-Peano example.   In Section \ref{sec: results}, we present seven theorems, two propositions and a corollary: rather than repeat and summarize them -- they are already clean and clear enough in their statements\fn{We note for the reader that we use both propositions and theorems in this paper, and clarify our usage.  We adapt the following convention. We  use the {\it theorem} designation only for the equivalence results in   Section 3  and the {\it proposition} designation for partial results that provide only sufficient conditions.   In  Section 4  we do not use any theorem on the grounds that the application results presented in this section draw on those    in the theory section. } --  we can invite the reader to quickly see how they articulate of the basic motivations delineated above, and how they relate to previous work in economic theory. 
   In Section \ref{sec: discussion} we present seven propositions and three corollaries, and further discuss the implications of the results to the antecedent literature: these applications of the theory constitute the deliverables of the theory, applied theory so to speak.   Section  \ref{sec:conc} concludes the paper with a summary and some observations regarding future and open direction.   
   We reserve  Appendix \ref{sec: proofs} to the proofs of the results and the technical lemmas  they require and  Appendix \ref{sec_examples} to a battery of nine examples that close lacunae that may suggest themselves to an interested reader.

\section{Mathematical Antecedents} \label{sec:math}

\subsection{Eilenberg's Paper: A Conspicuous Omission}

 We begin with the omission:   rather than  the issue of priority, the importance of the Eilenberg paper lies in  its  substantial  consequences for  economic theory. First, as delineated in some detail in Khan-Uyanik (\citeyear{ku21et},  \citeyear{ku21ch}),  Eilenberg is the first to emphasize the behavioral consequences of what subsequently came to be regarded as mere technical assumptions made to ensure  tractability, and thereby a professional missing-out of  considerations  later seminally considered by   \cite{so65, so67} and \cite{sc71}. To elaborate,  \cite{ei41} showed that a decision-maker (DM) with a continuous preference relation defined over a topologically connected choice set, must of necessity be consistent if he or she  is strictly decisive, consistency being formalized by {\it transitivity,}  and strict decisiveness by {\it completeness} of a binary relation that is assumed in addition to be  anti-symmetric.   Sonnenschein  delineated circumstances  under which  Eilenberg's theorem can be generalized to binary relations that are not necessarily anti-symmetric. In a parallel result, Schmeidler  showed that under topological  circumstances identical to Eilenberg's, which is to say continuity and connectedness, a DM  must of necessity be  decisive if he or she is consistent.    Under the rubric of what is referred to as  the Eilenberg-Sonnenschein research program, these results have been comprehensively generalized and  integrated by the authors:\fn{In this, see Khan-Uyanik (\citeyear{ku21et}, \citeyear{ku20a}) Uyanik-Khan \citeyearpar{uk19b}, Galaabaatar-Khan-Uyanik \citeyearpar{gku19}  
and also  Giarlotta-Watson  \citeyearpar{gw20}.}  they underscore the behavioral consequences of technical topological properties, and most particularly, the behavioral consequences of the continuity postulate.

The second contribution of  Eilenberg's  paper is his theorem    
 that a continuous, anti-symmetric,  complete and transitive relation on a connected space can be represented by a continuous real-valued function.  Prompted by Halmos, \cite{de54, de64} reached back to Eilenberg in choosing as his general frame of reference a  topological space, albeit a preordered rather than an ordered one, but  in which the preorder was not restricted by any ancillary monotonicity assumptions particular to the Euclidean setting. 
He  extended Eilenberg's result to relations that are not necessarily anti-symmetric, and sharpened it to highlight connectedness,  separability and second-countability  assumptions on the set over which the given relation is defined, formally defining a notion of a {\it natural } topology as being one in which the weak sections of the order are closed, and focusing  on the quotient topology for  his generalization.\fn{It is worth pointing out that this generalization is straightforward observation, if not a trivial one: it is the second theorem, and the alternative proof through the ``open-gap lemma" that is the contribution of \cite{de54}; see \cite{de64} and the definitive  analysis of    
Beardon-Mehta \citeyearpar{bm94} and Beardon's expository essay in Bosi-Campi{\'o}n-Candeal-Indurain  \citeyearpar{betal20}.}   Debreu's extension  of Eilenberg's result has a clear and evident  parallel to Sonnenschein's extension of the first result, and has since emerged as a central  preoccupation of decision theory as pursued in economic theory literature.  To sum up, 
  Eilenberg puts his entire emphasis on the topological register,\fn{This topological aspect was zeroed in by  \cite{pe70} and subsequently by  \cite{le72},  \cite{me83b} and \cite{he95}. Indeed, the work has been integrated into mainstream mathematics through a series of equivalence theorems that trace  Cantor's fundamental papers as the source of rich  stream.  See \cite{he95} for an omnibus result that involves an equivalence of eleven theorems one being used to provide a proof of the other; also see the texts  Bridges-Mehta \citeyearpar{bm95} and Aleskerov-Monjardet \citeyearpar{ab02}. } -- what Arrow-Hahn refer to as the \lq\lq deeper methods"\fn{In their masterly overview of the literature, Arrow-Hahn refer to the \lq\lq very considerable generalization, based on a mathematical paper by Eilenberg, [that] was achieved with deeper methods by Debreu" in his reliance only on the topological notions of  the continuity of preferences and the connectedness of the choice space on which the preferences were defined.} --  and unlike Koopmans, metric considerations  are totally bypassed and rendered inconsequential.\fn{A tracking of the word \lq\lq distance" in \cite{ko57} yields rewards in this connection.} Second,   the fact that these, and other issues, are considered in a single work allows one to see how the meaning given to the  continuity postulate  migrates from issue to issue and problem to problem.

\subsection{Rosenthal's Theorem:  Suggested Directions} \label{sec: rosenthal}  A point worthy of appreciation is the fact that  even  the  move from a binary relation to the more primitive setting of a real-valued function does not yield an unambiguously single continuity postulate. The distinction 
between joint and  separate continuity in each individual  variable is to be sure a staple of the first course in real analysis, but {\it linear} continuity is perhaps less a part of the vernacular, at least in so far it is current in economics curriculum.  Defining a   real valued function on the plane to be {\it linearly} continuous when its restriction to a line is continuous,    Genocchi-Peano \citeyearpar[pp. 173--174]{gp84} provide the following example of a discontinuous function which is linearly continuous. 
 \medskip
 
\nt {\bf Example (Genocchi-Peano).} {\it 
The function $f:\Re^2\ra \Re$ defined below is linearly continuous but not jointly continuous. 
$$f(x)=\frac{2x_1x_2^2}{x_1^2+x_2^4} ~\text{ for } x\neq 0 \text{ and } f(0)=0.$$}
\vspace{-15pt}

\nt This basic  example    underscores  the fact that linear  continuity of a function  is not enough to obtain ``full continuity" on the entire domain.\fn{The example,  and along with the other  GP  examples more generally,   is an important benchmark of a rich trajectory dating to Cauchy in the early part of the $18^{th}$-century. The fact that  joint continuity is stronger than separate continuity was, even then in the time of Cauchy,  standard material in textbooks on multivariate calculus, but an investigation of the  relationship between  joint continuity and  restricted  continuity properties of a function  constituted a rich development to which many mathematicians, including Heine, Baire and Lebesgue, contributed; see for example \cite{gl13} and the  recent survey Ciesielski-Miller \citeyearpar{cm16}. }  
It  led one to ask for a concrete characterization of  family of subsets in the domain of a function that ensure full continuity from 
 continuity on the restricted domain.   The answer to this question was given by a remarkable result of  \citet{ro55} whose special case for the plane reads as follows:
 
 \begin{customthm}{(Rosenthal)}[{}]
 Every real-valued function is jointly continuous iff it is restrictedly continuous,  the restriction being any subset of the domain that can be represented as the graph of a smooth  function of   one variable in terms of the other. 
\label{thm: ro}
\end{customthm}

\nt In terms of a general Euclidean space, the theorem established that the joint continuity of a real-valued function follows from the continuity of the restriction of the function to the graphs of all smooth 
  curves in its domain.\fn{Note that in the functional form neither of the variables   have to be fixed and the dependent variable can be one or the other. Furthermore, we take this continuity idea of Rosenthal and present an analogous definition for binary relations; see Definition \ref{dfn: arccontinuity} below.  
  This is still admittedly loose in that we defer a formal definition of smoothness to the subsequent section and to the references in Footnote~\ref{fn:ro2} below.  Rosenthal restricts himself, for the plane, to the graph of a function that is both smooth and convex, and for the relation of convexity, see \citet[Theorem 1.3.4]{gl13}.   \label{fn:ro1}} 
  
    In any case, Rosenthal's theorem  is  the second important benchmark of the trajectory under consideration, and  it motivates  the deconstruction of the continuity postulate  pursued here. 
 Taking our cue from Rosenthal's invocation of a restricted domain to generate an equivalence theorem,  we  invoke algebraic,  order-theoretic and analytical structures to 
investigate  the relationships between the myriad variety of  continuity assumptions. 
  This is to say,  to investigate how  convexity, monotonicity and differentiability structures, separately and together, render  seemingly unrelated continuity assumptions equivalent for functions {\it and} relations. 
   While of intrinsic interest, it bears emphasis that our attempt at a comprehensive and unified treatment is motivated primarily for the  sharpening and the generalizing of  the behavioral consequences that are already scattered  in the literature: to identify resemblances and basic patterns obscured in a non-holistic view of the subject of choice theory.

\section{The Theory} \label{sec: results}

In this section we present  the pure theoretical part of the work in the form of seven theorems and two propositions collected in two subsections.  It provides results on the relationship between different continuity postulates on  functions and binary relations.    The three categorizations of the continuity of preferences we mention in the introduction are a first exploratory overview, bringing out the important that  the boundaries between them are not clear-cut and precise, and merit precise delineation.

\subsection{On the Continuity of Functions} \label{sec_func}

As emphasized in the introduction, \cite{ro55} is the first to face the full consequences of the GP examples and ask for a strengthening of the linear continuity postulate on a function that would guarantee its full continuity: it offers a remarkable and elegant  solution in terms of continuity of an {\it arbitrary} function on every smooth curve in a Euclidean space. In this paper we offer two results inspired by Rosenthal's theorem. The first goes in a direction opposite to his by asking for restrictions on the class of functions rather than the subsets of their domains,  for which linear and full continuity are identical; we present this result in this subsection.  The second proceeds in a direction identical to his but for binary relations rather than a function; we present this result in the next subsection. It is a little surprising that these questions have not been asked before even in the mathematical literature, but their interest for us lies in that they set the stage for the results to follow. 

We shall need the following notation for our first result. 

\smallskip

\nt {\bf (C)} {\it A set $X\subseteq \Re^n$ satisfies property {\bf C} if it is either  open or a polyhedron where a {\it polyhedron} is a subset of $\Re^n$ that  is an intersection of a finite number of closed half-spaces. }

\smallskip

We can now present our first result. 


\thm
Let $X\subseteq \Re^n$ be a convex set with property {\bf C}. Then every real-valued quasi-concave (or quasi-convex) function defined on $X$ is linearly continuous iff~\!\fn{The expression ``iff" denotes ``if and only if" hereafter.} it is jointly continuous. 
\label{thm: quasi-concave}
\thmm

\nt We note for emphasis that the theorem is false without the sufficient conditions it  assumes:
\ben[{\nf (i)}, topsep=3pt]
\setlength{\itemsep}{-1pt} 
\ml the quasi-concavity assumption cannot be dropped by virtue of the GP example;  
\ml   the polyhedron assumption in property {\bf C} cannot be replaced by an arbitrary convex set by  virtue of Example \ref{exm_polyhed} in Appendix \ref{sec_examples}.  
\een
 Property {\bf C} covers a wide range of applications in economics and mathematics, but it is still restrictive.  
 Hence, we provide the following  weaker property   and leave it to the reader to check that  the proof of any result in this paper  using property  {\bf C}  carries through verbatim  for a reformulated result using the weaker property  {\bf C}$'$.    
 \smallskip
 
\nt  {\bf (C$'$)}  {\it A set   $X\subseteq \Re^n$  satisfies property {\bf C$'$} if there 
exist finitely many  hyperplanes $\{h_j\}_{j=1}^m$  
in {\nf aff}$X$ such that   $X\backslash \text{\nf ri}X \subseteq \bigcup_{j=1}^m h_j$, where {\nf aff}$X$ denotes the affine hull of $X$ and {\nf ri}$X$ denotes  its relative interior.\fn{See Appendix \ref{sec: proofs} for the definitions of  the  affine hull and the relative interior of a set.} 
}
\smallskip

\nt Property {\bf  C$'$}  imposes restrictions only on those points which are both in $X$ and on its boundary in the affine hull of $X$. For example,  the following sets satisfy  property {\bf  C$'$} but fail  property {\bf C}: $X=\{x\in \Re^3|x_3=0, x_1^2+x_2^2<1\}$ and $X'=\{x\in \Re^2_+| x_1^2+x_2^2<1\}$. Note that $X$ is not open in $\Re^3$ but it is open in the affine hull of $X$. Similarly, $X'$ is neither open nor a polyhedron, but the points that are both in $X$ and on its (relative) boundary  are contained in the union of two hyperplanes in $X'$.

\begin{figure}[htb]
\begin{center}
 \includegraphics[width=5.5in, height=1.1in]{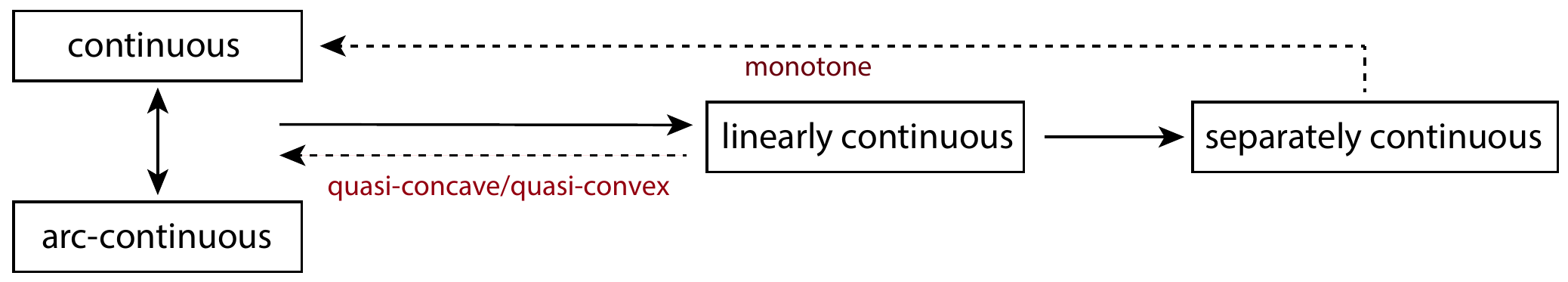}
\end{center}  
\vspace{-15pt}   
    
\caption{The Continuity Postulates on Functions}
\vspace{-5pt}   
   
  \label{fig: function}
\end{figure}

We now illustrate myriad relationships among the different continuity postulates on functions in Figure \ref{fig: function}. 

 \nt  The equivalence of joint-ontinuity and arc-continuity is used in \cite{ro55}; the first red implication using quasi-convexity (or quasi-concavity) of functions is proved in Theorem \ref{thm: quasi-concave} above, the second red implication using monotonicity of functions is due to \cite{yo10qjpam} and Kruse-Deely \citeyearpar{kd69amm}.\fn{See  also  Ciesielski-Miller \citeyearpar{cm16}   and  Ghosh-Khan-Uyanik \citeyearpar{gku20b} for implications of the monotonicity assumption for the relationship between separate and full continuity of functions and binary relations.} 
 
It is well-known that a concave (or a convex)  function on a non-empty, open and convex subset of $\mathbb R^n$ is  continuous. We next show that Theorem \ref{thm: quasi-concave} above can be used to provide an alternative proof of this result.\fn{We thank an anonymous referee for pointing out this direction.}    First, we show that 

\prp
Every concave (or a convex) real-valued  function on a non-empty, open and convex subset of $\mathbb R^n$ is linearly continuous. 
\label{thm_conc_lc}
\prpp

\nt Since every concave (convex) function is quasi-concave (quasi-convex),  the following is a direct corollary of Theorem \ref{thm: quasi-concave} and Proposition \ref{thm_conc_lc}. 

\cor
Every  real-valued  concave (or a convex) function on a non-empty, open and convex subset of $\mathbb R^n$ is  continuous. 
\label{thm_conc_c}
\corr

  We end this subsection by a remark on mid-point convexity for functions and   relations; see for example Herstein-Milnor \citeyearpar{hm53} and \cite{je67} for preferences, and \cite{be92op} and \citet{ku09book} for functions. The first two references already testify to the fact that the continuity and the convexity postulates have a certain complementary.  
It is also well-known that mid-point convexity of a function is equivalent to full convexity under the continuity postulate.  It is an interesting question whether,   under various versions of continuity  mid-point convexity  (mid-point versions of  concavity,  quasi-convexity and  quasi-concavity) and its generalizations  are equivalent to convexity (concavity,  quasi-convexity and  quasi-concavity). Further complementarities between various versions of convexity and  continuity  postulates are interesting and worth investigating. 


Our results require that function to be quasi-concave or quasi-convex on the entire domain. Index-wise convexity (concavity) of functions has important applications; for example Fan in his famous minimax theorem assumes the function of two variables  satisfies a version of convexity in one index and a version of concavity on the other.\fn{We are indebted to a referee for this interesting and important connection.}    This type of convexity postulate can be considered as  {\it separate-convexity}, analogous to the separate-continuity concept. Taking our cue from the distinction between continuity and separate continuity, global and individual continuity, it is natural to ask global and individual convexity. We can further ask, analogous to the mid-point vs full convexity, when separate convexity is equivalent to full convexity.   We show by providing two counterexamples  that index-wise quasi-convexity, or quasi-concavity, is not enough to obtain the equivalence of linear and full continuity of a function or a binary relation; see Examples  8 and 9 in the Appendix.  It is an interesting question to further investigate this direction, especially studying conditions guaranteeing full convexity of separately convex functions and topological properties of functions  that  are  quasi-convex in some indices and quasi-concave in the others.  

%

\subsection{On the Continuity of Relations} \label{sec: relc}

 
 It is by now well understood that  graph continuity  implies section continuity; for complete and transitive relations, they are equivalent, however, for incomplete and non-transitive relations, further technical or behavioral assumptions on preferences are required in order to obtain the equivalence.\fn{See  \citet{sc69} and  Bergstrom-Parks-Rader \citeyearpar{bpr76}  
   for equivalence results for strict preferences; \citet{wa54a}, \citet{de68}, \citet[Lemma]{sh74} and \citet{ge15} for weak preferences; and  \citet{ge13} for a comprehensive treatment for both weak and strict preferences.}   In this paper our emphasis is not on the investigation on this relationship.   Pioneering treatment on the continuity assumptions are introduced in \cite{bi48}, \cite{na65} and \cite{vi64}.  
 Section continuity  of preferences implies certain scalar continuity postulates;  \citet{in10} illustrates that the example of GP shows that mixture-continuity is weaker than section continuity.\fn{Similarly, Neuefeind-Trockel \citeyearpar{nt95} provide an example for infinite-dimensional spaces; see Footnote 2 in their Section \ref{sec: results} and the text it footnotes. Also, see Footnote \ref{fn_nt} below for details.}  The relationship between the two are recently picked up in decision theory.  It has been shown that under independence, or betweenness, assumptions, scalar continuity of a transitive preference relation is equivalent to its section continuity.\fn{See, for example,  Dubra-Maccheroni-Ok \citeyearpar[Proposition 1]{dmo04},  Gilboa-Maccheroni-Marinacci-Schmeidler \citeyearpar[Lemma 3]{gmms10}, \cite{du11} and Karni-Safra \citeyearpar{ks15}. We return to these references below.} We investigate the relationship between section and scalar continuity postulates and introduce new continuity notions that are motivated by the literature in mathematics on the continuity of functions. 
 
 Let $X$ be a set. A subset $\succcurlyeq$ of $X\times X$ denote a {\it binary relation} on $X.$ We denote an element $(x,y)\in ~\!\!\! \succcurlyeq$ as $x\succcurlyeq y.$ The {\it  asymmetric part} $\succ$ of $\succcurlyeq$ is defined as $x\succ y$ if $x\succcurlyeq y$ and  $y\not\succcurlyeq x$, and its {\it symmetric part} $\sim$ is defined as $x\sim y$ if $x\succcurlyeq y$ and $y\succcurlyeq x.$   We call $x\bowtie y$ if $x\not\succcurlyeq y$ and $y\not\succcurlyeq x$. The inverse of $\succcurlyeq$ is defined as  $x\preccurlyeq y$ if $y\succcurlyeq x$. Its asymmetric part $\prec$ is defined analogously and its symmetric part is $\sim$.  We provide the descriptive adjectives pertaining to a relation in a tabular form for the reader's convenience in the table below. 
\vspace{5pt}

\begin{table}[ht]
\begin{center}
\begin{tabular}{lll} 
\hline  \noalign{\vskip 1mm}     
{\it reflexive}   &$x\succcurlyeq x$ $\forall x\in X$\\
{\it non-trivial}   &  $\exists x,y\in X$ such that $x\succ y$\\  
{\it complete}  & $x\succcurlyeq y$ or $y\succcurlyeq x$ $\forall x,y\in X$\\ 
{\it symmetric}  & $x\succcurlyeq y \Rightarrow y\succcurlyeq x$  $\forall x,y\in X$\\ 
{\it asymmetric} & $x\succcurlyeq y\Rightarrow y\not\succcurlyeq x$ $\forall x,y\in X$\\ 
{\it anti-symmetric}  &  $x\succcurlyeq y\succcurlyeq x \Rightarrow x=y$ $\forall x,y\in X$ \\ 
{\it transitive}  & $x\succcurlyeq y\succcurlyeq z \Rightarrow x\succcurlyeq z$ $\forall x,y,z\in X$\\ 
{\it negatively transitive} & $x\not\succcurlyeq y\not\succcurlyeq z \Rightarrow x\not\succcurlyeq z$  $\forall x,y,z\in X$ \\
  {\it semi-transitive}  & $x\succ y\sim z \Rightarrow x\succ z$ and  $x\sim y\succ z \Rightarrow x\succ z$ $\forall x,y,z\in X$ \\ 
\hline
\end{tabular}
\end{center}
\vspace{-14pt}

\caption{Properties of Binary Relations}
\label{tbl: relation}
\end{table}  

We now define ``section continuity,'' ``graph continuity'' and ``restriction continuity'' of a binary relation. 
Let $\succcurlyeq$ be a binary relation  on a set $X$ that is endowed with a topology. For any $x\in X$, let    
 $
A_\succcurlyeq(x)=\{y\in X| y\succcurlyeq x\}$ denote the {\it upper section} of $\succcurlyeq$ at $x$,  and  $A_\preccurlyeq(x)=\{y\in X| y\preccurlyeq x\}$ its {\it lower section} at $x$.    
%
 The relation $\succcurlyeq$ has {\it closed (open) graph} if it is closed (open) in the product space $X\times X$; has {\it closed (open) upper sections} if it has closed upper sections; has {\it closed (open) lower sections} if it has closed lower sections;  and has {closed (open) sections} if it has closed (open) upper and lower sections.  
Moreover, $\succcurlyeq$  is {\it graph continuous} if it has closed graph and its asymmetric part $\succ$ has open graph, and it is    
{\it continuous} if it has closed sections and its asymmetric part $\succ$ has open sections.   
 %
 For any subset $S$ of $X$, let  
 $\succcurlyeq \upharpoonright \!\! S $
 denote the {\it restriction} of $\succcurlyeq$ to $S$ and defined as follows: for all $x\in X$, not just $x\in S$, 
 $$A_{\succcurlyeq \upharpoonright S}(x)=A_{\succcurlyeq}(x)\cap S,  A_{\preccurlyeq \upharpoonright  S}(x)=A_{\preccurlyeq}(x)\cap S, A_{\succ \upharpoonright  S}(x)=A_{\succ}(x)\cap S, A_{\prec \upharpoonright  S}(x)=A_{\prec}(x)\cap S.
 $$  Notice that the restricted relation excludes the complement of $S$ from the upper and lower sections of the relation. 
 This definition of  restricted relation is different than the alternative requirement that $\succcurlyeq \upharpoonright \!\! S  =~\!\!\!\! \succcurlyeq \cap ~(S \times S)$,  and is natural in our context since the usual continuity of a binary relation is defined by using its upper and lower sections. Given a set $S\subseteq X$, we want to   focus on the restriction of the better-than- and worse-than-sets of  {\it all}  $x$ in $X$ on the given set $S$, not only of  those $x$ contained in $S$.\fn{Example \ref{exm_restriction} in Appendix \ref{sec_examples} shows the crucial aspect of our definition. Moreover,   both versions of restricted relations are used in mathematics; see for example \citet[Section 3]{re08bulletin}.  \label{fn_restricted}  }
 Finally, for any $S\subseteq X$, the relation $\succcurlyeq$ is {\it restricted continuous with respect to $S$} if $\succcurlyeq \upharpoonright \!\! S$ is continuous, that is for all $x\in X$,  $A_\succcurlyeq(x)\cap S$ and $A_\preccurlyeq(x)\cap S$ are closed (in $S$) and $A_\succ(x)\cap S$ and $A_\prec(x)\cap S$ are open (in $S$).


%
%

We now turn to ``scalar continuity'' properties of a  relation $\succcurlyeq$ on a convex subset $X$  of a vector space  which is based on the  topology of the unit interval.       
 For all $x,y\in X$ and $\lambda\in [0,1]$, let $x\lambda y$ denote $\lambda x+ (1-\lambda)y$.  
%
 The relation $\succcurlyeq$  is  
{\it upper  $\left(\text{lower}\right)$ mixture-continuous} if $x,y,z\in X$ implies the set $\{\lambda| x\lambda y\succcurlyeq z\}$  $\left(\{\lambda| x\lambda y\preccurlyeq z\} \right)$ is closed;  
    {\it upper $\left(\text{lower}\right)$ Archimedean} if $x,y,z\in X, x\succ y$ implies that there exists $\lambda\in (0,1)$ $\left( \delta\in (0,1)\right)$  such that $x\lambda z\succ y$ $\left(x\succ y\delta z\right)$; and  
{\it upper  $\left(\text{lower}\right)$ strict-Archimedean} if $x,y,z\in X$ implies the set $\{\lambda| x\lambda y\succ z\}$  $\left(\{\lambda| x\lambda y\prec z\} \right)$ is open.   
Moreover, $\succcurlyeq$ is {\it mixture-continuous} if it is upper and lower mixture-continuous. The {\it Archimedean}\fn{The Archimedean property in this paper is slightly stronger than the standard Archimedean property used in the literature: {\it for all $x,y,z\in\CS$ with $x\succ y \succ z$, there exists $\lambda, \delta \in (0,1)$  such that $x\lambda z\succ y\succ x\delta z$ (or there exists $\lambda \in (0,1)$  such that $x\lambda z\sim y$ {\nf as presented in Yokoyama)}.}   However, for complete and transitive relations,  any of independence or strong monotonicity is enough to establish the equivalence between these two conditions. }  and {\it strict-Archimedean} properties are defined analogously.\fn{ Strict-Archimedean property is stronger than Archimedean property, and the two properties are equivalent under mixture-continuity; see Galaabaatar-Khan-Uyanik \citeyearpar[Proposition 1]{gku19}. } 
 The relation $\succcurlyeq$ is  {\it order dense} if $x\succ y$ implies that there exists $z$ such that $x\succ y\succ z$.  
 The relation $\succcurlyeq$ is  {\it weakly Wold-continuous} if {\nf (i)} it is order dense and {\nf (ii)} $x\succ z\succ y$ implies that the straight line\fn{A {\it straight line in $X$} is defined as the intersection of a one dimensional subset of aff$X$ and $X$. 
 \label{fn: straight_line}} joining $x$ to $y$ meets the indifference class of $z$.\fn{Note that the second part of weak Wold-continuity is equivalent to the {\it solvability} assumption in measurement theory; see for example \cite{ma50} and Krantz-Luce-Suppes-Tversky \citeyearpar{klst71}.} Now, assume $X$ contained in a Euclidean space.  The relation $\succcurlyeq$ is  {\it Wold-continuous}  if {\nf (i)} it is order dense and {\nf (ii)} $x\succ z\succ y$ implies that any curve\fn{A {\it curve} on $X$ is the range of a continuous injective function $m:[0,1]\ra X$; see page 14 for details.} joining $x$ to $y$ meets the indifference class of $z$.  

We now define a restriction continuity concept for preferences that is analogous to linear continuity of a function.

\df
A binary relation on a convex subset $X$ of a Euclidean space is linearly continuous if its restriction to any straight line in $X$ is continuous. 
\label{dfn: lcontinuity}
\dff


\nt The following result illustrates that linear continuity of a relation is not a mere mathematical curiosity  by showing that it is equivalent to mixture-continuity and Archimedean postulates.

\thm		
A binary relation on a convex subset of a Euclidean space	 is linearly continuous iff it is mixture-continuous and Archimedean.
\label{thm: linearcontinuous}
\thmm

Next we provide convexity  properties of a binary relation.     
Let $\succcurlyeq$ be a binary relation on a  convex subset $X$ of a vector space.\fn{Diewert-Avriel-Zang \citeyearpar{daz81} discuss different convexity assumptions on functions.}  The relation $\succcurlyeq$ is  {\it convex} if it has convex upper sections, and it {\it  demonstrates convex indifference} if for all $x,y\in X$   and all $\lambda\in (0,1)$  $~x\sim y$ implies $x\sim x\lambda y$. 
 %
A subset $Y$ of a topological vector space is {\it locally convex} if each point of the closure of $Y$, denoted by cl$Y$,   has a convex open  neighborhood whose intersection with $Y$ is convex. It is clear that every convex set is locally convex. However, the converse is not true since every finite subset of the space is locally  convex but it is not convex.\fn{See Example \ref{exm: localconvexity} and the paragraph previous to it in Appendix \ref{sec_examples} for a discussion of  local convexity.}  A relation $\succcurlyeq$ on $Y$ is {\it locally convex} if it has locally convex upper sections.

 We now provide monotonicity properties of a binary relation. 
 Let $\succcurlyeq$ be a  binary relation on a subset $X$ of $\Re^n$. The relation  $\succcurlyeq$ is {\it strongly monotonic} if   $x> y$ implies\fn{For vectors $x$ and $y$, ``$x\geq y$" means $x_i\geq y_i$  in every component;  ``$x> y$" means $x\geq y$ and  $x\neq y$; and ``$x\gg y$" means $x_i>y_i$  in every component.} $x\succ y$ and {\it weakly monotonic} if for all $x,y\in X$, $x>y$ implies  $x\succcurlyeq y$.  We call a subset $A$ of $X$ is {\it bounded by $\succcurlyeq$}   if for all $x,y\in A$ there exists $a,b\in X$ such that $a\succcurlyeq x,y$ and $x,y\succcurlyeq b$.

\smallskip

\nt {\bf (B)} {\it A set $X\subseteq \Re^n$ satisfies  property {\bf B} if it is  bounded by the usual relation $\geqq$.} 

\smallskip


%

%
%
%

We now present an equivalence result among seven of the continuity postulates we introduced above. It is an analogue of Theorem \ref{thm: quasi-concave}  above for preferences.  


\thm
Let $\succcurlyeq$ be a complete and transitive relation  on a convex subset of a Euclidean space with (i) property {\bf C} and  $\succcurlyeq$ is convex, or (ii) property {\bf B}  and $\succcurlyeq$ is weakly monotonic. Then the following  seven continuity postulates for $\succcurlyeq$ are equivalent: graph continuous, 
continuous,  
linearly continuous,  
mixture-continuous, 
Archimedean,   
Wold-continuous, and    
weakly Wold-continuous.  
\label{thm: ccontinuous}
\thmm

\nt  We note for emphasis that the theorem is false if properties {\bf B} and {\bf C} fail to hold;  see Examples \ref{exm_bdd} and \ref{exm_polyhed}  in Appendix \ref{sec_examples}.  
 Property {\bf C} holds for a  simplex and the non-negative orthant in $\Re^n$, the former is pervasive in decision theory and the latter in consumer theory, with convexity of preferences a common assumption in  both literatures.   
 As to property {\bf B},     when monotonicity of preferences is imposed (with or without convexity), the lattice structure is also assumed;  see \cite{gu92} in decision theory and \cite{sc69} in Walrasian economies where property {\bf B} holds.    
Since properties {\bf B} and {\bf C} are used as substitutes in the theorem, and are not required to hold simultaneously, and the theorem can cover  a wide range of models in economics.  
This being said, we note that  properties {\bf B}, {\bf C}, and {\bf C$'$}, the weakening of the latter introduced in Section \ref{sec_func} above, are still restrictive. For example, a simplex is a common assumption in decision theory which satisfies property {\bf C} but fails property {\bf B}. If a version of the  convexity assumption, such as independence and  betweenness, is not imposed in these models,  then the theorem above does not apply even if  the preference relation is assumed to be monotone.\fn{This paragraph is in response to a comment of an anonymous referee.}  Hence, the generalization of the results by weakening the assumptions on the choice sets is worthwhile and interesting.\fn{A referee provided an interesting example showing how the convexity assumption can rule out  first-order stochastic dominance, surely another pervasive criterion  in decision theory. Insofar  as our discussion reads  as an appeal to generality, our claims must be muted by this fact. We leave  further consideration of the convexity  postulate to future work.  }


%

Theorem \ref{thm: ccontinuous} is a strong integration of continuity postulates. We next provide a result that first completely deconstruct the postulate by dropping all assumptions  on preferences, and then a partial integration of the postulates by adding  completeness and transitivity assumptions. It shows that some of the relationships in Theorem \ref{thm: ccontinuous} can be recovered.

\prp
For any relation on a convex subset of a Euclidean space, only the following relationships hold among the following continuity postulates. 
 \ben[{\nf (a)}, topsep=3pt]
\setlength{\itemsep}{-1pt} 
\ml    Graph continuity $\Rightarrow$ continuity $\Rightarrow$ linear continuity $\Leftrightarrow$ mixture-continuity {\nf \&} Archimedean $\Rightarrow$ strict-Archimedean $\Rightarrow$Archimedean.  
\ml  Wold-continuity $\Rightarrow$ weak Wold-continuity.
\een
If the relation is complete and transitive, then only the following additional relationships hold. 
 \ben[{\nf (c)}, topsep=3pt]
\setlength{\itemsep}{-1pt} 
\ml    Continuity $\Rightarrow$ graph continuity   {\nf \&}  Wold-continuity.  

\ml[{\nf (d)}]  Strict-Archimedean $\Rightarrow$ mixture-continuity  $\Rightarrow$ Archimedean {\nf \&} weak Wold-continuity.  

\ml [{\nf (e)}]  Weak Wold-continuity $\Rightarrow$  Archimedean.  

\een
\label{thm: continuity0}
\prpp 

\nt Please note the use of the word ``only" in Proposition \ref{thm: continuity0}.\fn{Note that the Archimedean postulate or the other assumptions do  not imply the continuity postulate.  To belabour the point, there are no other relationships in addition to those illustrated in Proposition  \ref{thm: continuity0} without imposing additional  assumptions on the choice set or on the binary relation.  }    The thrust of this paper is the complementarity between different postulates, especially  between convexity, monotonicity, and the continuity postulates. 
In principle, by using  the nine continuity postulates  individually and collectively  presented in Proposition \ref{thm: continuity0}, one can obtain many  results  on the relationship between different collections of these continuity postulates. These results follow from this proposition.    

%
%
%
%
%
%
%
We next present a result establishing the equivalence between continuity and linear continuity of a relation which is not necessarily complete or transitive under different convexity assumptions.      


\thm
Let $\succcurlyeq$ be a binary relation on a convex set in $\Re^n$ with property {\bf C}.  
 Then $\succcurlyeq$ is linearly continuous iff it is continuous,  provided that any of the following holds. 
\ben[{\nf (a)}, topsep=3pt]
\setlength{\itemsep}{-1pt} 

\ml $\succcurlyeq$ is complete, and $\succcurlyeq$ and $\succ$ have locally convex upper sections,  \label{it: completeconvex}
\ml  $\succcurlyeq$ and $\succ$ have locally convex sections,  \label{it: convexconcave}
\ml  $\succcurlyeq$ is reflexive and demonstrates convex indifference, and  $\sim$ is transitive.  \label{it: reflexivelinear}

\een
\label{thm: linearcontinuity}
\thmm

 Figure   \ref{fig: relation_full} below illustrates the relationship among different continuity postulates on binary  relations.  Panel (a) illustrates the  partial relationships among the continuity postulates based on Proposition \ref{thm: continuity0}, and  Panel (b)    further  relationships in the presence of convexity or monotonicity assumptions based on Theorems \ref{thm: ccontinuous} and \ref{thm: linearcontinuity}. 


\begin{figure}[ht!]
\begin{subfigure}{1\textwidth}
  \centering
  \includegraphics[width=1\linewidth]{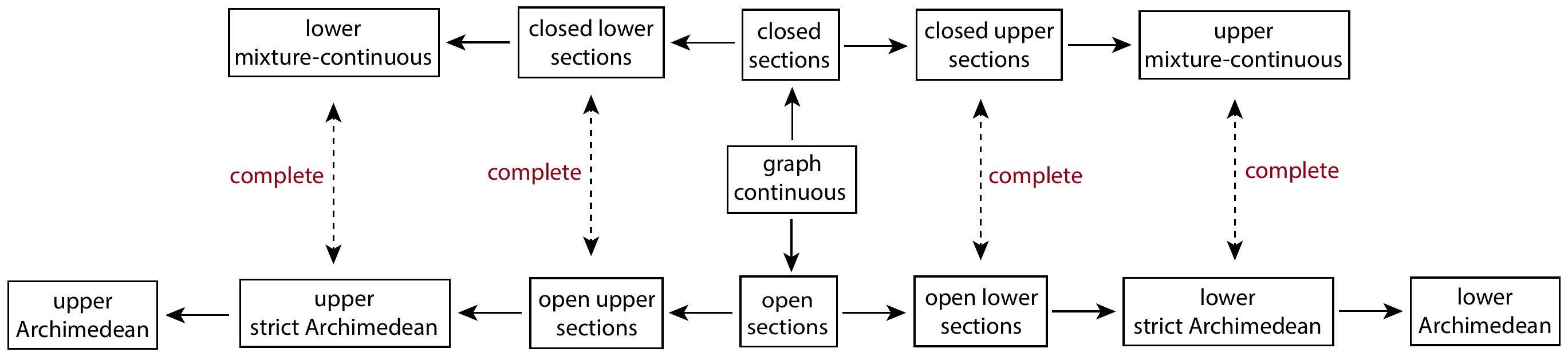}  
  \caption{}
  \label{fig:sub-first}
\end{subfigure}
\newline
\newline

\begin{subfigure}{1\textwidth}
  \centering
  \includegraphics[width=0.87\linewidth]{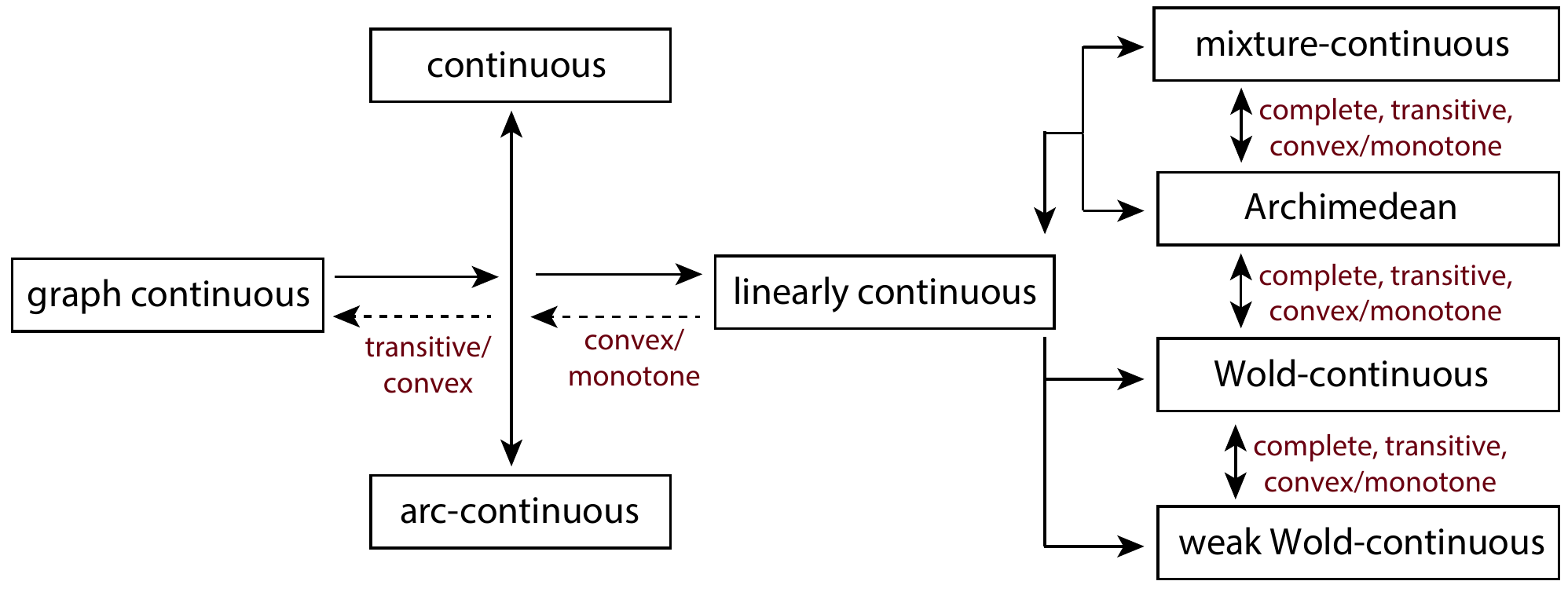} 
  \caption{}
  \label{fig:sub-second}
\end{subfigure}
\caption{The Continuity Postulates on  Relations}
  \label{fig: relation_full}
\end{figure}


%



Although independence axiom has been commonly used in decision theory,  additivity is also an important convexity assumption. For example \cites{de74} axiomatic decision theory model assumes the decision-maker has additive preference relation; see also Blackwell-Girshick \citeyearpar{bg54}, \citet{wa88a, wa88b, wa89, wa93} and Neuefeind-Trockel \citeyearpar{nt95}.   
%
  %
We next present a result which provides a further deconstruction of the continuity postulate by showing that upper and lower scalar continuity properties are equivalent for additive relations.\fn{Theorem \ref{thm: mcarc} below shows that lower  scalar continuity is equivalent to upper scalar continuity under the additivity assumption on a convex cone.    
 McCarthy-Mikkola  \citeyearpar{mm18} provide analogous equivalence results for an independent, reflexive and transitive relation on a convex set. Theorem \ref{thm: mcarc} generalizes their result by dropping the transitivity and reflexivity assumption and weaken the independence assumptions.       
 The works of  Neuefeind-Trockel  \citeyearpar[Proposition]{nt95}, Candeal-Haro-Indur\' ain-Eraso \citeyearpar[Theorem]{ci95} and \citet[Theorem 1]{ge15} provide equivalence results between lower and upper section continuity of a relation.  Theorem \ref{thm: mcarc} provides an analogous result for scalar continuity. Example \ref{exm: ac}  in Appendix \ref{sec_examples} illustrates that neither our result implies the results in  Neuefeind-Trockel and Gerasimou, nor their results imply ours.}   
  Before that we need the following concepts. A {\it convex cone} is a subset $X$ of a vector space such that $\alpha x + \beta y\in X$ for all scalars $\alpha,\beta>0$ and all $x, y \in X$.  
A binary relation  $\succcurlyeq$ on a convex cone $X$ is   {\it additive} if for all $x,y,z\in X$,  $x\succcurlyeq y$ iff\fn{When the set $X$ is a vector space, then ``iff" can be equivalently replaced by ``if".} $x+z\succcurlyeq y+z$.  
%


\thm
Let $\succcurlyeq$ be an additive binary relation on a convex cone. Then
\ben[{\nf (a)}, topsep=3pt]
\setlength{\itemsep}{-1pt} 

\ml $\succcurlyeq$ is upper mixture-continuous iff it is lower mixture-continuous iff it  is mixture-continuous,
 \ml $\succ$ is upper strict-Archimedean iff it is lower strict-Archimedean iff it is strict-Archimedean.
\een
\label{thm: mcarc}
\thmm


The following result shows that some of the equivalence results pertaining to scalar continuity postulates in Theorem \ref{thm: ccontinuous} do not require finite-dimensionality or a topological structure on the choice set.  

\thm
For every convex, complete and transitive relation on a convex subset of a vector space, the following continuity postulates are equivalent: Archimedean,  
strict Archimedean, mixture-continuous and weakly Wold-continuous. 
\label{thm: scontinuous}
\thmm

It is easy to show that the result above is true for mixture sets. 


Next, we proceed in a direction identical to Rosenthal's by asking for a condition on an {\it arbitrary} binary   relation that guarantees  its continuity. We show that  continuity along every smooth curve is enough for full continuity.  The question of course is the kind of continuity that is at issue. 
 The GP example was recently picked up (without citation) by \citet[Example 1]{in10} in order to illustrate that mixture-continuity is weaker than the continuity of a complete and transitive relation by showing that the relation induced by the function in GP example is mixture-continuous but not continuous.  The interesting question of course is what condition works? Before we develop this, it is well to put down some basic terminology.

We now define continuity of preferences restricted to continuously differentiable curves.      
An {\it arc} in $\Re^n$ is a continuous injective function $m:[0,1]\ra \Re^n$ where $m(\lambda)=(m_1(\lambda), \ldots, m_n(\lambda))$. An arc is called {\it smooth} if  $m_i$ is continuously differentiable for all $i$ and  $m'(\lambda)=(m_1'(\lambda), \ldots, m_n'(\lambda))\neq 0$ for all $\lambda\in [0,1]$. A {\it curve} in $\Re^n$ is the range of an arc and a {\it smooth curve} is the range of a smooth arc.      
 Since an arc $m$ is continuous and injective, it is a bijection from $[0,1]$ to its image $m([0,1])$. Since $[0,1]$ is compact and $m([0,1])$ is Hausdorff, $m$ is an homeomorphism between $[0,1]$ and $m([0,1])$; see for example \citet[Theorem 2.1, p. 226]{du66}. Therefore, 
 $[0,1]$ and the curve induced by an arc  are homeomorphic.\fn{Note that an arc induces a unique curve but a curve can be induced by distinct arcs. For example, any closed segment $[x,y]$ of the diagonal in $\Re^2$ is induced by every arc with $m_1(\lambda)=m_2(\lambda)$ for all $\lambda\in [0,1]$ where $m_1(0)=x, m_1(1)=y$. } 
 %

  %
%

\df
A binary relation on a subset $X$ of a Euclidean space is arc-continuous if its restriction to any smooth curve in $X$ is continuous. 
\label{dfn: arccontinuity}
\dff

We next introduce stronger versions of mixture-continuity and Archimedean properties by using smooth arcs instead of straight lines.    
The following notation for subsets of $[0,1]$ is useful. Let $\CM$ be the set of all smooth arcs. For all $m\in \CM$, when the values $m(1)=x$ and $m(0)=y$ are of interest,  we use the notation $m_{xy}(\lambda)$ for $m(\lambda)$.   
 Define the set of all smooth curves induced by the set of smooth arcs as $\CC_\CM$.          For any binary relation   $\succcurlyeq$ on  a convex set $X$, any $x\in X$ and any function $f:[0,1]\ra X$, define   
\begin{equation}
\begin{array}{c}
I_\succcurlyeq(f,x)=  \{\lambda~|~f(\lambda) \succcurlyeq x\} ~\text{ and }~ I_\preccurlyeq(f,x)= \{\lambda~|~x\succcurlyeq f(\lambda) \}.
\end{array}
\label{eq: Iset}
\end{equation}
%
A binary relation $\succcurlyeq$ on a convex subset $X$ of $\Re^n$ is {\it strongly mixture-continuous} if for all $x,y,z\in X$ and all $m_{xy}\in \CM$, the sets $I_\succcurlyeq(m_{xy},z)$ and $I_\preccurlyeq(m_{xy},z)$ are closed,  
 is {\it strongly Archimedean} if for all $x,y,z\in X$ with $x\succ y$ and all  $m_{xz}, m_{yz}\in \CM$, there exists $\lambda, \delta \in (0,1)$  such that $m_{xz}(\lambda)\succ y$ and   $x\succ m_{yz}(\delta)$ and  
is {\it strongly strict-Archimedean} if for all $x,y,z\in X$ and all $m_{xy}\in \CM$, the sets $I_\succ(m_{xy},z)$ and $I_\prec(m_{xy},z)$ are open.    
 %
 
 
\thm
Let $\succcurlyeq$ be a binary relation on a convex set in $\Re^n$ with property {\bf C}.  Then $\succcurlyeq$ is arc-continuous  iff  it is continuous iff it is strongly mixture-continuous and strongly Archimedean. 
\label{thm: curverelation}
\thmm
\vspace{-15pt}

\nt    Rosenthal's theorem characterizes the family of subsets such that restriction continuity of an arbitrary function is equivalent to its continuity.   Theorem \ref{thm: curverelation} generalizes Rosenthal's result to binary relations by showing that continuity of an arbitrary  binary relation is equivalent to its restriction continuity on smooth curves.\fn{Rosenthal's theorem uses simple arcs which is a subset of the class of smooth curves. It is easy to observe that our results hold for simple arcs. We present our result by using smooth curves for the exposition purposes which is by now standard in this literature.  See Young-Young \citeyearpar{yy10}, \citet{ke43},  \citet{ro55} 
 and Ciesielski-Miller  \citeyearpar{cm16
 } for the deconstruction of the continuity postulate for functions; also Footnote~\ref{fn:ro1} above.    \label{fn:ro2}}

\section{The  Theory Applied}  \label{sec: discussion}
This section takes to opportunity  to connect and discuss the contributions of our results on the continuity of preferences and  functions on the antecedent literature in economics.
 Theorem \ref{thm: ccontinuous} of the previous section brought out  the many different meanings -- seven to be precise -- that can be given to the 
 continuity postulate, and that in so doing,  give it various colorings  that, in conjunction with the eight other results,  thereby also  impinge on the postulate's  form and content.  We now ask for the deliverables of these results in so far  as applications are concerned. Our discussion is sectioned into four parts: ranging from representability, to  neoclassical consumer theory and its incorporation into Walrasian equilibrium theory, to the behavioral implications of the transitivity and continuity postulates and finally,  some of what still remains to be done.

%

\subsection{Representation of Preferences: Benchmark Results}

 In applying the theorems presented in Section \ref{sec: results}, we make three introductory observations: (i) whereas scalar continuity assumptions have been mainly used in choice theory under uncertainty regarding  cardinal representation of preferences, Wold had already relied on them  in his treatment of non-stochastic consumer theory, (ii) in Walrasian general equilibrium theory, scalar continuity is dropped and section continuity is used.  The scalar continuity assumptions, by themselves, are weaker than section continuity. In decision theory, consumer theory and Walrasian economies, either convexity or monotonicity assumptions are imposed on preferences. We show in this section that the distinct use of continuity postulates  is inessential -- in the presence of the other assumptions of the model, scalar and section continuity assumptions are identical. It is interesting that this relationship is not asked in consumer theory and Walrasian economies. There are some scattered special partial equivalences in decision theory,\fn{See Gilboa-Maccheroni-Marinacci-Schmeidler \citeyearpar{gmms10}, \citet{du11} and Karni-Safra \citeyearpar{ks15}.} yet a comprehensive analysis, which connects different literatures and results,  is missing.  We wish to emphasize that the following categories are not mutually exclusive.

In this subsection, we present three corollaries and three propositions. The first corollary  illustrates that any of the eight continuity postulates can be equivalently used in Wold's representation theorem, the second shows that in the expected utility theorem of  Herstein-Milnor \citeyearpar{hm53}, mixture continuity can be replaced with weak Wold-continuity, and the third obtains the expected utility theorem of Neuefeind-Trockel  \citeyearpar{nt95} as a corollary of our results.  The  first proposition establishes the link between additivity and independence postulates, the second obtains an expected utility representation theorem under additivity postulate and and the last shows that the section continuity assumption of \cite{gu92} can be replaced by mixture-continuity.

%
%
%

\subsubsection{An Euclidean Space: \cite{wo43}}

\cite{wo43} in his seminal paper proved a utility representation theorem. The following is a direct corollary of Wold's theorem and Theorems \ref{thm: ccontinuous} and \ref{thm: curverelation}. 

\cor
 Every complete, transitive, weakly monotonic preference relation on $\Re_+^n$ is representable by a continuous utility function iff the preference relation satisfies any of the following eight continuity postulates: graph continuous, 
continuous,  
arc-continuous, 
linearly continuous,  
mixture-continuous, 
Archimedean,   
Wold-continuous, and    
weakly Wold-continuous.  
\label{thm: wold}
\corr

\nt  Note that this result provides  a  novel form to a  classical theorem by providing equivalences under eight different continuity postulates; hence in addition to  a consolidation of the antecedent literature, it unmasks new connections.  
 In particular, this result  shows that we can  replace Wold-continuity with any of the continuity properties listed in Theorem \ref{thm: ccontinuous}.   
 The precise continuity hypotheses assumed by Wold in his theorems are  fundamentally germane to this  paper.  In Wold's proof, monotonicity of the relation plays a crucial role.   
  Wold's representation theorem  in his seminal work \citet{wo43} provides an equivalence relationship between Wold-continuity and section continuity of a relation on a Euclidean space  under a weak monotonicity postulate. Almost ten years after his papers, Wold provides separate sufficient conditions for the existence of a continuous utility representation in Wold-Jureen \citeyearpar{wj53}. The new theorem replaces weak monotonicity with strong monotonicity,  and Wold-continuity with weak Wold-continuity.\fn{Even though Wold assumes only part (ii) of weak Wold-continuity, it is easy to show that its part (i) follows from strong monotonicity under the transitivity postulate; see Example \ref{exm: wold} and the discussion following it in Section \ref{sec_examples}.  \label{fn: wj}}  There is a dissonance in 
the  literature in the way that these two theorems were received, but this need not concern us here.\fn{\citet{yo56}, Arrow-Hahn \citeyearpar{ah71} and Mitra-Ozbek \citeyearpar{mo13} mainly picked the set of assumptions listed in Wold's book.   \citet{yo56} provides a direct proof of the relationship between  weak Wold-continuity and continuity of a relation on $\Re_+^n$ under strong monotonicity postulate. On the other hand, Beardon-Mehta \citeyearpar{bm94} pick up the hypothesis of Wold's theorem as stated in his papers and  show that Wold's method-of-proof can be used in order to obtain a simple alternative proof of Debreu's theorem.  Moreover,  Lloyd-Rohr-Walker \citeyearpar{lrw67} provide an explicit proof of Wold's theorem he presented in his papers, but under strong monotonicity. See also \citet{yo54} for a discussion on Wold's strong continuity assumption in  consumer theory and Banerjee-Mitra \citeyearpar{bm18a} for a recent reference to Wold's work.}     
The  point is that the  two separate continuity assumptions of Wold can be considered as a link between the scalar continuity assumption  of Rosenthal which is based on smooth curves and of Herstein-Milnor which is based on straight lines.   We elaborate this link in this paper by showing that under a weak monotonicity assumption, commonly used scalar continuity assumptions on preferences are equivalent to each other, and also each is equivalent to the continuity of the relation.

As we mention in the Introduction, Eilenberg's and Debreu's representation result is more general than Wold's theorem since they simply drop the monotonicity assumption.\fn{Even though Eilenberg's  result is for anti-symmetric relations, it is trivial to observe that by using the tools of the quotient space, the anti-symmetry assumption can simply be dropped; see for example \citet{de54}. Of course this does not devalue Debreu's or Wold's contributions -- Debreu's celebrated open gap lemma and Wold's simple method-of-proof have substantial effect in economic theory; see Beardon-Mehta \citeyearpar{bm94} for a discussion.} And, even though Wold's continuity assumption is by itself weaker than the section continuity property used in Eilenberg and Debreu, we showed that they are equivalent in the presence of other assumptions of Wold.\fn{For ordial representation of preferences, see Herv{\'e}s-Beloso-del
  Valle-Incl{\'a}n~Cruces \citeyearpar{hd19} and its reference to Hausdorff's work.} 
Moreover, Arrow-Hahn \citeyearpar{ah71} in their representation theorem do not impose convexity, or monotonicity, assumptions on preferences, hence our results are silent about the possibility of replacing the continuity assumption with mixture-continuity. In fact, as we already noted in the Introduction, the Genocchi-Peano example provides an example which satisfy the assumption of Arrow-Hahn except section continuity. Hence, it is indeed not possible to replace the continuity assumption with scalar continuity in Arrow-Hahn's theorem.\fn{Beardon-Mehta \citeyearpar{bm94} has an excellent   
survey and extension on the method-of-proof of Arrow-Hahn and its relation to other method-of-proofs  including Debreu's and Wold's methods.   
 In particular, they consider the unit interval  endowed with an equivalence relation, and under the assumption of closed equivalence classes, show  the existence of a non-decreasing real-valued continuous function on the interval that is constant on each equivalence class. Ascribing this result of real analysis to Wold, they use it to prove Debreu's representation theorems as well as his \lq\lq open-gap lemma"  which he used to prove his theorems.   
(As is well-known, this lemma was originally formulated in \cite{de54} and fully proved in \cite{de64}. For its importance in many areas of the mathematical sciences,  see, for example,   Beardon-Mehta \citeyearpar{bm94}.)  
Furthermore, they go beyond this equivalence and a corresponding revisionary re-framing of Wold's work to   prove a   generalized version of  Arrow-Hahn's representation theorem. However, the point of departure for the  investigation pursued and reported here in this paper is not this question of whether one theorem can be used to prove, and be proved by, another, but rather the conceptual move that 
Wold makes  from the topology on the choice space to that of the topology on the operation under which these objects of choice can be combined. Finally, see also Bosi-Campi{\'o}n-Candeal-Indurain  \citeyearpar{betal20} on recent results on representation of preferences and in particular Bosi-Zuanon \citeyearpar[Theorem 2.23]{bz20} for equivalence results on section and graph continuity postulates.} 
Note, however, that the fact that the utility representation results of AH and Debreu does not use convexity, or monotonicity assumptions in their utility representation theorems, in many of their applications they assume continuous and convex preferences, which have quasi-concave  utility representations. In these results we can replace the continuity assumption with various continuity assumptions we listed in Theorem \ref{thm: ccontinuous}. This observation also connects us to our Theorem \ref{thm: quasi-concave}.

%

\subsubsection{A Mixture Space: Herstein-Milnor \citeyearpar{hm53}}

We now provide a corollary of Herstein-Milnor \citeyearpar{hm53}, von Neumann-Morgenstern \citeyearpar{vm47}, \cite{je67} and  Theorem \ref{thm: scontinuous}. Before that we need the following definition: a  relation $\succcurlyeq$ on a mixture set $X$ is {\it independent}  if for all $x,y,z\in X$  and all $\lambda\in (0,1)$,  $~x\succcurlyeq y$ implies $x\lambda z \succcurlyeq y\lambda z$.

\cor
A binary relation defined on a mixture set is representable by a mixture-linear\fn{A function $l:X \ra \Re$ defined on a convex subset of a  vector space $X$ is {\it mixture-linear} if $l(x\lambda y)=\lambda l(x)+(1-\lambda)l(y)$ for all $x,y\in X$ and $\lambda\in [0,1]$.} function iff it is  complete, transitive, independent and satisfies any of the following three continuity postulates:   mixture-continuous,     Archimedean, and weakly Wold-continuous.   
\label{thm: hmrepresentation}
\corr

\nt  As we mention above, this result provides  a  novel form to the classical expected utility theorem and connects it to Wold's work. 
Note that the necessity and sufficiency of mixture continuity is proved by Herstein-Milnor \citeyearpar{hm53}, of Archimedean by von Neumann-Morgenstern \citeyearpar{vm47} and \cite{je67},\fn{See \citet{fi82} for a detailed discussion on this. Moreover, note that there are different versions of the independence assumption in the literature.} and of weak Wold-continuity follows from Theorem \ref{thm: scontinuous}.

\subsubsection{A Finite State Space: \cite{gu92}} 

\citet{gu92} re-does Savage's work for finite state space by imposing topological structure on the set of outcomes.\fn{See also \cite{wa88a, wa89} and Gravel-Merchant-Sen \citeyearpar{gms12}.}  
 His framework is as follows.     
 Let $N=\{1, \cdots, n\}$, $\CN=2^N\backslash \emptyset$ and $X=[0,1]^n$. Let $\Delta=\{x\in X|x_i=x_j \text{ for all }i,j\in N\}$.    
For $x\in \Delta$, we use $x$ and $x_i$  interchangeably for any $i\in N$.   For $I\in \CN$ and $x,y\in \Delta$, $xIy$ denotes $z$ such that $z_i=x_i$ for $i\in I$ and $z_i=y_i$ for $i\notin I$.   Let $\succcurlyeq$ be a preference relation on $X$.  A set $I\in \CN$ will be called {\it null} if $x_i=y_i$ for all $i\in I^c$ (i.e., $x=y$ on $I^c$) implies $x\sim y$. For simplicity, we assume no $i\in N$ is null. 
  Consider the following axioms. 

{\it 
\ben[{\nf }, topsep=3pt]
\setlength{\itemsep}{-1pt} 

\ml[{\bf (A1)}]  $\succcurlyeq$ is complete and transitive. 

\ml[{\bf (A2)}]  For $I\in \CN$, $x'_i \sim x_i I z_i, y'_i \sim y_i I z_i$  for all $i\in N$ 
 implies $x\succ y$ if and only if $x'\succ y'$.

\ml[{\bf (A3)}]  For all $x,y\in \Delta$,  $ x> y$ implies $x\succ y$. Furthermore, there exist $I^*\subseteq N$ such that for all $x, y\in \Delta$, $xI^* y\sim yI^* x$.

\ml[{\bf (A4)}]  $\succcurlyeq$  is continuous. 
\een
}

The following result and Theorem \ref{thm: ccontinuous} imply that {\nf A4} and mixture-continuity of $\succcurlyeq$ are equivalent, hence A4 in \citet[Theorem]{gu92} can be replaced by mixture continuity.\fn{Similar results can be obtained in   Chew-Karni \citeyearpar{ck94}, which generalize \citet{gu92} and  \citet{na90}; \citet{de86}, Ahn-Ergin \citeyearpar{ae10}, and  Maccheroni-Marinacci-Rustichini \citeyearpar{mmr06}. }

\prp
Under {\nf A1--A3}, if $\succcurlyeq$ is mixture-continuous, then it is strongly monotone.  
\label{thm: gul}
\prpp


%
%
%

\subsubsection{A Topological Vector Space: Neuefeind-Trockel \citeyearpar{nt95}  }
 We now present the relationship between additivity and the following two convexity assumptions on binary relations.  A relation $\succcurlyeq$ on a convex cone $X$ is {\it homothetic} if for all $x,y\in X$ and all $\lambda\in \Re_{++}$ with $\lambda x,\lambda y\in X$,  $x\succcurlyeq y$ implies $\lambda x\succcurlyeq \lambda y$, and recall that it is  independent if for all $x,y,z\in X$  and all $\lambda\in (0,1)$,  $~x\succcurlyeq y$ implies $x\lambda z \succcurlyeq y\lambda z$.  
Independence and additivity assumptions are used commonly in decision theory, homotheticity is used mainly in consumer theory. The following result provides an equivalence relationship between these three common convexity postulates in decision theory as well as in consumer theory: additivity, homotheticity  and independence.

\prp
The following are true for any transitive relation $\succcurlyeq$ defined on a vector space. 
\ben[{\nf (a)}, topsep=3pt]
\setlength{\itemsep}{-1pt} 
\ml $\succcurlyeq$ is  additive and homothetic iff it is independent. \label{it: ahit}
\ml If $\succcurlyeq$ is complete and mixture-continuous, then it is additive iff it is independent. \label{it: ahict}
\een
\label{thm: ahi}
\prpp

 We next present a result that provides an expected utility representation theorem under  minimal (explicit) technical and behavioral assumptions on preferences.

\prp
The following are equivalent for any non-trivial relation  on a vector space. 
\ben[{\nf (a)}, topsep=3pt]
\setlength{\itemsep}{-1pt} 
\ml The relation is non-trivial, semi-transitive, additive, upper mixture-continuous and upper  Archimedean.  
\ml The relation is representable by a mixture-linear 
 function. 
\een
\label{thm: representation}
\prpp

\nt It should be noted that this is a new representation theorem by virtue of the fact that it replaces section continuity assumptions of 
Neuefeind-Trockel \citeyearpar[Proposition]{nt95}  by weaker scalar continuity assumptions.  
 Note also that Proposition \ref{thm: representation} does not explicitly assume  completeness, transitivity, independence and full mixture-continuity of the relation. However, if a mixture-linear representation exists, then all of these properties hold by necessity. Therefore, these assumptions are  {\it hidden} in the hypotheses of this result. Since Proposition  \ref{thm: representation} and Herstein-Milnor's theorem  have {\it essentially} the same hypotheses, it is an {\it inessential} generalization of the theorem of Herstein-Milnor  in the sense of Kim-Richter \citeyearpar{kr86}.    

The following is a corollary of Proposition \ref{thm: representation} and the representation theorem of \citet[Theorem I]{de54}. 

\cor
 Every non-trivial, semi-transitive and additive relation $\succcurlyeq$ on a topological vector space with $A_\succcurlyeq(0)$ is closed and $A_\succ(0)$ is open, is complete, transitive, continuous and representable by a continuous linear utility function. 
 \label{thm: ntrepresentation}
\corr

\nt This corollary is presented first in Neuefeind-Trockel \citeyearpar[Proposition]{nt95}.  Proposition \ref{thm: representation} replaces closed upper sections of the weak relation and open upper sections of the strict relation with the weaker upper mixture-continuity and upper Archimedean properties, respectively. Hence, we obtain a representation theorem under a weaker continuity assumption. Of course the utility function  Proposition \ref{thm: representation}  yields may not be continuous. However, since the preference relation is complete, the preference relation has closed upper and lower sections.  Then, Debreu's representation theorem imply that there exists a continuous utility representation which is also linear, due to additivity.

\subsection{Neoclassical Consumer Theory}


In this subsection, we present two observations and two propositions. The first observation illustrates an ex-post {\it inessential} remark of Debreu, and the second illustrates that and of the eight continuity postulates above can be equivalently  used for the results in the Walrasian equilibrium theory and neoclassical consumer theory. The two propositions show that some of the equivalences survive even if completeness or transitivity assumptions are dropped.


\subsubsection{\cites{de59} Remark}

We have already related our results to the antecedent literature, but in a sharp way that may be missing the outlines of the forest in its focus on its individual trees. In this section, we step  back and  provide an overview of 
 the results from the criterial perspective of  {\it inessentiality} in Kim-Richter \citeyearpar{kr86}, and its invocation in \cite{ep87}, and those of   {\it hiddenness} and {\it redundancy} explicated in  Khan-Uyanik \citeyearpar{ku21et}.

We already referred to Debreu's  generalization of Wold's representation theorem  from Euclidean spaces by using the section continuity assumption to eliminate monotonicity. We have now already seen in  Theorem \ref{thm: ccontinuous} that under the hypothesis of Wold's theorem Debreu's section continuity assumption is not only equivalent to Wold-continuity but also the commonly used mixture-continuity, Archimedean and strong Wold-continuity properties.\fn{To be precise, the equivalence of parts (a) and (d) of Theorem \ref{thm: ccontinuous} under completeness, transitivity and weak monotonicity assumptions.}  If we drop the monotonicity assumption, these scalar continuity properties are weaker than section continuity. However, as we presented above, monotonicity can be compensated by convexity in order to obtain the equivalences. Our results in Section \ref{sec: results} illustrate the fact that seemingly unrelated topological, order and convexity properties together have behavioral topological implications on preferences.  Hence, a scalar continuity assumption on preferences is strictly weaker than the section continuity property, however, in the presence of additional assumptions on preferences they may be equivalent. We illustrate this point by providing an example from the classic work of Debreu. He writes 
\bqu
Certain theorems whose statements list (a) of 4.6 [closed upper sections] among their hypotheses can, in 
fact, be proved by using weaker continuity assumptions on preferences [upper mixture-continuity] inspired by I. N. Herstein and J. Milnor.  \hfill \citet[p.73 ]{de59} 
\equ
He then notes that he can replace the closed upper sections assumption with the weaker upper mixture-continuity property in order to obtain a relationship between certain convexity assumptions on preferences and in the statement of the second welfare theorem.  Lemma \ref{thm: convexall} implies that under upper mixture-continuity assumption, Debreu's convexity assumption (star-convexity of $\succ$)  is equivalent to convexity of $\succcurlyeq$. And  Lemma \ref{thm: ccontinuity}  implies that upper mixture-continuity and closed upper sections are equivalent properties under the hypothesis of Debreu's results.    

\obs
Debreu's remark about replacing closed sections assumption with the upper mixture-continuity property is  ex-post  inessential. 
\label{thm: dremark}
\obss

Wold (1943), \citet{wo43} provides  a model of consumer theory which assumes the consumers have  complete, transitive, weakly  monotonic and  Wold-continuous preference relations.  Wold-Jureen \citeyearpar{wj53}  replaces the last two assumptions with strong monotonicity and weak Wold-continuity assumptions. Our results show that we can replace his continuity assumption with any of the scalar, section or graph continuity assumptions listed in Theorem \ref{thm: ccontinuous}.  We get back to this in Section 4.2.4 below.



 \subsubsection{\cites{sc69} Non-decisive Consumer}

 \cite{sc69} generalizes  \cites{au66} work on the existence of a general equilibrium with continuum of players by dropping the completeness assumption. Schmeidler's  consumer has a {\it strict preference relation defined on $\Re_+^n$ which is irreflexive, strongly monotonic, transitive with open sections.} The following result shows that Schmeidler's continuity assumption can be replaced with the Archimedean property.\fn{Note that we define Archimedean property for an arbitrary relation $\succcurlyeq$ where its asymmetric part is $\succ$. In Schmeidler's model, the primitive is a transitive and irreflexive relation $\succ$. Then it is asymmetric. Hence, its  asymmetric part is equal to the relation itself.}  
 
 \prp
The preference relation of Schmeidler's consumer is Archimedean iff it is strict-Archimedean iff it has open sections. 
 \label{thm: scmonotone}
 \prpp

 Note that in the classical consumer theory, completeness and transitivity are assumed. We show that different continuity assumptions are equivalent under these assumptions.  Our discussion of the work of Shafer-Schmeidler show that when one of these two behavioral assumptions, we can still have equivalence results.   
 
 In his result, Schmeidler does not assume full continuity, instead he assumes only that the strict relation has open sections.  With the added full continuity and transitivity assumptions, however, Schmeidler's result would be a straightforward corollary of Aumann's theorem.  Therefore, Schmeidler's result  is an {\it essential} generalization as  a result of weaker continuity and transitivity assumptions -- without such a weakening, completeness and full transitivity would already be hidden in his seemingly more general assumptions.   The recent work on Walrasian general equilibrium and game theory with discontinuous and/or non-ordered  preferences is in line with this observation; see for example  \cite{re20are}, \cite{ku21ettribute} and \cite{adku21et} for details. 

\subsubsection{\cites{sh74} Non-transitive Consumer}   

\citet{sh74} re-works the neoclassical theory of demand for a  consumer with possibly non-transitive preferences defined on $\Re_+^n$.  Shafer's consumers have {\it continuous, complete and convex preferences whose asymmetric part is also convex.}  
The following result shows that it is possible to replace the section continuity assumption of Shafer with and of the mixture-continuity or  Archimedean properties.

 \prp
The preference relation of Shafer's consumer is Archimedean iff it is mixture-continuous iff  it is continuous. 
 \label{thm: shafer}
 \prpp

Moreover, Khan-Uyanik \citeyearpar{ku21et} noted that  if the preference relation $\succcurlyeq$ of Shafer's non-transitive consumer satisfies (i) semi-transitivity ($x\sim y\succ z$ implies $x\succ z$, or equivalently, $x\succ y\sim z$ implies $x\succ z$), or (ii) transitive indifference ($\sim$ is transitive), or (iii) transitive strict relation ($\succ$ is transitive), then the preference relation $\succcurlyeq$ is transitive. That is to say, a \lq\lq little'' bit of consistency implies \lq\lq fully'' consistency in Shafer's model of non-transitive consumer, thereby destructive of ``{all}'' non-transitivity.  Therefore, Shafer's  non-transitive consumer has to be, by necessity, a  ``fundamentally'' non-transitive agent.     
That is, the preferences of Shafer's consumer cannot satisfy any of the three transitivity postulates (i), (ii), (iii). If any of these three transitivity postulates is desirable in a specific model, but $\succcurlyeq$ being non-transitive,  then either the convexity (in fact, path connectedness) or continuity assumptions on preferences should be relaxed. In the light of the proposition above, the relaxation of the continuity assumption cannot be one of mixture continuity or Archimedean. To be sure, this does not rule out transitivity on a subset of the domain of the preference relation, or does not contradict with existence results in consumer theory and Walrasian equilibrium; the theory is by now well-established under very weak assumptions on preferences.

\subsubsection{Hidden Assumptions in Walrasian Equilibrium Theory}

 The classical Walrasian equilibrium theory and neoclassical consumer theory  assume that the preferences of the consumers are complete, transitive and convex (and/or monotone); see for example \cite{de59}, \cite{au66}, Arrow-Hahn \citeyearpar{ah71} and \cite{mc02}.  For suitable consumption sets, it follows from Theorem \ref{thm: ccontinuous} above  that 

\obs
The continuity assumptions in the Walrasian equilibrium theory and neoclassical consumer theory can be replaced by any of the eight continuity postulates above.
\label{thm: eqm}
\obss

In many results in consumer theory and Walrasian equilibrium, our results imply that  completeness and full transitivity properties follow from the other assumptions; see Khan-Uyanik\citeyearpar{ku21et} for an extended discussion on the  Walrasian equilibrium theory and consumer theory in the  above-mentioned work.   We will return to the hiddenness of completeness and full transitivity properties in the subsequent subsection.

\subsection{Consistency and Decisiveness: A Reconsideration  } 


In this subsection, we present two propositions on the  behavioral implications of  continuity postulates. The first  proposition uses the {\it Dubra-method} to show that convexity and scalar continuity imply completeness and full-transitivity of preferences, and the second shows that additivity, a stronger convexity property, allows weakening of the continuity postulates in the first proposition. 

%
%
 %
%
We start with 

\prp
Let $X$ be a convex set in $\Re^n$ with property {\bf C}.   Then every non-trivial, reflexive,   
 mixture-continuous and Archimedean binary relation on $X$ which demonstrates convex indifference and has a transitive symmetric part, is continuous, complete and transitive. 
\label{thm: ctdubra}
\prpp

\nt  
 Scalar continuity, such as Archimedean or mixture-continuity assumptions, is used extensively in decision theory. An equivalence of this type has useful implications.     
%
 Dubra-Maccheroni-Ok \citeyearpar[Proposition 1]{dmo04} shows that for a reflexive, transitive and independent relation, a stronger version of mixture-continuity, which is due to Shapley-Baucells \citeyearpar{sb98}, is equivalent to closed graph property on a finite simplex.  Gilboa-Maccheroni-Marinacci-Schmeidler \citeyearpar[Lemma 3]{gmms10}  shows that if a reflexive and transitive relation on a normed space satisfies independence and monotonicity properties, then its mixture-continuity is equivalent to the closed graph  property.\fn{See also Chateauneuf-Cohen-Jaffray \citeyearpar{ccj09} and Abdellaoui-Wakker \citeyearpar{aw20}.}   \cite{du11} and Karni-Safra \citeyearpar{ks15} show that if a reflexive and transitive relation on a simplex satisfies any of the independence, betweenness, and cone monotonicity properties, then mixture-continuity and Archimedean properties imply continuity if the relation on a finite simplex.   Proposition \ref{thm: ctdubra} in  generalizes these results by weakening, or dropping, the convexity, monotonicity, reflexivity and transitivity postulates, and by expanding the choice set in the context of Euclidean spaces.\fn{Note that the result presented in Gilboa-Maccheroni-Marinacci-Schmeidler \citeyearpar[Lemma 3]{gmms10} 
is for infinite dimensional spaces, hence our result is not a full generalization of their result. Moreover,   Karni-Safra  
 \citeyearpar{ks15} replace the independence assumption in Dubra's theorem with the betweenness, or cone-monotonicity, assumption.    
  Proposition \ref{thm: ctdubra} generalizes their results by weakening, or dropping, the convexity, reflexivity and transitivity postulates, by expanding the choice set,  and more importantly by obtaining {\it both} completeness and transitivity as necessary conditions.  Note that Galaabaatar-Khan-Uyanik \citeyearpar[Theorem 1]{gku19} provide more general results than Proposition \ref{thm: ctdubra}. However, our method-of-proof is different than theirs -- their method-of-proof is direct and also their choice set does not have a topological structure.  We, on the other hand, first obtain the continuity of a linearly continuous relation, and then apply a result due to Khan-Uyanik \citeyearpar[Theorem 2]{ku21et} in order to obtain completeness and transitivity postulates. This method-of-proof is in line with the method-of-proof of Dubra which is then picked up by Karni-Safra. } 

 We next provide a result for arbitrary vector spaces illustrating the behavioral implications of scalar continuity in the presence of additivity.  

\prp
Every non-trivial, semi-transitive, additive, upper mixture-continuous and upper  Archimedean  relation on a vector space, is complete and transitive. 
\label{thm: ast}
\prpp

\nt  Neuefeind-Trockel  \citeyearpar[Proposition]{nt95}  prove a special case of  this result for topological vector spaces under stronger section continuity postulates.   
%
Proposition \ref{thm: ast}  provides a considerable generalization of Neuefeind-Trockel's result by replacing their  continuity assumption with the weaker linear continuity postulate and allowing the space to be an arbitrary vector space.\fn{This generalization requires a proof technique which is based on a closure concept for subsets of a vector space by using the topological structure of the unit interval; see the proof in Appendix \ref{sec: proofs} for details.   
Moreover, Neuefeind-Trockel in Footnote 2 in their Section 3  note the fact that scalar continuity assumption is weaker than section continuity by providing an example of a preference relation defined on an  infinite-dimensional space which satisfies the assumptions of Herstein-Milnor's theorem yet it is not continuous. Inoue's illustration of GP's example is on a simplest possible choice set: on $\Re^2$. Our results above show that the infinite dimensionality in Neuefeind-Trockel's    example is essential since under the hypothesis of the Herstein-Milnor's theorem, continuity and linear continuity postulates are equivalent in finite dimensional spaces. \label{fn_nt} }  Lastly, this result illustrates the strength of the additivity assumption by showing that it allows us to obtain both completeness and transitivity of a relation under weak continuity and transitivity assumptions. Note that the transitivity of $\sim$ plays a crucial role in the antecedent results in  the Eilenberg-Sonnenschein research program.  This result, by replacing the transitive indifference assumption with the semi-transitivity postulate under the additivity assumption complements this research program.

\subsection{Other Potential Applications}

%

 Our equivalence results have limited themselves to finite-dimensional spaces when they involve continuity postulates based on the topological structure of the choice set. As such, for the equivalence results among the scalar continuity postulates, which do not require any topological structure on the choice set, the dimension of the choice set is not of any consequence.  Therefore, our investigation asks for a natural extension to an infinite-dimensional setting that brings topological and other algebraic structures to bear on the discussion.  
Since finiteness of the dimension of the choice set is used crucially in some of our results, it is not a routine exercise to generalize them to infinite-dimensional spaces.\fn{Theorem \ref{thm: rockafellar} and Lemma \ref{thm: locallyconvex} are not true for infinite-dimensional spaces; see Borwein-Lewis  \citeyearpar[p. 20]{bl92}.}   
Moreover, taking $X$ as a convex subset of $L^p$ space, $0<p<1$, and $\succcurlyeq$ a complete, transitive, non-trivial, independent and mixture continuous binary relation imply, by Herstein-Milnor's theorem, $\succcurlyeq$ has an expected utility representation. Since the dual of the $L^p$ space is $\{0\}$ for $0<p<1$, the only continuous function is the zero-function; see for example \cite{da40bams}. Therefore, $\succcurlyeq$ cannot have a continuous representation, hence cannot be continuous, by Debreu's \citeyearpar{de54} representation theorem. This example illustrates that some of our equivalence results fail for infinite dimensional spaces. Of course, it may be possible to establish equivalence results for a restricted class of infinite dimensional spaces or under stronger convexity and/or monotonicity assumptions.\fn{For example, Gilboa-Maccheroni-Marinacci-Schmeidler \citeyearpar[Lemma 3]{gmms10} establish a partial equivalence result for infinite dimensional spaces under independence and monotonicity assumptions.}  The following references may be of use in order to generalize the results we present in this paper to infinite dimensional spaces.  Generalization of the results in the literature in mathematics pertaining to locally convex sets are provided in  \citet{sc42} and \citet{kl51}, and their followers. Similarly, Borwein-Lewis \citeyearpar{bl92} and Borwein-Goebel \citeyearpar{bg03} provide generalizations of some of the results pertaining to relative interior to infinite-dimensional spaces.\fn{The non-emptiness of the relative interior is a crucial property (for convex optimization and general equilibrium) which we also use in the construction of the proofs of our results. There are various extensions of the concept of relative interior for infinite-dimensional topological vector spaces; see Borwein-Lewis \citeyearpar{bl92} and Borwein-Goebel \citeyearpar{bg03} for a detailed  reference. Moreover, \citet[p. 38]{sc66book} proves that ``{\it For any convex subset $A$ of a t.v.s, if $x$ is interior to $A$ and $y$ in the closure of $A$, then the open line segment joining $x$ and $y$ is interior to $A$.}" This result may be useful for the generalization of our results to infinite dimensional spaces, but a different method-of-proof is needed.}  To the best of our knowledge, there is no generalization of \citet{ro55} to infinite dimensional spaces, hence generalizations of our results on arc-continuity to infinite-dimensions seem even harder.





The recent book Bosi-Campi{\'o}n-Candeal-Indurain  \citeyearpar{betal20} contains representation theorems in abstract topological spaces, which contains finite and infinite dimensional spaces. The subtleties in the behavioral implications of continuity postulates reveal themselves. For example,   Theorem 4.2 of Bosi-Zuanon \citeyearpar{bz20} is, as stated false, since it lacks the open sections assumption. The usual order $\leqq$ on $\Re^2$ is a counterexample. It is reflexive, transitive, non-trivial has closed graph and closed sections, and the space is Hausdorff. It does not have the open sections. And the relation is not complete. As another example, Theorem 2.2(a) of Herves-Beloso-Monteiro in the same volume is false. It lacks the non-triviality assumption. The following relation on [0,1] is a counterexample: each point  is comparable itself and there are no other points that are  comparable. This relation  satisfies reflexivity, transitivity as well as closed and open sections (of the asymmetric part) assumptions. But it is clear that it is not representable.

 There is a rich literature on the structure of the discontinuity of linearly continuous functions; see Young-Young \citeyearpar{yy10},  \citet{ke43} and Ciesielski-Miller \citeyearpar{cm16}. This direction has been neglected for preferences in both mathematics and in economics literature.\fn{Banerjee-Mitra  \citeyearpar{bm18a} provide some preliminary results on the structure of discontinuity of a utility function representing a scalarly continuous relation.} It is possible to study this problem not only for linear continuity but also for graph and section continuity properties. This may be  relevant to the {\it essentiality} concept of Kim-Richter \citeyearpar{kr86}: if the set of discontinuities is not rich, then  the weakening of the continuity assumption may  be  inessential. 

 Finally, in this paper we provide a deconstruction and an integration of the continuity postulate based on section,  scalar and restricted continuity of a binary relation. First, there is a rich literature taking a binary relation with some desirable properties on a space  and studies the extension of it to a larger space by keeping the desirable properties of the relation; see for example \cite {yi93jme}, \cite{du99jet},  \citet{ko16jme} and Evren-H\"{u}sseinov \citeyearpar{eh21mor}. Second, there is an active literature taking two binary relations as primitive; see for example   \cite{gg13} and  \cite{co18}.\fn{See also  \cite{go71}, \cite{ch71}, \cite{gi19} and \cite{uk19b}.}   
 It is an interesting direction to investigate the relationship between the results in the current paper to those in these literatures.

%

\section{Concluding Remarks} \label{sec:conc}
We conclude this essay with two remarks,  both of an epistemological  nature, and ones  that underscore the methodological preconceptions of the continuity postulate, and also in so far as it feeds into theorization.  The first remark supplements the Leibnizian 
dictum that \lq\lq nature does not make leaps" by noting with \cite{ar05}  that he \lq\lq also viewed it as a carrier of sets of points denser than the set of reals, sets including ideal infinitely small elements greater than zero and smaller than any positive real number."\fn{A discussion of Leibniz on the continuity postulate, and more generally the relevance of his thought to {\it nonstandard analysis} is beyond the scope of this essay. This is even more true of the rich philosophical issues stemming from his capacious {\it oeuvre}. In addition to \cite{ar05}, we refer the reader to \cite{ku91},  \cite{ma96book},  \cite{re04}, \cite{jo09jhp}  and their references.}   The basic issue hinges on {\it perception} and {\it barely} perceptible differences.\fn{For the rich philosophical literature associated with {\it apperception} and these Leibnizian terms; see, for example, \cite{ku74, ku91}, \cite{jo09jhp} and their references.}     An extended quotation  from    \cite{ar66} goes to this.

\bqu The assumption of Continuity seems, I believe correctly, to be the harmless simplification almost inevitable in the formalization of any real-life problem. It is sometimes held that certain possible consequences, such as death, are incommensurably greater than others, such as receiving one cent. Let action $a^1$ involve receiving one cent with no risk of life, $a^2$ receiving nothing with no risk of life, and $a^3$ receiving one cent with an exceeding small probability of death. Clearly, $a^1$ is preferred to $a^2$.
Continuity would demand that $a^3$ be preferred to $a^2$ if the probability of death under $a^3$ is sufficiently small. This may sound outrageous at first blush, but I think a little reflection will demonstrate the reasonableness
of the result. The probability in question may be $10^{-6}$ or  $10^{-10}$, inconceivably small magnitudes. Also, if in the above example, one cent were replaced by one billion dollars, on would hardly raise the same argument, and yet to go from one cent to one billion dollars certainly involves no discontinuity, however big the difference in scale may be. ``Every journey,
no matter how long, begins with a single step."\fn{Arrow continues, \lq\lq  
Blaise Pascal, or perhaps one of his co-authors of the {\it Port-Royal
Logic}, indeed suggested that the salvation of the soul or the avoidance 
of eternal damnation might be of infinitely greater value than any earthly regard; but the humble economist may be excused for regarding such choices as beyond the scope of his theories."}
\equ

\nt  So it all hinges on \lq\lq inconceivably small magnitudes" and some \lq\lq incommensurably greater than others."  It leads to a requirement of asymptotic implementation of the theory, and leads Le Cam \citeyearpar{le86} to write:

\bqu  Indeed, limit theorems \lq\lq as $n$ tends to infinity" are
logically devoid of content about what happens at any particular $n.$ All they
can do is suggest certain approaches whose performance must then be checked
on the case at hand. Unfortunately the approximation bounds we could get
were too often too crude and cumbersome to be of any practical use. Thus we
have let $n$ tend to infinity, but we would urge the reader to think of the material
in approximation terms.\fn{Since this essay is dedicated to Tjalling Koopmans, it is worth pointing out that he was very strict in his requirement of  asymptotic implementation of any result for an idealized model; see Brown-Robinson \citeyearpar{br72-ns} and \cite{kh73}.  To return to the first epigraph, a referee writes, \lq\lq There is now very strong evidence that this is not true at small enough scales of measurement, indeed, time itself may be discrete. This is an argument that nature is nothing but jumps, but that this happens at so small a scale that none of our human sensory apparatus can distinguish the differences, and this is the source of the intuition that the quote is correct."  \label{fn:ko}}    \equ

Our second observation concerns our second epigraph:  a reading of Koopmans' 1957 magisterial overview of the state of economic science and in terms of a summarizing overview, one that brings out  the following points into salience: 

\ben[{\nf (i)}, topsep=3pt]
\setlength{\itemsep}{-1pt} 
\ml   an {\it omission} of a pioneering  paper,  \cite{ei41},   that can now be seen as  inaugurating an important aspect of the modern neoclassical theory of individual choice, one relating to ``nice" preference relations and their representation as well as their behavioral underpinnings, 
\item  a {\it masking} of the continuity postulate that led to the obscuring of the connection between axiomatic decision theory and the neoclassical theory  of choice in both its deterministic and stochastic modes. 
\een 

 It is then a  contention of this reading that both aspects have had profound consequences on the reception and direction of subsequent work, and even the most perfunctory examination of these origins allows one to obtain both   a systematic and comprehensive reading of the current    literature, and also to highlight connections that go into shaping it. It is also the  contention of this work that the history of the problematic is relevant to a successful resolution of the problematic itself, and that the line between exposition and research may have been overplayed in mainstream decision theory: Abdellaoui-Wakker \citeyearpar{aw20} and \cite{ha20} have furnished embarrassing instances of this disregard.   A successful theory must keep in the foreground the economic phenomenon to be explained;  a successful generalization of a theorem needs orientation to the direction of the generalization and it   cannot be oblivious of the history of the theorem.\fn{This sentence is in response to an anonymous referee who  asked for  implications that stem from the  claim in a sentence previous to it.}  But beyond this, surely our investigation raises questions beyond the finite-dimensional case to which it has been confined:\fn{Of course to the extent that our results rely on a Euclidean space structure and not only on a vector space one, as for example in Theorems  \ref{thm: mcarc} and \ref{thm: scontinuous}.} 
 Koopman's seminal characterizations of impatience,  axiomatization of the overtaking criterion, axiomatization of the Suppes-Sen grading principle and Diamond's  impossibility theorem are all germane to our work and concern infinite-dimensional spaces.\fn{See \citet[387--428]{ko70} and \citet[Chapters 4 and 5]{ko85}, and his references. For extensions, see 
 Kettering-Kochov \citeyearpar{kk20};  for the overtaking criterion, see \cites{br70}; for  the extension of Diamond's  impossibility theorem and axiomatization of the Suppes-Sen grading principle, see Basu-Mitra \citeyearpar{bm03, bm07b}. \label{fn:koop}
 }




\appendix

\section{Appendix}

In this Appendix, we present the proofs of the results and nine technical examples.

\subsection{Proofs of the Results} \label{sec: proofs}



The proofs of the results we present in this paper use an important result in convex analysis due to Rockafellar \citeyearpar[p. 45]{ro70} which, in Rockafellar's words,   ``provides a fundamental relationship between the closure and relative interior of convex sets."  
Let $X$ be a subset of $\Re^n.$   
 Since any lower dimensional subset of $\Re^n$ has empty interior, it is more convenient to work with the concept of relative interior.   
  A subset $X$ of a (real) vector space is called {\it affine} if for all $x,y\in X$ and $\lambda\in \Re,$ $\lambda x+ (1-\lambda)y\in X.$ It is clear that $A$ is affine if and only if $A-\{a\}$ is a subspace of $X$ for all $a\in A.$ The {\it affine hull of $X,$} aff$X,$ is  the smallest affine set containing $X$.  
The {\it relative interior} of a subset $X$ of $\Re^n$ is defined as 
$$
\text{\nf ri} X=\{x\in \mbox{\nf aff}X~|~\exists N_\varepsilon,	\mbox{ an } \varepsilon \text{ neighborhood of } x, \text{ such that }  N_\varepsilon\cap \text{\nf aff}X\subseteq X\}.
$$ 
That is, the relative interior of $X$ is the interior of $X$ with respect to the smallest affine subspace containing $X$.
%

\thm[Rockafellar]
Let $X$ be a non-empty and convex subset of $\Re^n.$ Then $\mbox{\nf ri}X$ is non-empty, and for all $x\in \mbox{\nf ri}X, y\in \mbox{\nf cl}X$ and all $\lambda\in [0,1),$ $y\lambda x \in \mbox{\nf ri}X.$ 
\label{thm: rockafellar}
\thmm

We are now ready to present the proof of Theorem \ref{thm: quasi-concave}. 

\prf[Proof of Theorem \ref{thm: quasi-concave}]
Let $X$ be a non-empty and convex subset of $\Re^n$ and $f:X\ra \Re$ a quasi-concave function. Note that $f$ is continuous if and only if it is both upper and lower semi-continuous, i.e., for all $\alpha\in \Re$, the set $A_\alpha=\{x\in X|f(x)\geq \alpha\}$ is closed  (in the subspace $X$) and the set $B_\alpha=\{x\in X|f(x)> \alpha\}$ is open (in the subspace $X$). Quasi-concavity of $f$ implies, by definition, $A_\alpha$ is convex for all $\alpha\in \Re$. We now show that $B_\alpha$ is convex for all $\alpha\in \Re$. Towards this end, assume $B_\alpha$ is not convex for some $\alpha\in \Re$. Then, there exists $x, y\in X$ such that $f(x), f(y)> \alpha$ and there exists $\lambda\in (0,1)$ with $f(x\lambda y) \leq \alpha$. Pick $\beta\in \Re$ such that $f(x), f(y)> \beta > \alpha$. Then, $x,y\in A_\beta$. Since $A_\beta$ is convex, $f(x\lambda y)\geq \beta$. This furnishes us a contradiction. Therefore, $B_{\alpha'}$ is convex for all $\alpha'\in \Re$. 

Pick $\alpha\in \Re$. If $A_\alpha$ is empty or a singleton,  then it is closed. Otherwise, pick $x\in \text{cl}A_\alpha$. Since $A_\alpha$ is convex, Theorem \ref{thm: rockafellar} implies that its relative interior is non-empty and for all $y\in \text{ri}A_\alpha$ and all $\lambda\in [0,1)$, $x\lambda y\in \text{ri}A_\alpha$, hence  for all $\lambda^n\ra 1$, $f(x\lambda^n y)\in A_\alpha$ for all $n$. Pick $y\in \text{ri}A_\alpha$ and let $L$ denote the straight line in $X$ passing through $x$ and $y$.  Linear continuity implies that $f\uhr L$ is continuous, hence $x\in A_\alpha$. Then  $A_\alpha$ is closed. Therefore, $f$ is upper semi-continuous. 

If $B_\alpha$ is empty or is equal to $X$, then it is open. Assume $B_\alpha$ is not open. Then there exists $x$ in $B_\alpha$ which lies on the boundary of $B_\alpha$, i.e., $x\in B_\alpha\cap \text{cl} B_\alpha^c  \cap \text{cl}B_\alpha$.  Let $\mathcal H_x$ denote the set of all supporting hyperplanes at $x$, and for all $h\in \mathcal H_x$, let $H$ denote the closed half space determined by $h$ which contains $B_\alpha$. First, assume there exists $h\in \mathcal H_x$ such that $C=H^c\cap X\neq \emptyset$. Since $H^c$ and $X$ are convex, Theorem \ref{thm: rockafellar} implies that the relative interior of $C$ is non-empty and for all $y\in \text{ri}C$ and all $\lambda\in [0,1)$, $x\lambda y\in \text{ri}C$.  Pick $y\in \text{ri}C$ and let $L$ denote the straight line in $X$ passing through $x$ and $y$. Note that $f(z)\leq \alpha$ for all $z\in C$. Linear continuity implies that  $f\uhr L$ is continuous, hence $f(x)\leq \alpha$, i.e., $x\in C$. This furnishes us a contradiction with $x\in B_\alpha$.   
Second, assume  for all $h\in \mathcal H_x$, the set $C=H^c\cap X= \emptyset$. This case happens only if $x$ lies on one of the hyperplanes that determines the space $X$.\fn{This step of the proof  uses the polyhedron  assumption. 
 Example \ref{exm_polyhed} below shows that the polyhedron assumption is not redundant.  
We thank two anonymous referees for  questioning the need for the polyhedron assumption. Also see Section \ref{sec_func} for a discussion on property {\bf C} and on a weaker property {\bf C$'$}, and  Footnote \ref{fn_polyR} on the role of the polyhedron assumption in another context. 
\label{fn_poly}} 
 (Hence, when $X$ is open (in $\Re^n$), the first case is exhaustive.) Let $\mathcal H'_x$ be the set of all such hyperplanes. Recall that a hyperplane is an affine and closed subspace of $\Re^n$. Moreover, $B_\alpha\cap h$ is non-empty and convex for all $h\in \mathcal H'_x$. Furthermore, for all $h\in \mathcal H'_x$, let $H'$ denote the half space in the affine subspace $h$ that contains $B_\alpha\cap h$. It follows from  $x\in B_\alpha\cap \text{cl} B_\alpha^c  \cap \text{cl}B_\alpha$ that there exists $h\in \mathcal H'_x$ such that $(H')^c\cap X\neq\emptyset$. Then, an argument analogous to the first case furnishes a contradiction. Hence, $B_\alpha$ is open.   Therefore, $f$ is lower semi-continuous.   
%
%
\prff


\prf[Proof of Proposition \ref{thm_conc_lc}]
Let $X$ be a non-empty, open and convex subset of $\mathbb R^n$ and $f: X\ra \Re$ a convex function on $X$. Pick a straight line $L \subseteq X$ and $x_0\in L$.    Since $X$ is open, there exist $a,b\in L$, $a\neq b$,   
such that $x_0=(a+b)/2.$

We next show that $f$ is bounded on $L_{ab}=\{a\lambda b ~ | ~\lambda\in [0,1]\}$.  By definition, for any $x\in L_{ab}$, there exists a unique $\lambda_x\in [0,1]$ such that $x=a\lambda_x b$ (where $\lambda_x=(b_i-x_i)/(b_i-a_i)$ for all $i=1,\ldots, n$). Let $M_0=\mmax\{f(a), f(b)\}$. Then, by the convexity of $f$, 
$$
f(x)\leq f(a)\lambda_x f(b) \leq M_0\lambda_x M_0\leq M_0. 
$$
Hence, $f$ is bounded above on $L_{ab}$. For all $y\in L_{\frac{a+b}{2} b}$,  there exists a unique $\lambda_y\in [0.5,1]$ such that $(a+b)/2=a\lambda_y y$ (where $\lambda_y=(2y_i-a_i-b_i)/(2(y_i-a_i))$  for all $i=1,\ldots, n$). Then, by the convexity of $f$, 
$$
f\left(\frac{a+b}{2}\right)\leq f(a)\lambda_y f(y).  
$$
Then, $f(y)\geq f\left((a+b)/2\right)$ for $y=(a+b)/2$.  It follows from  $\lambda_y=(2y_i-a_i-b_i)/(2(y_i-a_i))$  for all $i=1,\ldots, n$ that  for all  $y\in L_{\frac{a+b}{2} b}$, $y\neq (a+b)/2$, 
$$
f(y)\geq \frac{1}{1-\lambda_y} \left( f\left(\frac{a+b}{2}\right) -\lambda_y f(a) \right)=\frac{2(y_i-a_i)}{b_i-a_i}   f\left(\frac{a+b}{2}\right) - f(a) \text{ for all } i=1,\ldots, n. 
$$
It follows from $(y_i-a_i)/(b_i-a_i)=(y_j-a_j)/(b_j-a_j)\in [0,1]$ for all $i, j=1,\ldots, n$ that  for all  $y\in L_{\frac{a+b}{2} b}$, 
$$
f(y)\geq -2 \left|  f\left(\frac{a+b}{2}\right)\right| -\left|f(a)\right|. 
$$
Hence, $f$ is bounded below on $L_{\frac{a+b}{2} b}$. The proof of $f$ is bounded below on $L_{a\frac{a+b}{2}}$ is analogous. Hence, $f$ is bounded on $L_{ab}$. Therefore, there exists $M>0$ such that  $|f(x)|\leq M$ for all $x\in L_{ab}$.

Let $\delta=\norm{a-x_0}=\norm{b-x_0}$ and $L^\circ_{ab}=L_{ab}\backslash \{a,b\}$.  Hence, for all $x\in L^\circ_{ab}$, $\norm{x-x_0}<\delta$.    
%
%
%
Pick $x\in L^\circ_{ab}\backslash \{x_0\}$. Let $y\in L^\circ_{ab}$ and $\lambda\in (0,1)$ be such that 
\begin{align*}
\lambda\delta \in \left( \norm{x-x_0}, 2\norm{x-x_0} \right) \text{ and }  x=y\lambda x_0. 
 \end{align*} 
 Pick $z\in L^\circ_{ab}$ such that $(y+z)/2=x_0$. Then,  
$x_0=x(1+\lambda)^{-1} z$.  
By convexity of $f$, 
\begin{align*}
f(x)\leq f(y) \lambda f(x_0) \text{  and }  f(x_0)\leq f(x) (1+\lambda)^{-1} f(z),
\end{align*}   
hence,  $\lambda ( f(x_0) - f(z))\leq f(x)-f(x_0)\leq \lambda (f(y) -f(x_0))$.   
Then, by $x,y,z\in L^\circ_{ab}$ and $|f(x')|\leq M$ for all $x'\in L_{ab}$,   
$
-2M\lambda \leq f(x)-f(x_0) \leq 2M\lambda. 
$   
Since $\lambda<(2\norm{x-x_0})/\delta$, 
\begin{align*}
|f(x)-f(x_0)| < \frac{4M}{\delta}\norm{x-x_0}. 
\end{align*}
Since $L$ and $x_0$ are arbitrarily chosen and  $|f(x_0)-f(x_0)| =\norm{x_0-x_0}$, $f$ is linearly continuous.

The proof is analogous for a concave function. 
\prff



We now show that the definition of linear continuity of a relation can be equivalently stated  by using straight line segments instead of  straight lines.   
Let $X$ be a convex set and $\succcurlyeq$ a binary relation on it.  For any $x,y\in X$, the set $L_{xy}=\{x\lambda y| \lambda\in [0,1]\}$ denotes the {\it straight line segment in $X$ connecting $x$ and $y$}.

\lm
  A relation on a convex subset of a Euclidean space is  linearly-continuous iff its restriction to any straight line segment in $X$ is continuous. 
\label{thm: lsegment}
\lmm

\prf[Proof of Lemma \ref{thm: lsegment}]
Let $X$ be a convex subset of $\Re^n$ and $\succcurlyeq$ a binary relation on it. Assume $\succcurlyeq$ is linearly continuous. Pick $x,y\in X$. Let $L$ be the straight line containing $x$ and $y$. Then $L_{xy}\subseteq L$. Hence, the restriction of $\succcurlyeq$ on $L_{xy}$ is continuous. 

  Now assume the  restriction of $\succcurlyeq$ to any straight line segment in $X$ is continuous. Assume there exists a straight line $L$ in $X$ such that $\succcurlyeq ~\!\!\! \uhr L$ is not continuous. First, assume $A_\succcurlyeq(x)\cap L$ is not closed in $L$ for some $x\in X$. Then there exists $y\in \text{cl}A_\succcurlyeq(x)$ and $y\not\succcurlyeq x$. Since $L$ is a straight line, therefore there exists $z\in L$ and $\lambda^t\ra 0$ such that $z\lambda^t y  \in A_\succcurlyeq(x)\cap L$ for all $t$. This furnishes us a contradiction with the continuity of $\succcurlyeq ~ \!\! \!\uhr L_{zy}$.   Second, assume $A_\succ(x)\cap L$ is not open in $L$  for some $x\in X$. Then there exists $y\in A_\succ(x)\cap L$ which is not an interior point of $A_\succ(x)\cap L$. Pick $z\in L$ such that $z\neq y$. Assume there exists $z'\in L$ such that $y=z\delta z'$ for some $\delta\in (0,1)$. Then $\succ ~\!\! \!\uhr L_{zz'}$ does not have open sections (in $L_{zz'}$).  Now assume there does not exist $z'\in L$ such that $y=z\delta z'$ for some $\delta\in (0,1)$. Then $\succ ~\!\! \! \uhr L_{zy}$ does not have open sections (in $L_{zy}$).     These contradict  the continuity of $\succcurlyeq ~ \!\! \!\! \uhr L_{zz'}$ and $\succcurlyeq ~ \!\! \!\! \uhr L_{zy}$, respectively.  
 Analogous arguments yield contradictions if for some $x\in X$, $A_\preccurlyeq(x)\cap L$ is not closed in $L$ or $A_\prec(x)\cap L$ is not open in $L$.
\prff


\prf[Proof of Theorem \ref{thm: linearcontinuous}]
Assume $\succcurlyeq$ is linearly continuous. It follows from Lemma \ref{thm: lsegment} that the restriction of $\succcurlyeq$ to any straight line segment is continuous. Pick $x,y,z\in X$ and let $L_{xy}\subseteq X$ be the straight line segment connecting $x$ and $y$.  Then linear continuity implies that $\{x'\in L_{xy}|x'\succcurlyeq z \}$ is closed in $L_{xy}$. Let $l_{xy}$ denote the mixture-linear function passing through $x$ and $y$. Since $l_{xy}$ is a homeomorphism between $[0,1]$ and $L_{xy}$, therefore the set  
  $l_{xy}^{-1}(\{x'\in L_{xy}|x'\succcurlyeq z \})=I_\succcurlyeq(l_{xy},z)$ is closed in $[0,1]$. Hence, $\succcurlyeq$ is upper mixture-continuous. Similarly, it follows from linear continuity that  $\{x'\in L_{xy}|x'\succ z \}$ is open in $L_{xy}$. Then  $l_{xy}^{-1}(\{x'\in L_{xy}|x'\succ z \})=I_\succ(l_{xy},z)$ is open in $[0,1]$. Hence, $\succcurlyeq$ is upper  strict-Archimedean which implies it is upper Archimedean.   Analogous arguments imply that $\succcurlyeq$ is lower mixture-continuous and lower Archimedean. 

Now assume $\succcurlyeq$ is mixture-continuous and Archimedean. Assume for some straight line 
$L$ in $X$ and some $x\in X$, the set $\{x'\in L|x'\succcurlyeq x\}$ is not closed in $L$. Then  there exist  a sequence $\{y^n\}$ and a point $y$ in $L$ such that $y^n\succcurlyeq x$ for all $n$, $y^n\rightarrow y$ and  $y\not\succcurlyeq x$.  
 Pick $k$ such that the straight line segment connecting $y^k$ and $y$ contains infinitely many members of the sequence. Then, since the mixture-linear function $l_{y^ky}$ passing through $y^k$ and $y$ is an isomorphism between [0,1] and the straight line segment  $L_{y^ky}$ connecting $y^k$ and $y$, the set $I_{\succcurlyeq}(l_{y^ky}, x)$ is not closed. This contradicts mixture-continuity. An analogous argument shows that for all straight line $L$, the lower sections of $\succcurlyeq  ~\!\! \! \uhr L$ is closed in $L$.   
  
It remains to prove that for all straight lines $L$ in $X$,   $\succ~\!\! \!\! \uhr L$ has open sections in $L$. It follows from  of  Galaabaatar-Khan-Uyanik \citeyearpar[Proposition 1]{gku19} that mixture-continuity and Archimedean properties imply strict-Archimedean property. Pick a straight line $L$ in $X$,  $x\in X$ and $y\in A_\succ(x)\cap L$. Pick an $\varepsilon$-neighborhood $N_\varepsilon(y)$ of $y$ in the subspace $L$. Let $\underline y, \bar y$ are the boundary points of $N_\varepsilon(y)$ in $L$. Then strict-Archimedean property implies that $I_\succ(l_{\underline y \bar y}, x)$ is open in $[0,1]$.  Note that there exists $\lambda_y\in [0,1]$ such that $y=l_{\underline y \bar y}(\lambda_y)$.  It is clear that $\lambda_y\in I_\succ(l_{\underline y \bar y}, x)$. For some small $\delta>0$, the set $V(y)=l_{\underline y \bar y}\left(I_\succ(l_{\underline y \bar y}, x)\cap N_\delta(\lambda_y)\right)$ is an open neighborhood of $y$ in the subspace $L$. It is clear that $V(y)\subseteq A_\succ(x)\cap L$. Therefore, $\succ~\!\! \!\! \uhr L$ has open upper sections in $L$. An analogous argument implies that the restricted relation has open lower sections. 
\prff


\prf[Proof of Proposition \ref{thm: continuity0}]
{\nf (a)} The relationship between graph continuity and continuity follows from their definitions.  The relationship between   continuity and linear continuity follows from taking the restriction of the sections on straight lines. The equivalence between linear continuity and mixture continuity \& Archimedean follows form Theorem \ref{thm: linearcontinuous}. Moreover,  Proposition 1 in  Galaabaatar-Khan-Uyanik \citeyearpar{gku19}, and the discussion following it, imply the relationship among mixture continuity, Archimedean and strict-Archimedean.

\smallskip

\nt {\nf (b)}  The relationship between Wold-continuity and weak Wold-continuity follows from their definitions.  

Before moving to part (c), we show that there is no further relationship between these continuity postulates in the absence of additional assumptions. It follows from the examples in Galaabaatar-Khan-Uyanik \citeyearpar[Section 7]{gku19} that there is no further relationship between mixture continuity, Archimedean and strict-Archimedean postulates. The fact that linear continuity does not imply continuity follows from  Theorem  \ref{thm: linearcontinuous} and \citet[Example 1]{in10}. For the fact that graph continuity is stronger than continuity, see Bergstrom-Parks-Rader \citeyearpar{bpr76}. The following example shows that Wold continuity is stronger than weak Wold-continuity. Let $\succcurlyeq$ be a binary relation of $X=\Re^2$ such that $x\succcurlyeq y$ iff $f(x)\geq f(y)$ for all $x,y\in X$, where $f$ is defined as in Example (Genocchi-Peano). It is easy to show that $\succcurlyeq$ is order-dense and weakly Wold-continuous. Pick $x\in \Re^2$ with $x_2=1$ and $x_1\in (0,1)$. Then, $f(x)\in (0,1)$. Hence, $(1,1)\succ x\succ (0,0)$. However, for the unbroken curve connecting (0,0) and (1,1) and consisting of points $\hat x$ such that $\hat x_1^2=\hat x_2$, all points except (0,0) is indifferent to $(1,1)$. Hence, $\succcurlyeq$ is not Wold-continuous.  
We finally illustrate that Wold continuity or weak Wold-continuity neither implies nor implied by any of the continuity postulates in Part (a). Define a relation $\succcurlyeq$ on $[0,1]$ as $x\succ y$  if $x\in (0.5,1], y\in [0, 0.5)$, and $x\sim y$ if  $x,y\in [0,0.5)$ or $x,y\in (0.5,1]$ or $x=0.5, y\in [0,1]$. It is easy to check that $\succcurlyeq$ has closed graph, but it is not weakly Wold-continuous since $1\succ 0$ but there is no $z\in [0,1]$ such that $1\succ z\succ 0$.\fn{Note that this counterexample hinges on the order-denseness property in the definition of  Wold-continuity. It is not difficult to show that for order-dense preferences, continuity implies Wold-continuity and mixture continuity implies weak Wold-continuity.}  
  %

\smallskip

Now assume the ration is complete and transitive. 
\smallskip

\nt {\nf (c)}  When the relation is complete, graph continuity is equivalent to continuity, see \citet[Lemma 1]{wa54a}. 
We next show that  mixture-continuity implies order-denseness. Let $X$ be a convex subset of $\Re^n$ and  $\succcurlyeq$ a complete, transitive and mixture-continuous relation on it.  Pick $x,y\in X$ such that $x\succ y$. Assume for all $z\in X$, $x\nsucc z$ or $z\nsucc y$. Then by completeness, either $z\succcurlyeq x$ or $y\succcurlyeq z$. Then by transitivity, $\{\lambda | x\lambda y\succcurlyeq x\}$ and $\{\lambda | x\lambda y\preccurlyeq y\}$ constitute a partition of $[0,1]$. This contradicts mixture-continuity of $\succcurlyeq$ and connectedness of $[0,1]$.  

Now we show that continuity implies Wold-continuity. 
Let $X$ be a convex subset of $\Re^n$ and  $\succcurlyeq$ a complete, transitive and continuous relation on it.  Pick $x,y\in X$ such that $x\succ y$. 
 Then part (a) implies $\succcurlyeq$ is mixture-continuous, hence it is order-dense. Therefore, there exists $z$ such that $x\succ z\succ y$. By continuity, the sets $R(z)$ and $R^{-1}(z)$ are closed, hence their restrictions to any unbroken curve $C_{xy}$ connecting $x$ and $y$ are closed in the subspace. By completeness $C_{xy}\cap (R(z)\cup R^{-1}(z))=C_{xy}$. Then, connectedness of  $C_{xy}$ implies that  $C_{xy}\cap (R(z)\cap R^{-1}(z))\neq \emptyset$.  Hence,  there exists $z'\in C_{xy}$ such that $z'\sim z$. Therefore, $\succcurlyeq$ is Wold-continuous.


%
\smallskip

\nt {\nf (d)}  The equivalence between strict-Archimedean and mixture-continuity follows from the completeness of the relation. The fact that mixture continuity implies Archimedean follows from Galaabaatar-Khan-Uyanik \citeyearpar[Proposition 1]{gku19}. We next show that mixture-continuity implies weak Wold-continuity. 
Let $X$ be a convex subset of $\Re^n$ and  $\succcurlyeq$ a complete, transitive and continuous relation on it.  Pick $x,y\in X$ such that $x\succ y$.  Mixture-continuity implies order-denseness, hence there exists $z$ such that $x\succ z\succ y$. By mixture-continuity, the sets $\{\lambda | x\lambda y\succcurlyeq z\}$ and $\{\lambda | x\lambda y\preccurlyeq z\}$ are closed, and by completeness $[0,1]=\{\lambda | x\lambda y\succcurlyeq z\} \cup \{\lambda | x\lambda y\preccurlyeq z\}$. Then, connectedness of $[0,1]$ implies that  $\{\lambda | x\lambda y\succcurlyeq z\} \cap \{\lambda | x\lambda y\preccurlyeq z\}\neq \emptyset$.  Hence,  $x\lambda y\sim z$ for some $\lambda\in (0,1)$.   
%
Therefore $\succcurlyeq$ is weakly Wold-continuous.

\smallskip

\nt {\nf (e)}\fn{The proof of this claim is presented in \cite[Theorem 1]{gku20b}. We present a proof here for completeness.}  Assume there exists $x,y,z\in X$ such that $x\succ y$ but for all $\lambda \in (0,1)$, $x\lambda z\nsucc y$ (the proof of the case where $x\nsucc y\lambda z$ is analogous). By completeness, $y\succcurlyeq x\lambda z$. Pick $\lambda \in (0,1)$. If $y\succ x\lambda z$, then $x\succ y\succ  x\lambda z$ and weak Wold-continuity imply there exist $\delta \in (0,1)$ such that $y\sim x\delta z$.   If $y\sim  x\lambda z$, then set $\delta=\lambda$.  Then by transitivity, $x\succ x\delta z$. By weak Wold-continuity, there exists $\gamma\in (0,1)$ such that $x\succ x\gamma z\succ x\delta z\sim y$. Hence, by transitivity, $x\gamma z\succ y$. This furnishes us a contradiction with the assumption that for all $\lambda' \in (0,1)$, $x\lambda' z\nsucc y$. 

\smallskip

The binary relation in the example we provide above for the relationship between linear continuity and continuity is complete and transitive, hence the fact that  continuity is stronger than linear continuity still holds under completeness and transitivity assumptions. The fact that Archimedean property does not imply any other continuity postulates listed in the proposition, see Ghosh-Khan-Uyanik \citeyearpar[Example 5, Appendix B]{gku20b}.

%
%
%
%

Now consider the following example. Let $X=\Re_+$ and $f(x)=\text{sin}(1/x)$ if $x>0$ and $f(0)=1$. Then, it is clear that $f(x)\in [-1,1]$ for all $x\in X$. Define a binary relation $\succcurlyeq$ on $X$ as  $x\succcurlyeq y$ is and only if $f(x)\geq f(y)$. Pick $\bar x$ such that $f(\bar x)\in (0,1)$. The set $A_\preccurlyeq(\bar x)=\{x'\in X|\bar x\succcurlyeq x'\}$ is not closed since it contains a sequence $x_n\ra 0$ but $0\succ \bar x$. Therefore, $\succcurlyeq$ is not continuous. Moreover, since $X$ is one dimensional, $\succcurlyeq$ is not mixture-continuous. 

We next show that $\succcurlyeq$ satisfies Wold-continuity. Pick $x,y\in X$ such that $x\succ y$. Assume without loss of generality $x>y$. Since $[f(y),f(x)]\subseteq f([y,x])$, therefore there exists $z$ such that $x\succ z\succ y$. Analogously, for all $z$ such that $x\succ z\succ y$, there exists a point $w$ on the line connecting $x$ and $y$ such that $w\sim z$. Therefore, $\succcurlyeq$ is weakly Wold-continuous. Since $X$ is one dimensional, $\succcurlyeq$ is   Wold-continuous. Hence, Wold-continuity does not imply continuity. 
  
We finally show that $\succcurlyeq$ satisfies Archimedean property.  Pick $x,y,z\in X$ such that $x\succ y$.  Assume without loss of generality $x>y$. Recall that  $[f(a),f(b)]\cup [f(b), f(a)]\subseteq f([a,b])$ for all $a<b$.  Therefore, there exists $\lambda, \delta\in (0,1)$ such that $f(x\lambda z)>f(y)$ and $f(x)>f(y\delta z)$.  Hence, Archimedean and weak Wold-continuity postulates do not imply mixture-continuity. 
%
%
\prff


We next turn to the proof of Theorem \ref{thm: linearcontinuity}  and then to the proof of Theorem \ref{thm: ccontinuous}. Before that,  we present a generalization of  Rockafellar's theorem to locally convex sets, and three preliminary results on continuity and convexity of a binary relation.       We start with the  generalization of  Rockafellar's theorem.

\lm
Let $X$ be a nonempty locally convex subset of $\Re^n$. Then for all $y\in \mbox{\nf cl}X$ there exists an open neighborhood $U^y$ of $y$ such that $\text{\nf ri}(X\cap U^y)$ is non-empty, and  for all $x\in  \mbox{\nf ri} (X\cap U^y)$ and all $\lambda\in [0,1),$ $y\lambda x \in \mbox{\nf ri}(X\cap U^y).$ 
\label{thm: locallyconvex}
\lmm


\prf[Proof of Lemma \ref{thm: locallyconvex}]
Pick $y\in \mbox{\nf cl}X$  and let $U(y)$ be the open neighborhood of $y$ such that $U(y)\cap X$ is convex.  
 Note that $y\in \text{\nf cl}(X\cap U(y))$. Otherwise,  there exists an open neighborhood $V(y)$ of $y$ such that $(X\cap U(y))\cap V(y)=\emptyset$. However, $U(y)\cap V(y)$ is an open neighborhood of $y$, hence $(X\cap U(y))\cap V(y)=X\cap( U(y)\cap V(y))$ furnishes us a contradiction with  $y\in \text{\nf cl}X$.  
 Then, $U(y)\cap X$  is non-empty as its closure is non-empty. The conclusion follows from 
 Theorem \ref{thm: rockafellar}.
\prff


 The following two lemmas provide partial equivalences between scalar and section  continuity postulates in the presence of a convexity assumption without the completeness and transitivity assumptions. 

\lm
Let $\succcurlyeq$ be a binary relation on a convex subset $X$ of a Euclidean space.     
\ben[{\nf (a)}, topsep=1pt]
\setlength{\itemsep}{-1pt} 
\ml  If $\succcurlyeq$ has locally convex upper  sections, then it has closed upper sections iff it is upper mixture-continuous.     
\ml  If $\succcurlyeq$ has locally convex lower sections, then it has closed lower sections iff it is lower mixture-continuous. 
\een
%
%
\label{thm: ccontinuity}
\lmm

\prf[Proof of Lemma \ref{thm: ccontinuity}]
Assume $\succcurlyeq$ has closed upper sections. Pick $x,y,z\in X$ and a sequence $\{\lambda^t\}$ such that  $x\lambda^t y\succcurlyeq z$ for all $t$ and $\lambda^t\rightarrow \lambda$. Define $w^t=(x_1\lambda^t y_1, \ldots, x_n\lambda^t y_n)$. Then $w^t\rightarrow w=x\lambda y$. Since $\succcurlyeq$ has closed upper sections, $x\lambda y\succcurlyeq z$. Hence, $\succcurlyeq$  is upper mixture-continuous. (Note that this direction does not require the local convexity assumption.)

 In order to prove the forward direction, assume $\succcurlyeq$ is upper mixture-continuous but it does not have closed upper sections. Then there exists $x_0\in X$ such that  $A_\succcurlyeq(x_0)$ is not closed, hence there exists $x\in \text{cl}A_\succcurlyeq(x_0)$ such that $x\not\succcurlyeq x_0$. Since its closure is nonempty, $A_\succcurlyeq(x_0)$ is non-empty. Then local convexity property and Lemma \ref{thm: locallyconvex} implies that there exists  $y\succcurlyeq x_0$ such that $x\lambda y\succcurlyeq x_0$ for all $\lambda\in [0,1)$. Since $x1y=x\not\succcurlyeq x_0$, this furnishes us a contradiction with upper mixture-continuity. Therefore, $\succcurlyeq$ has closed upper sections.
 
 The proof of the equivalence between the closed lower sections and lower mixture-continuity properties is analogous. 
\prff


\lm
Let $\succcurlyeq$ be a binary relation on a set $X$ with property {\bf C}.\fn{This proposition is false for an arbitrary convex subset of a Euclidean space; see Example \ref{exm: weakstrict} in Appendix  \ref{sec_examples}.} 
\ben[{\nf (a)}, topsep=3pt]
\setlength{\itemsep}{-1pt} 
\ml If $\succ$ has locally convex upper   sections, then it has open upper sections iff $\succcurlyeq$ is upper strict-Archimedean. \label{it: ousections}

\ml If $\succ$ has locally convex lower  sections, then it has open lower sections iff $\succcurlyeq$ is lower strict-Archimedean. \label{it: olsections}
\een
\label{thm: ocontinuity}
\lmm


%
%

%

\nt The construction in the proof of Lemma \ref{thm: ocontinuity} is similar to that of Theorem \ref{thm: quasi-concave}. 

\prf[Proof of Lemma \ref{thm: ocontinuity}]
We provide a detailed proof of part \ref{it: ousections}, the  proof of part \ref{it: olsections} is analogous. Assume $\succ$ has open upper sections. Pick $x,y,z\in X$ and $\lambda$ such that $x\lambda y\succ z$. Then open upper sections property implies that there exists $\varepsilon >0$ such that for all $w$ in the $\varepsilon$ neighborhood $ N_{\varepsilon}$ of $x\lambda y$, $w\succ z$. Since  there exists $\epsilon>0$ such that $x\delta y\in N_\varepsilon$ for all $\delta\in (\lambda -\epsilon, \lambda+\epsilon)\cap [0,1]$, the relation $\succcurlyeq$ satisfies the upper strict-Archimedean property.  (Note that this direction does not require the local convexity assumption.)

 In order to prove the backward direction, assume $\succcurlyeq$ is upper strict-Archimedean but it does not have open upper sections. Then there exists $x_0\in X$ such that  $A_\succ(x_0)$ is not open (in  $X$), hence there exists $x\succ x_0$ such that $x$ is not an interior point of $A_\succ(x_0)$. It follows from Lemma \ref{thm: locallyconvex}  that $x$ has a neighborhood $V\subseteq X$ which is open in $X$ such that $V\cap A_\succ(x_0)$ is convex and has non-empty relative interior.    
 Let $\mathcal H_x$ denote the set of all supporting hyperplanes of $V\cap A_\succ(x_0)$ at $x$, and for all $h\in \mathcal H_x$, let $H_h$ denote the closed half space determined by $h$ which contains  $V\cap A_\succ(x_0)$. 
 
 First, assume there exists $h\in \mathcal H_x$ such that $C=H^c_h\cap V\neq \emptyset$. Since $H^c_h\cap V$ is non-empty and convex, Theorem \ref{thm: rockafellar} implies that $x\delta w\nsucc x_0$ for all $\delta\in [0, 1)$ and all $w$ in the relative interior of $H^c_h\cap V$. Since $x\succ x_0$, therefore $I_\succ(l_{xw},x_0)=\{1\}$, which is not open in $[0,1]$. This furnishes us a contradiction with the upper strict-Archimedean property.

Second, assume  for all $h\in \mathcal H_x$, the set $H^c_h\cap V= \emptyset$. This case happens only if $X$ is a polyhedron and  $x$ lies in one of the hyperplanes that determines the set $X$. (Hence, when $X$ is open, the first case is exhaustive.) 
Let $\mathcal H'_x$ be the set of all such hyperplanes.  It is easy to show that there exists $h\in \mathcal H'_x$ and $y\in h$ such that for all $\delta\in [0, 1)$, $x\delta y\notin V\cap A_\succ(x_0)$. (Otherwise, there exists an open ball around $x$ that is contained in $V\cap A_\succ(x_0)$.) This furnishes us a contradiction with the upper strict-Archimedean property. Therefore, $\succ$ has open upper sections.  
\prff 



\lm
Let $\succcurlyeq$ be a reflexive, mixture-continuous and Archimedean  binary relation on a convex subset $X$ of $\Re^n$ which demonstrates convex indifference and has a transitive  symmetric part. Then, $\succcurlyeq$ and its asymmetric part $\succ$ have convex sections.  
\label{thm: fullconvexity}
\lmm

\prf[Proof of Lemma \ref{thm: fullconvexity}]
The convexity of the sections of $\succcurlyeq$ follows from Lemma 2 of Galaabaatar-Khan-Uyanik \citeyearpar{gku19}. The following claim implies that  in order to show $\succ$ is convex, it is enough to prove that $I_\succ(l_{xy},z)$ and $I_\prec(l_{xy},z)$ are convex for all $x,y,z\in X$. 
\cl
$\succ$ is convex $\Longleftrightarrow$ $I_\succ(l_{xy},z)$ and $I_\prec(l_{xy},z)$ are convex for all $x,y,z\in X$. 
\label{thm: sconvex}
\cll


Pick $x,y,z\in X$. We next show that $I_\succ(l_{xy},z)$ is convex. Note that  Theorem 2 and Proposition 1 of Galaabaatar-Khan-Uyanik \citeyearpar{gku19} imply  that $\succcurlyeq$ is semi-transitive and strict-Archimedean.   If $I_\succ(l_{xy},z)$ is empty, then it is convex. If $I_\sim(l_{xy},z)$ is empty, then $I_\succ(l_{xy},z)=I_{\succcurlyeq}(l_{xy},z)$, hence it is convex. Now assume  $I_\sim(l_{xy},z)$ is non-empty.  Then $I_\succcurlyeq(l_{xy},z)$  is non-empty.  
 It follows from the fact that $\succcurlyeq$ demonstrates convex indifference and is mixture-continuous that both  $I_\sim(l_{xy},z)$ and $I_\succcurlyeq(l_{xy},z)$ are closed intervals in $[0,1]$.  
 
 Now assume $I_{\succ}(l_{xy},z)$ is not convex. The  convexity and strict-Archimedean property of $\succcurlyeq$ imply that there exist two nonempty, disjoint, open intervals $I_1, I_2$ such that $I_\succ(l_{xy},z)=I_1\cup I_2$ where the intervals of type $[0,\alpha)$ or  $(\alpha, 1]$ are allowed for $\alpha\in (0,1)$. Note that $I_\succcurlyeq(l_{xy},z)=I_1\cup I_{\sim}(l_{xy},z)\cup I_2$ and $I_\sim(l_{xy},z)=[\lambda_1,\lambda_2]$ for some  $\lambda_1 \leq \lambda_2$.   
Therefore, $I_1\cup I_2=[0,\lambda_1)\cup (\lambda_2,1]$. Without loss of generality assume $I_1=[0,\lambda_1)$. 
 
 Assume there exists $\lambda \in I_1$ such that $I_2\not\subseteq I_\prec(l_{xy},x\lambda y)$, that is, there exist $\lambda\in I_1,\delta \in I_2$ (hence, $x\lambda y\succ z, x\delta y\succ z$)  such that  $x\lambda y\nsucc x\delta y$.  It follows from $x\lambda y\succ z$ and  semi-transitivity of $\succcurlyeq$ that $[\lambda_1, \lambda_2]\subseteq I_\prec(l_{xy},x\lambda y)$. Then it follows from $\succ$ has open sections that there exist $\delta_1\in I_1$ and $\delta_2\in I_2$ such that $\delta_1, \delta_2\in I_\prec(l_{xy},x\lambda y)$. Reflexivity of $\succcurlyeq$ implies that $\lambda\in I_\sim(l_{xy},x\lambda y)\cap I_1$. Let $\gamma_1$ be the smallest scalar $\lambda'$ such that $x\lambda' y\sim x\lambda y$. It is clear that $\gamma_1\leq \lambda$, hence $\gamma_1\in I_1$. We next show that there exists $\gamma_2\in I_2$ such that $x\gamma_2 y\sim x\lambda y$. Assume such $\gamma_2$ does not exist. Then $I_\preccurlyeq(l_{xy},x\lambda y)\cap I_2=I_\prec(l_{xy},x\lambda y)\cap I_2$. Convexity of the set on the left hand side of the equation,  the openness and the non-emptiness  of the set on the right hand side and $I_2\not\subseteq I_\prec(l_{xy},x\lambda y)$ imply that those sets are equal to $(\lambda_2, t)$ for some $t<1$. Since $I_\preccurlyeq(l_{xy},x\lambda y)$ is convex and $I_2=(\lambda_2, 1]$, therefore $I_\preccurlyeq(l_{xy},x\lambda y)=[\gamma_1, t)$. This furnishes us a contradiction with mixture-continuity. Hence, there exists $\gamma_2\in I_2$ such that $x\gamma_2 y\sim x\lambda y$. 
  Then $\gamma_1, \gamma_2\in I_\succcurlyeq(l_{xy},x\lambda y)$. Since $\succcurlyeq$ is convex, therefore $[\gamma_1, \gamma_2]\subseteq I_\succcurlyeq(l_{xy},x\lambda y)$. This furnishes us a contradiction with $[\lambda_1, \lambda_2]\subseteq I_\prec(l_{xy},x\lambda y)$.
  
Now assume $I_2\subseteq I_\prec(l_{xy},x\lambda' y)$ for all $\lambda'\in I_1$. Pick $\lambda\in I_2$. An argument analogous  to the one presented in the previous paragraph yields a contradiction. 

Therefore, $I_{\succ}(l_{xy},z)$ is convex. The proof of the convexity of $I_{\prec}(l_{xy},z)$ is analogous. Then, Claim \ref{thm: sconvex} implies that $\succ$ is convex.

\prf[Proof of Claim \ref{thm: sconvex}]

 Assume $\succ$ is convex.    Pick $x,y,z\in X$ and $\lambda, \delta\in I_\succ(l_{xy},z)$ and $\beta\in [0,1].$ Define $w_\lambda=x\lambda y$ and $w_\delta=x\delta y.$ By construction, $w_\lambda\succ z$ and $w_\delta\succ z.$ It follows from $\succ$ is convex  that $w_\lambda \beta w_\delta\succ z.$ A simple algebra implies $w_\lambda \beta w_\delta=x(\beta\lambda+(1-\beta)\delta)y.$  Therefore, $\beta\lambda+(1-\beta)\delta\in I_\succ(l_{xy},z).$ Hence, $I_\succ(l_{xy},z)$ is convex.  An analogous argument proves the convexity of $I_\prec(l_{xy},z)$
Now assume $I_\succ(l_{xy},z)$ is convex for all $x,y,z\in X.$  Pick $x,y,z\in X$ and $\lambda\in [0,1]$ such that $x\succ z$ and $y\succ z.$ Then $0,1\in I_\succ(l_{xy},z).$ Since $I_\succ(l_{xy},z)$ is convex, therefore  $I_\succ(l_{xy},z)=[0,1].$ Hence, $x\lambda y\succ z.$ Therefore,  $\succ$ has convex upper sections. An analogous argument implies the convexity of lower sections of $\succ$. 
\prff  

The proof of Lemma \ref{thm: fullconvexity} is complete. 
\prff


We are now ready to prove Theorem  \ref{thm: linearcontinuity}. 

\prf[Proof of Theorem \ref{thm: linearcontinuity}] For each part, the backward direction is clear. It remains to prove that linear continuity implies continuity.
\smallskip

\nt  $\mathbf{\ref{it: completeconvex}}$  Theorem \ref{thm: linearcontinuous} implies that linear continuity of $\succcurlyeq$ is equivalent to its mixture-continuity and  its Archimedean. It follows from the local convexity assumption and Lemmas  \ref{thm: ccontinuity} and \ref{thm: ocontinuity} that the upper sections of $\succcurlyeq$ are closed  and of $\succ$ are open. Then completeness of $\succcurlyeq$ implies the upper sections of $\succcurlyeq$ are closed  and of $\succ$ are open. Therefore, $\succcurlyeq$ is continuous. 
\medskip

\nt $\mathbf{\ref{it: convexconcave}}$ The proof follows from Theorem \ref{thm: linearcontinuous}  and Lemmas  \ref{thm: ccontinuity} and \ref{thm: ocontinuity}.
\medskip

\nt $\mathbf{\ref{it: reflexivelinear}}$ Lemma  \ref{thm: fullconvexity} implies that the sections  of $\succcurlyeq$ and of $\succ$ are convex. Then part \ref{it: convexconcave} above completes the proof. 

The proof of Theorem  \ref{thm: linearcontinuity} is complete. 
\prff

%


We now define additional 
 convexity postulates and provides a result on their relationship to those we defined in Section \ref{sec: results}.  
The relation $\succcurlyeq$ is {\it star-convex} if for all distinct $x, y\in X$  and all $\lambda\in (0,1),$  $x\succcurlyeq y$ implies $x\lambda y \succcurlyeq y$; and {\it strictly star-convex} if for all distinct $x, y\in X$  and all $\lambda\in (0,1),$  $x\sim y$ implies $x\lambda y \succ y.$  
  
\lm
Let $\succcurlyeq$ be a complete and transitive relation on a convex subset $X$ of a vector space. Then the following are equivalent: 
\begin{enumerate*}[label=(\roman*)]
\ml[(C1)] $\succcurlyeq$ is convex, \label{it: wconvex}
\ml[(C2)] $\succ$ is convex, \label{it: sconvex}
\ml[(C3)] $\succcurlyeq$ is star-convex. \label{it: wsconvex}
\end{enumerate*}
\vspace{-17pt} 

\nt  If $\succcurlyeq$ is 
 lower Archimedean, then star-convexity of $\succ$ implies the convexity properties (C1)--(C3). Moreover,  if $\succcurlyeq$ is weakly Wold-continuous, then its  strict star-convexity implies $\succ$ is star-convex, hence the convexity properties (C1)--(C3).\fn{Some of the relationships can be found elsewhere; see for example \cite{de59}. We provide the proofs of them also for completeness.} 
\label{thm: convexall}
\lmm
 

\prf[Proof of Lemma \ref{thm: convexall}]
\nt $\text{\nf (C1)} \Rightarrow \text{\nf (C2)}$ Pick $x\in X$ and $y,z\in A_{\succ}(x)$ and $\lambda\in (0,1)$. Assume without loss of generality that $z\succcurlyeq y$. Since $A_\succcurlyeq(y)$ is convex, therefore $y\lambda z\succcurlyeq y$.    Therefore, transitivity of $\succcurlyeq$  and  $y\succ x$ imply\fn{See  Khan-Uyanik \citeyearpar[Proposition 2, Theorems 5 and 6 and Figure 3]{ku21et} for the relationship among different transitivity postulates.} that $y\lambda z\succ x$. 
\smallskip

\nt $\text{\nf (C2)} \Rightarrow \text{\nf (C3)}$ Assume there exists $x,y\in X$ and $\lambda\in (0,1)$ such that $y\succcurlyeq x$ and $y\lambda x\prec x$. Transitivity of $\succcurlyeq$ implies $y\lambda x\prec y$. Then convexity of $\succ$ implies $y\lambda x\succ y\lambda x$. This furnishes us a contradiction. 

\smallskip

\nt $\text{\nf (C3)} \Rightarrow \text{\nf (C1)}$ Pick $x\in X$ and $y,z\in A_{\succcurlyeq}(x)$ and $\lambda\in (0,1)$. Assume without loss of generality that $z\succcurlyeq y$. Then  star-convexity of $\succcurlyeq$ implies $y\lambda z \succcurlyeq y$. It follows from transitivity of $\succcurlyeq$ that $y\lambda z \succcurlyeq x$. 

Now assume $\succcurlyeq$ is 
lower Archimedean and $\succ$ is star-convex. Assume there exist $x,y\in X$ and $\lambda\in (0,1)$ such that $y\succcurlyeq x$ and  $x\succ x\lambda y$. Then 
lower Archimedean property and completeness of $\succcurlyeq$ imply that there exists $\delta\in (0,1)$ such that  $x\succ (x\lambda y)\delta y=x(\lambda\delta)y$. 
Then star-convexity of $\succ$ implies that $x\lambda y\succ x(\lambda\delta) y$.  
It follows from transitivity and $y\succcurlyeq x\succ x\lambda y$ that $y\succ x\lambda y$.  Then star-convexity of $\succ$  implies that $x(\lambda\delta) y\succ x\lambda y$.   
This furnishes us a contradiction. Therefore $\succcurlyeq$ is star-convex.\fn{Note that, the lower Archimedean property is not redundant in this result. In order to see this, consider the following example: $X=[0,1]$, $\succcurlyeq$ is a reflexive preference relation on $X$ such that for ll $x,y\neq 0.5$, $x\sim y$ and $x\succ 0.5$. It is clear that $\succcurlyeq$ is complete and transitive, not lower Archimedean and $\succ$ is star-convex. However, $\succcurlyeq$ is not star-convex since $0\succcurlyeq 1$ and $0\lambda 1\prec 1$ for $\lambda=0.5$. } 

Lastly assume $\succcurlyeq$ is 
%
%
%
 weakly Wold-continuous. 
Now let   $\succcurlyeq$ be strictly star-convex.  
Assume there exist $x,y\in X$ and $\lambda\in (0,1)$ such that  $y\succ x$ and  $x\lambda y\preccurlyeq x$. Let $z=x\lambda y$. If $z\prec x$, then weak Wold-continuity implies that there exists $\delta\in (0,1)$ with $x\succ x\delta z\succ z$. If $z\sim x$, then strict star-convexity implies that for all $\delta\in (0,1)$, $x\delta z\succ z$. Pick $\delta\in (0,1)$ such that $x\delta z \succ z$. Then, it follows from weak Wold-continuity  and $y\succ x\delta z\succ z$ that there exists $\gamma\in (0,1)$ such that $y\gamma z \sim  x\delta z$. Since $z=(y\gamma z)\mu (x\delta z)$ for some $\mu\in (0,1)$, strict star-convexity implies $z\succ x\delta z$. Then, $x\delta z\succ z$ and transitivity of $\succcurlyeq$ yield a contradiction.  
\prff



\prf[Proof of Theorem \ref{thm: ccontinuous}]
Proposition \ref{thm: continuity0}  provides partial relationships among the continuity postulates without the convexity or monotonicity postulates. 
%
We first show that the remaining  relationships hold under convexity of the relation, and then we replace the convexity assumption with weak monotonicity. 

In the first part of the proof, let $\succcurlyeq$ be a convex, complete and transitive binary relation on a convex subset of $\Re^n$ which satisfies property {\bf C}.  

First, by Lemma \ref{thm: convexall}, convexity of $\succcurlyeq$ implies convexity of $\succ$. Therefore, by Theorem \ref{thm: linearcontinuity},  linear continuity implies continuity.   

Second, we show that Archimedean property implies mixture-continuity.\fn{Note that the second and third parts of the proof, that is the proof of mixture continuity follows from each of Archimedean and weak Wold-continuity properties, are independent of the dimension of the space. We return to this in the subsequent results. \label{fn: vectorspace}} To this end pick $x,y,z\in X$. Assume there exists a convergent sequence $\lambda^n$ in $[0,1]$ such that $x\lambda^n y\succcurlyeq z$ for all $n$ and $z\succ x\lambda y$. Since $A_\succcurlyeq(z)$ is convex, if $\lambda^k \leq \lambda \leq\lambda^m$ for some $k,m$, then $x\lambda y\succcurlyeq z$. Therefore, either $\lambda^n>\lambda$ for all $n$ or $\lambda^n<\lambda$ for all $n$. Assume wlog $\lambda^n>\lambda$ for all $n$. Then convexity of $\succcurlyeq$ implies that $x\lambda' y\succcurlyeq z$ for all $\lambda'\in (\lambda, \lambda^1]$.   
 %
It follows from Archimedean property and $z\succ x\lambda y$ that  there exists $\delta\in (0,1)$ such that $z\succ (x\lambda y)\delta(x\lambda^1 y)$. Since $(x\lambda y)\delta(x\lambda^1 y)=x(\delta \lambda+ (1-\delta)\lambda^1)y$ and $\delta \lambda+ (1-\delta)\lambda^1\in (\lambda, \lambda^1]$, this furnishes us a contradiction. Therefore, $\succcurlyeq$ is upper mixture-continuous.  

Now assume there exists a convergent sequence $\lambda^n$ in $[0,1]$ such that $z\succcurlyeq x\lambda^n y$ for all $n$ and $x\lambda y\succ z$. It follows from Archimedean property that  $x\delta (x\lambda y)\succ z$ and $(x\lambda y)\gamma y \succ z$ for some $\delta, \gamma\in (0,1)$. Note that $x\delta (x\lambda y)=x(\lambda+ \delta(1-\lambda))$ and $(x\lambda y)\gamma y=x(\lambda\gamma)y$, and $\lambda\gamma\leq \lambda\leq \lambda+ \delta(1-\lambda)$. Note that $\lambda\in (0,1)$ implies both inequalities are strict, $\lambda=1$ implies $\lambda\gamma<\lambda$ and $\lambda=0$ implies that $\lambda< \lambda+ \delta(1-\lambda)$.  
These furnish us a contradiction with the convexity assumption and $x\lambda^n y \succcurlyeq z$ for all $n$.  Therefore, $\succcurlyeq$ is lower mixture-continuous.   

Third, we prove that weak Wold-continuity implies mixture-continuity.  Assume there exist $x,y,z\in X$ and a convergent sequence $\lambda^k\ra \lambda$ such that $x\lambda^k y\succcurlyeq z$ for all $k$ and $z\succ x\lambda y$.   
If $\lambda^k \leq \lambda\leq \lambda^m$ for some $k,m$, then convexity of $\succcurlyeq$ implies that $x\lambda y\succcurlyeq z$. Therefore, either $\lambda^n>\lambda$ for all $n$ or $\lambda^n<\lambda$ for all $n$. Assume without loss of generality that $\lambda^n>\lambda$ for all $n$. Then convexity of $\succcurlyeq$ implies that $x\lambda' y\succcurlyeq z$ for all $\lambda'\in (\lambda, \lambda^1]$.  
We next show that there exists $\bar \lambda \in (\lambda, \lambda^1]$ such that $z\sim x\bar \lambda y$.  
If $x\lambda'y\succ z$ for some $\lambda'\in (\lambda, \lambda^1]$, then $z\succ x\lambda y$ and weak Wold-continuity imply that there exists $\delta\in (0,1)$ such that  $z\sim (x\lambda' y)\delta (x\lambda y)=x(\delta\lambda'+(1-\delta)\lambda)y$. Otherwise, $x\lambda' y\sim z$ for all $\lambda'\in (\lambda, \lambda^1]$. 
Now, it follows from transitivity and weak Wold-continuity of $\succcurlyeq$, and  $x\bar\lambda y\sim z  \succ x\lambda y$ that there exists $\hat \lambda\in (\lambda, \bar \lambda)$ such that $x\bar\lambda y\succ x\hat\lambda y\succ x\lambda y$. The transitivity of $\succcurlyeq$ and $z\sim x\bar \lambda y\succ x\hat \lambda y$ imply that $z\succ x\hat\lambda y$. Then, $\hat \lambda \in (\lambda, \lambda^1]$ contradicts convexity of $\succcurlyeq$. Therefore, $\succcurlyeq$ is upper mixture-continuous. 

Now assume there exist $x,y,z\in X$ and a convergent sequence $\lambda^k\ra \lambda$ such that $z\succcurlyeq x\lambda^k y$ for all $k$ and $x\lambda y\succ z$.   
Assume without loss of generality that there exists a subsequence $\lambda^{k_i}$ of $\lambda^k$ such that $\lambda^{k_i}>\lambda$ for all $i=1,2,\ldots$.   
We next show that there exists $\bar \lambda \in (\lambda, \lambda^{k_1}]$ such that $z\sim x\bar \lambda y$.  
If $z\succ x\lambda^{k_1} y$, then $x\lambda y\succ z$ and weak Wold-continuity imply that there exists $\delta\in (0,1)$ such that  $z\sim (x\lambda^{k_1} y)\delta (x\lambda y)=x(\delta\lambda^{k_1}+(1-\delta)\lambda)y$. Otherwise, $z\sim x\lambda^{k_1} y$. 
Then, it follows from transitivity and weak Wold-continuity of $\succcurlyeq$, and  $x\lambda y\succ z\sim x\bar\lambda y$ that there exists $\hat \lambda\in (\lambda, \bar \lambda)$ such that $x\lambda y\succ x\hat\lambda y\succ x\bar\lambda y$. The transitivity of $\succcurlyeq$ and $z\sim x\bar \lambda y\prec x\hat \lambda y$ imply that $z\prec x\hat\lambda y$. Moreover,  convexity of $\succcurlyeq$ implies that for all $\lambda'\in [\lambda, \hat\lambda]$, $x\lambda'y\succcurlyeq x\hat\lambda y$. Since $\lambda^{k_i}\ra \lambda$, there exists $j$ such that $\lambda^{k_j}\in  (\lambda, \hat\lambda)$. Then, $z\succcurlyeq x\lambda^{k_j} y\succcurlyeq x\hat \lambda y \succ z$ furnishes us a contradiction.   
 Therefore, $\succcurlyeq$ is lower mixture-continuous.

In the second part of the proof, assume $\succcurlyeq$ is a weakly monotonic, complete and transitive binary relation on a convex subset $X$ of $\Re^n_+$ which satisfies property {\bf B}.  It remains to show that   each of   Archimedean  and weak Wold-continuity postulates implies the continuity postulate. 
  %
  %
  %
Towards this end, assume $\succcurlyeq$ is Archimedean. 

Assume there exists $x\in X$ such that $A_\preccurlyeq(x)$ is not closed, i.e., there exists $y^k\ra y$ such that $y^k\in A_{\preccurlyeq}(x)$ for all $k$ and   $y\succ x$. By property {\bf B}, there exists $b\in X$ such that $x,y\geq b$. The  Archimedean property implies that  there exists $\lambda\in (0,1)$ such that $y\lambda b\succ x$. Define $z=y\lambda b$. Since $y>b$, therefore $z>b$. Define $\varepsilon=\mmin_{z_i\neq b_i}(y_i-z_i)$. For any $y'$ in the $\varepsilon$ neighborhood of $y$, $y'>z$.  Then weak monotonicity implies that $y'\succcurlyeq z$. Since $z\succ x$, it follows from transitivity that $y'\succ x$ for all $y'$ in the $\varepsilon$ neighborhood of $y$. This furnishes us a contradiction with $y^k\ra y$. 

Now assume  there exists $x\in X$ such that $A_\succcurlyeq(x)$ is not closed, i.e.,  there exists $y^k\ra y$ such that $y^k\in A_{\succcurlyeq}(x)$ for all $k$ and   $x\succ y$. By property {\bf B}, there exists $x'\in X$ such that  $x'\geq x, y$. Then Archimedean property implies that  there exists $\lambda\in (0,1)$ such that $x\succ x' \lambda y$. Note that $x'>  y$. Define $z=x'\lambda y$. Since $x'> y$, therefore $z> y$. Define $\varepsilon=\mmin_{i}(z_i-y_i)$. For any $y'$ in the $\varepsilon$ neighborhood of $y$, $y'< z$.  Then weak monotonicity implies that $z\succcurlyeq y'$. Since $x\succ z$, it follows from transitivity that $x\succ y'$. This furnishes us a contradiction with $y^k\ra y$. Therefore, $\succcurlyeq$ is continuous.   
%
%


  
 Finally, we show that continuity follows from weak Wold-continuity. To this end, assume $\succcurlyeq$ is weakly Wold-continuous.  
Assume there exists $x\in X$ such that $A_\preccurlyeq(x)$ is not closed, i.e., there exists $y^k\ra y$ such that $y^k\in A_{\preccurlyeq}(x)$ for all $k$ and   $y\succ x$. By property {\bf B}, there exists $b\in X$ such that $x,y\geq b$.  
 %
 %
 Order denseness of $\succcurlyeq$ implies that there exists $z'$ such that $y\succ z'\succ x$.  It follows from $\succcurlyeq$ is transitive and weakly monotonic,  and $y,z'>b$ that $y\succ z'\succ b$. Then,  weak Wold-continuity implies that $z'\sim y\lambda b$ for some $\lambda\in (0,1)$. Define $z=y\lambda b$. Since $y>b$, therefore $z>b$. Define $\varepsilon=\mmin_{z_i\neq b_i}(y_i-z_i)$. For any $y'$ in the $\varepsilon$ neighborhood of $y$, $y'>z$.  Then weak monotonicity implies that $y'\succcurlyeq z$. Since $z\sim z'\succ x$, it follows from transitivity that $y'\succ x$ for all $y'$ in the $\varepsilon$ neighborhood of $y$. This furnishes us a contradiction with $y^k\ra y$. 
 

Now assume  there exists $x\in X$ such that $A_\succcurlyeq(x)$ is not closed, i.e.,  there exists $y^k\ra y$ such that $y^k\in A_{\succcurlyeq}(x)$ for all $k$ and   $x\succ y$.   
 Then order denseness of $\succcurlyeq$ implies that there exists $z'$ such that $x\succ z'\succ y$.   By property {\bf B}, there exists $x'\in X$ such that  $x'\geq x, y$.   Therefore, weak monotonicity and transitivity properties imply that $x'\succ z'\succ y$.  
Then, weak Wold-continuity implies that $z'\sim x'\lambda y$ for some $\lambda\in (0,1)$.    
  Define $z=x'\lambda y$. Since $x'> y$, therefore $z> y$. Define $\varepsilon=\mmin_{i}(z_i-y_i)$. For any $y'$ in the $\varepsilon$ neighborhood of $y$, $y'< z$.  Then weak monotonicity implies that $z\succcurlyeq y'$. Since $x\succ z$, it follows from transitivity that $x\succ y'$. This furnishes us a contradiction with $y^k\ra y$ and $y^k\in A_{\succcurlyeq}(x)$ for all $k$ . Therefore, $\succcurlyeq$ is continuous. 
\prff


\prf[Proof of Theorem \ref{thm: mcarc}]
 %
 {\bf (a)}  Assume $\succcurlyeq$ is upper mixture-continuous. Pick $x,y,z,\in X$ and a sequence $\lambda^n\in [0,1]$ such that $\lambda^n\rightarrow \lambda$ and $z\succcurlyeq x\lambda^n y$ for all $n$. Since $X$ is a convex cone, $z+y\lambda^n x$ and $x+y$ are in $X$. Note that  $z+x+y-x\lambda^n y=z+ y\lambda^n x=(z+y)\lambda^n (z+x)$ for all $n$. Additivity implies that $(z+y)\lambda^n (z+x)=z+x+y-x\lambda^n y\succcurlyeq x+y$. Since $\succcurlyeq$ is upper  mixture-continuous, $(z+y)\lambda (z+x)=z+x+y-x\lambda y\succcurlyeq x+y$. Additivity implies that $z+x+y \succcurlyeq x+y+x\lambda y$. Since $z, x\lambda y, x+y\in X$, therefore additivity implies that $z\succcurlyeq x\lambda y$. Hence, $\succcurlyeq$ is lower mixture-continuous. The proof of the converse statement is analogous. Then, by definition, mixture continuity follows from any of the upper and lower mixture-continuity. 
\medskip

\nt {\bf (b)} Assume $\succcurlyeq$ is upper strict-Archimedean. Pick $x,y,z,\in X$ and $\lambda\in (0,1)$ such that\fn{When $\lambda=0,1$, then the proof is analogous.} $z\succ x\lambda y$. Additivity implies that $z+x+y-x\lambda y=z+y\lambda x\succ x+y$.
 Note that $z+y\mu x=(z+y)\mu (z+x)$ for all $\mu\in [0,1]$. Upper strict-Archimedean property implies that there exists $\underline \lambda<\lambda< \bar \lambda$ such that $z+y\delta x\succ x+y$ for all $\delta\in (\underline \lambda, \bar \lambda)$. Then additivity and $z, x\delta y\in X$ imply that $z\succ x\delta y$ for all $\delta\in (\underline \lambda, \bar \lambda)$. Hence, $\succcurlyeq$ is lower strict-Archimedean.  The proof of the converse statement is analogous. Then, by definition, strict-Archimedean property follows from any of the upper and lower strict-Archimedean property. 
\prff


\prf[Proof of Theorem \ref{thm: scontinuous}]

Proposition \ref{thm: continuity0} shows that mixture continuity is equivalent to the strict Archimedean postulate, and implies Archimedean and weak Wold-continuity postulates. The proof of mixture-continuity follows from  each of Archimedean and weak Wold-continuity postulates is identical to that provided in the proof of Theorem \ref{thm: ccontinuous} since the dimension of the space is not used in the proof; see Footnote \ref{fn: vectorspace} for details.  
\prff


We next turn to the proof of Theorem \ref{thm: curverelation}.  Before that we present an important result due to \citet[Theorem 3]{ro55} which is crucial for the proof of his remarkable result we provided in the Introduction,  and  a lemma on the continuity of  binary relations. We restate the theorem of Rosenthal with the notation of our paper. 

\thm[Rosenthal]
Any bounded infinite subset of set $X$ in $\Re^n$ with property {\bf C} contains an infinite subset through which a smooth curve can be laid.\fn{Rosenthal states the theorem for $X=\Re^n$, but it is easy to observe that his theorem is true for any subset $X$ of $\Re^n$ with property {\bf C}.}  
\label{thm: rosenthal}
\thmm

The following lemma provides a characterization of strong Archimedean property under strong mixture-continuity.

\lm
  Let $\succcurlyeq$ be a strongly mixture-continuous binary relation on a convex subset of $\Re^n$. Then $\succcurlyeq$ is strong Archimedean iff it is strongly strict-Archimedean. 
\label{thm: sarc}
\lmm
%

\prf[Proof of Lemma \ref{thm: sarc}]
 Assume $\succcurlyeq$ is strongly mixture-continuous and strongly Archimedean. Pick $x,y,z\in X$ and $m_{xy}\in \CM$. If $I_\succ(m_{xy},z)$ is empty, then it is open. Otherwise, pick $\lambda\in I_\succ(m_{xy},z).$ It follows from strong mixture-continuity and  $\lambda \notin I_\preccurlyeq(m_{xy},z)$ that there exists $t>0$ such that $N_t(\lambda)=\{\beta~|~|\beta-\lambda |<t\}$ is contained in the complement of $I_\preccurlyeq(m_{xy},z).$

Assume  there exists $\beta\in N_t(\lambda)\cap I_{\bowtie}(m_{xy},z).$ It follows from strong mixture-continuity that $I_{\bowtie}(m_{xy},z)$ is open. Therefore, as an open set in [0,1], $I_{\bowtie}(m_{xy},z)$ is union of at most countably many mutually disjoint open intervals such that the intervals of type $[0,\alpha)$ or  $(\alpha, 1]$ are allowed for any $\alpha\in (0,1)$.  
 Among these open intervals, there exists open interval $I$ such that $\beta\in I.$ Assume $\beta>\lambda$. Set $\delta=\text{inf } I$.  Then $\delta\in I_\succ(m_{xy},z).$ Note that there exists $m_{m_{xy}(\delta) m_{xy}(\beta)}\in \CM$  such that $m_{m_{xy}(\delta) m_{xy}(\beta)}((0,1))=m_{xy}((\delta,\beta))$.    
  Since $(\delta, \beta)\subseteq I,$ therefore $(\delta, \beta)\subseteq I_{\bowtie}(m_{xy},z)$.  This furnishes us a contradiction with $\succcurlyeq$ is strongly Archimedean. Now assume $\beta>\lambda$. Then setting $\delta=\text{sup } I$ and repeating the step above by replacing $(\delta,\beta)$ with $(\beta, \delta)$ furnishes us a contradiction with strongly Archimedean property. Since $\beta\neq \lambda,$ therefore, $I_\succ(m_{xy},z)$ is open. An analogous argument implies $I_\prec(m_{xy},z)$ is open. Therefore  $\succcurlyeq$ is strongly strict-Archimedean.  

 Now assume $\succcurlyeq$ is strongly mixture-continuous and strongly strict-Archimedean.   Pick $x,y,z\in X, m_{xz}, m_{yz}\in \CM$. Assume $x\succ y$.  Then $m_{xz}(1)=x$ implies that  $1\in I_\succ(m_{xz},y).$  Since $I_\succ(m_{xz},y)$ is open, therefore  there exists $\lambda<1$ such that $(\lambda, 1]\subseteq I_\succ(m_{xz},y).$ Similarly,  $1\in I_\prec(m_{yz},x).$ Since $A_\prec(m_{yz},x)$ is open, therefore there exists $\delta<1$ such that $(\delta, 1]\subseteq I_\prec(m_{yz},x).$  Hence, $\succcurlyeq$ is strongly Archimedean.
\prff




We are now ready to prove Theorem \ref{thm: curverelation}.

\prf[Proof of Theorem \ref{thm: curverelation}]
Lemma \ref{thm: sarc} and an argument analogous to the one presented in the proof of Theorem \ref{thm: linearcontinuous} establish the equivalence between arc-continuity postulate and  the postulates of strong mixture-continuity and strong Archimedean.  

It remains to prove the equivalence between arc-continuity and continuity postulates. The backward direction is obvious. In order to prove the forward direction assume $\succcurlyeq$ is arc-continuous and pick $x_0\in X$ and $x\in \text{cl}A_\succcurlyeq(x_0)$. If $A_\succcurlyeq(x_0)$ is empty or singleton, then it is closed. Otherwise pick a sequence $x_n$ in $A_\succcurlyeq(x_0)$ such that $x_n\rightarrow x$. It follows from Theorem \ref{thm: rosenthal} that there exists a smooth curve\fn{We note for the convenience of the reader that this step of the proof  uses the polyhedron  assumption. The existence of such a smooth curve is also  guaranteed   when the boundary of the domain  of the binary relation is locally smooth at every point.  A natural question arises whether these requirements can be dispensed with, and we leave this question for further investigation.  On this issue, also see Foothonte  \ref{fn_poly} above.  \label{fn_polyR}} containing $x$ and a subsequence $x_{n_k}$ of $x_n$.   Then
 arc-continuity implies that $x\in A_\succcurlyeq(x_0)$.  The closedness of the lower sections of $\succcurlyeq$ follows from an analogous argument. 
 
 Now assume $A_\succ(x_0)$ is not open for some $x_0\in X$. Then there exists $x\in A_\succ(x_0)\cap \text{cl}(A_\succ(x_0))^c$. Pick a sequence $x_n$ in $(A_\succ(x_0))^c$ such that $x_n\rightarrow x$.  It follows from Theorem \ref{thm: rosenthal} that there exists a smooth curve containing $x$ and a subsequence $x_{n_k}$ of $x_n$.   This  implies that $\succcurlyeq$ is not strongly strict-Archimedean. Then the first equivalence we proved above and Lemma \ref{thm: sarc} furnish us a contradiction.     The openness of the lower sections of $\succ$ follows from an analogous argument. 
\prff


Next, we present the proof of Proposition \ref{thm: gul} which uses the construction in \cites{gu92} proof.  

\prf[Proof of Proposition \ref{thm: gul}]  

Let $I^*\in \CN$ satisfying A3(ii), i.e., for all $x,y\in \Delta$, $x I^* y\sim y I^* x$.  

\cl
For all $x,y,z,z'\in \Delta$ with $x>y$ and $z\geq z'$,  {\nf (i)} $x\succ xI^*y \succ y$ and  {\nf (ii)} $xI^* z \succ yI^* z'$. 
\label{thm: monotone}
\cll

Note that Claim \ref{thm: monotone} and mixture continuity imply that for all $x,y\in \Delta$, there exists a unique $z\in \Delta$ such that $z\sim x I^* y$. 

 Let $S=(x^1,\ldots, x^m)$ where $x^t\in \Delta$ for all $t$. We say that {\it $y^0$ reaches $x$ through $S$} if $y^t\sim y^{t-1}I^* x^t$ for $t=1,\ldots, m$ and $y^m=x$.  
\cl
{\nf (i)} For any $x, y^0\in \Delta$, define $y^t\sim y^{t-1}I^* x$ where $y^t\in \Delta$ for $t\geq 1$. Then for all $t$ there exists a unique $\lambda^t\in [0,1]$ such that $y^t=1\lambda^t 0, \lambda^t\ra \bar\lambda$ and $1\bar \lambda 0=x$.   
\smallskip

\nt {\nf (ii)}  
 For $y^0\in \Delta$ and $x\in (0,1)^n$, there exists some finite $S$ such that $y^0$ reaches $x$ through $S$.  
\label{thm: reach}
\cll

Let $\bar S=(x^1,\ldots, x^m)$ where $x^t\in X$ for all $t$. We say that {\it $y^0\in X$ reaches $x\in X$ through $\bar S$} if for all $i\in N$, $y^t_i \sim y^{t-1}_i I^* x^t_i$ for $t=1,\ldots, m$ and $y^m=x$.

\cl

The following is true. 
\smallskip

\nt {\nf (i)} $y^0\in X$ and  $x\in (0,1)^n$ imply that there exists a finite $\bar S$ such that $y^0$ reaches $x$ through $\bar S$. 

\smallskip

\nt {\nf (ii)} If $y^0\in X$ reaches $x\in X$ through $\bar S$ and $\hat y^0$ reaches $\hat x$ through $\bar S$, then $y^0\succ \hat y^0$ iff $x\succ \hat x$ and for all $i\in N$,  $y^0_i\succ \hat y^0_i$ iff $x^t\succ \hat x^t$.  
\label{thm: reach2}
\cll

Now pick $x,y\in X$ such that $x>y$, i.e., $x_i\geq y_i$ for all $i$ and $x_j>y_j$ for some $j$.  We next show that $x\succ y$. We establish the result for the case in which $x_{-j}=y_{-j}$. Then the transitivity of $\succcurlyeq$ yields the desired conclusion. It follows from Claim \ref{thm: reach2}(i) that for $z\in \Delta$, there exists $\bar S$ such that   $x$ reaches $z$ through $\bar S$.   Define $\bar y\in X$ as $\bar y_{-j}=z$ and $\bar y_j=w$ where  $w\in \Delta$ with $w<z$.  
Then Claim  \ref{thm: reach2}(ii) implies that   $y$ reaches $\bar y$ through $\bar S$. By Claim \ref{thm: reach}(i),  we can choose $w$ arbitrarily close to $z$, hence $1\succ \bar y\succ 0$.  Then mixture continuity implies  there exists $\bar z$ such that $\bar z\sim \bar y$. Then A2 implies that $z\succ \bar z\sim \bar y$. Therefore Claim  \ref{thm: reach2}(ii) implies that $x\succ y$.

It remains to prove Claims \ref{thm: monotone}, \ref{thm: reach} and \ref{thm: reach2}.

\prf[Proof of Claim \ref{thm: monotone}]   
Pick $x,y,z,z'\in \Delta$ such that $x>y$ and $z\geq z'$. Then A3(i) implies that $x\succ y$ and $z\succcurlyeq z'$. 
\medskip

\nt (i) Assume towards a contradiction that $xI^* y\succcurlyeq x$. Then $y\prec x$, completeness and mixture-continuity of $\succcurlyeq$ imply that there exists $\lambda\in (0,1]$ such that $(x\lambda y)I^* y\sim x$. Moreover, A3(i) implies that $(x\lambda y)I^*(x\lambda y)\succ yI^*y$. Then $x \sim y I^*(x\lambda y), (x\lambda y)\succ y$ and A2 imply that $(x\lambda y)\succ x$. This contradicts A3(i). Hence,  $x\succ xI^* y$. An analogous argument implies that $xI^*y \succ y$. 
\medskip

\nt (ii)  It follows from mixture continuity, A3(i) and part (i) above that there exists $x', y'\in \Delta$ such that $x'\sim xI^* z$ and $y'\sim yI^* z$. Then $x\succ y$ and A2 imply that $x'\succ y'$, hence $xI^* z\succ yI^* z$.  If $z=z'$, then the proof is complete. If $z>z'$, then A3(i) implies that $z\succ z'$. Then A3(ii) and the argument in part (i) above imply that there exists $y''\in \Delta$ such that $y''\sim yI^* z'$. Hence, A2  imply that $xI^* z\succ yI^* z'$. 
\prff

\prf[Proof of Claim \ref{thm: reach}] 
(i) Assume wlog that $x> y^0$. Then A3(i) and Claim \ref{thm: monotone} imply that $y^t$ is a strictly increasing sequence and $y^t<x$ for all $t$. Since $1\geq x> y^t\geq 0$, therefore for all $t$ there exists a unique $\lambda^t\in [0,1]$ such that $y^t=1\lambda^t 0$. Assume   $1\bar \lambda 0 < x$ where $\bar\lambda = \text{lim }\lambda^t$. Define $\bar  y=1\bar \lambda 0$.    
 Let $\hat y\in \Delta$ be such that $\hat y\sim \bar y I^* x$. Define $\hat \lambda$ such that $1\hat\lambda 0=\hat y$. Claim \ref{thm: monotone} implies that $\bar y<\hat y < x$.  Hence, $\hat y>0.5(\hat y+\bar y)>\bar y > y^{t+1}\sim y^t I^* x$ for all $t$.  Then A3(i) implies that $\hat y \succ 0.5(\hat y+\bar y)\succ y^t I^* x$ for all $t$. Hence,   
 $\hat y \succ 0.5(\hat y+\bar y)\succ (1\lambda^t 0)I^* x$.  
 Note that $(1\bar \lambda 0)I^* x=\bar y I^* x\sim \hat y\succ  0.5(\hat y+\bar y)$.  This contradicts mixture-continuity of $\succcurlyeq$, i.e., the closedness of $\{\lambda\in [0,1] ~|~ 0.5(\hat y+\bar y)\succcurlyeq (1I^*x)\lambda (0I^* x)\}$.  
 \medskip

\nt (ii) Assume wlog $x>y^0$. Let $y^t\sim  y^{t-1}I^*1$. Part (i) implies that $y^t\rightarrow 1$. Let $\tau=\text{\nf inf}\{t~|~y^t>x\}-1$. Since $x<M$ and $y^t\rightarrow 1$, $\tau$ is well-defined. Hence,  $y^{\tau-1}\leq x<y^\tau\sim y^{\tau-1}I^*1$. By A3(i), $y^{\tau-1}I^* y^{\tau-1}\preccurlyeq x \prec  y^{\tau-1}I^*1$. Then mixture continuity implies that there exists $\lambda\in [0,1)$ such that $x\sim (y^{\tau-1}I^* 1)\lambda (y^{\tau-1}I^* y^{\tau-1})=y^{\tau-1}I^* (1\lambda y^{\tau-1})$. Define $z=1\lambda y^{\tau-1}$. Setting $x^i=M$ for $i=1,\ldots, n-1$ and $x^n=z$ establishes the desired set $S$.  
\prff

\prf[Proof of Claim \ref{thm: reach2}] 
(i) follows from a repeated application of Claim \ref{thm: reach}. 

 \medskip

\nt (ii) The first statement in (ii) follows from a repeated application of A2; the second statement follows from Claim \ref{thm: monotone} and A3. 
\prff
\prff


 We next turn to the proof of  Proposition \ref{thm: ahi}. 
 
\prf[Proof of Proposition \ref{thm: ahi}]
$\mathbf{\ref{it: ahit}}$  Note that independence property, $x\succcurlyeq y$ and $\lambda \in (0,1)$ imply that $\lambda x=\lambda x + (1-\lambda )0\succcurlyeq \lambda y+ (1-\lambda) 0=\lambda y$. Then $\succcurlyeq$ is   lower-homothetic.\fn{A binary relation is {\it lower-homothetic} if for all $x,y\in X$ and all $\lambda\in (0,1)$  with $\lambda x,\lambda y\in X$,  $x\succcurlyeq y$ implies $\lambda x\succcurlyeq \lambda y$.   }  The following result of  \citet[Claim 2]{ge15} finishes the proof except that independence implies  homotheticity.  

\cl
A lower-homothetic relation on a vector space is additive iff it is independent. 
\label{thm: ge15}
\cll

 Assume $\succcurlyeq$ is independent. 
  The discussion above and Claim \ref{thm: ge15} imply that the relation is additive and lower-homothetic. Additivity and transitivity imply that if $x\succcurlyeq y$, then $kx=\sum_{i=1}^k x\succcurlyeq \sum_{i=1}^k y=ky$ for all natural number $k$. In order to see this, $x\succcurlyeq y$ implies that $x+x\succcurlyeq x+y \succcurlyeq y+y$. It follows from transitivity that $2x\succcurlyeq 2y$. Repeating this step $k$-many times provides the desired result. Now pick $x\succcurlyeq y$ and $\lambda >1$. Let $k$ be the largest integer smaller than or equal to $\lambda$ and set $\lambda'=\lambda-k$. Note that $\lambda x=kx+\lambda' x$. It follows from additivity that $kx\succcurlyeq ky$ and from lower-homotheticity that $\lambda' x\succcurlyeq \lambda' y$. Then additivity implies that $\lambda x= kx+\lambda' x\succcurlyeq ky+ \lambda' y=\lambda y$. Hence, $\succcurlyeq$ is homothetic. 

\medskip

\nt $\mathbf{\ref{it: ahict}}$ Let $\succcurlyeq$ be a complete, transitive, mixture-continuous and additive relation on $X$. If we show that $\succcurlyeq$ is lower homothetic, then Claim \ref{thm: ge15} above shows that $\succcurlyeq$ is independent.   To this end assume it is not lower homothetic, i.e., there exist $x,y\in X$ with $x\succcurlyeq y$ and $\alpha\in (0,1)$ such that $\alpha x\prec \alpha y$. Additivity implies that $z=x-y\succcurlyeq 0$ and $\alpha z\prec 0$. Additivity and transitivity imply that $kz\succcurlyeq 0$ for any positive integer $k$. Completeness and mixture-continuity imply that $\succcurlyeq$ is strict-Archimedean. Therefore $I_{\prec}(l_{z0}, 0)$ is open. Then there exists an interval $(\alpha_1,\alpha_2)$ which contains $\alpha$ such that $\alpha' z\prec 0$ for all $\alpha'\in (\alpha_1,\alpha_2)$. Pick a rational $\lambda\in (\alpha_1,\alpha_2)$. Then there exist positive integers $m,n$ such that $\lambda=m/n$. Then, transitivity and additivity of $\succcurlyeq$ imply that $mz=\sum_{i=1}^n (m/n)z\prec 0$. This furnishes us a contradiction.    
Therefore, $\succcurlyeq$ is lower homothetic. 

The backward direction follows from part \ref{it: ahit}.  Therefore the proof of Proposition \ref{thm: ahi} is complete. 
\prff

\prf[Proof of Proposition \ref{thm: representation}]

It is clear that part (b) implies part (a) is true.  
Assume part (a).   It follows from  Proposition \ref{thm: ast} that the relation is complete and transitive, from Theorem \ref{thm: mcarc} that it is mixture-continuous, and from Proposition \ref{thm: ahi}  that it is independent. The celebrated representation theorem of Herstein-Milnor completes the proof.
\prff


 \prf[Proof of Proposition \ref{thm: scmonotone}] 
It is easy to show that open sections property implies that the relation has the strict Archimedean property which implies the relation has the Archimedean property.   Therefore, it remains to prove that Archimedean property implies open sections property. To this end,  assume $\succ$ is Archimedean. Pick $x\in \Re_+^n$ and $y\in A_\succ(x)$. Then Archimedean property implies that  there exists $\lambda\in (0,1)$ such that $y\lambda 0\succ x$. Define $z=y\lambda 0$. Since $y>0$, therefore $z>0$. Define $\varepsilon=\mmin_{z_i\neq 0}(y_i-z_i)$. For any $y'$ in the $\varepsilon$ neighborhood of $y$, $y'>z$.  Then strong monotonicity implies that $y'\succ z$. Since $z\succ x$, it follows from transitivity of $\succ$ that $y'\succ x$ for all $y'$ in the $\varepsilon$ neighborhood of $y$. Therefore, $\succ$ has open upper sections. 

Now pick $x\in \Re_+^n, y\in A_\prec(x)$ and  $x'\geq x, y$. Then Archimedean property implies that  there exists $\lambda\in (0,1)$ such that $x\succ x' \lambda y$. Note that $x'>  y$. Define $z=x'\lambda y$. Since $x'> y$, therefore $z> y$. Define $\varepsilon=\mmin_{i}(z_i-y_i)$. For any $y'$ in the $\varepsilon$ neighborhood of $y$, $y'< z$.  Then strong monotonicity implies that $z\succ y'$. Since $x\succ z$, it follows from transitivity of $\succ$ that $x\succ y'$. This furnishes us a contradiction with $y^k\ra y$. Therefore, $\succ$ has open lower sections. 
 \prff


\prf[Proof of Proposition \ref{thm: shafer}]
Continuity implies mixture-continuity, and the converse implication follows from completeness and convexity assumptions and Theorem \ref{thm: linearcontinuity}. It follows from completeness of the relation that mixture-continuity implies Archimedean property.  In order to show the converse implication pick $x,y,z\in \Re_+^n$ and $\lambda\in (0,1)$ such that $x\lambda y\succ z$. It follows from Archimedean property that there exists $\delta,\gamma\in (0,1)$ such that $x(\lambda\delta)y\succ z$ and $y((1-\lambda)\gamma)x\succ z$. Note that $y((1-\lambda)\gamma)x=x(1-(1-\lambda)\gamma)y$ and $\underline\lambda=\lambda\delta<\lambda <1-(1-\lambda)\gamma=\bar\lambda$.  Then the convexity assumption implies that for all $\lambda'\in (\underline \lambda, \bar \lambda)$, $x\lambda' y\succ z$. Hence, $\succcurlyeq$ is upper strict-Archimedean. Therefore, it is lower mixture-continuous.

Now assume that there exists $x,y,z\in \Re_+^n$ and $\lambda^k\in [0,1]$ such that $x\lambda^k y\succcurlyeq z$ for all $k$ with $\lambda^k\ra \lambda$ such that $x\lambda y\prec z$. If there exists $k,l$ such that $\lambda^k<\lambda<\lambda^l$, then the convexity assumption implies that $x\lambda y\succ z$. Then assume wlog that $\lambda^k>\lambda$ for all $k$. It follows from the convexity assumption that $x\lambda' y\succ z$ for all $\lambda'\in (\lambda^1, \lambda)$. Archimedean property and $z\succ x\lambda y$ implies that there exists $\delta\in (0,1)$ such that $z\succ (x\lambda y)\delta (x\lambda^1 y)=x(\delta\lambda+(1-\delta)\lambda^1)y$. Then $\lambda < \delta\lambda+(1-\delta)\lambda^1 <\lambda^1$ furnishes us a contradiction. Hence, $\succcurlyeq$ is upper  mixture-continuous. 
\prff


\prf[Proof of Proposition \ref{thm: ctdubra}]
It follows from Theorem \ref{thm: linearcontinuous} that the relation is linearly continuous, from  Theorem \ref{thm: linearcontinuity}  that it is continuous and from   Khan-Uyanik \citeyearpar[Theorem 2]{ku21et} that it is complete and transitive.   
\prff


We now turn to the proof of Proposition \ref{thm: ast}. The construction of our proof has some similarity to the proof of Neuefeind-Trockel \citeyearpar[Proposition]{nt95}, however their method-of-proof does not work since the topological structure of the space has a central role in their proof. We, instead, define a closure concept for subsets of vector spaces by using the topological structure of the unit interval and then obtain our result by using this closure concept.   
  Let $A$ be a subset of a vector space $X$. The {\it linear closure of $A$} is defined as   
$\overline A=\{z\in X| x,y\in X, \lambda^n\ra \lambda,  x\lambda^n y\in A, z=x\lambda y\}$.    
 In other words, for all $x,y\in X$ and $\lambda^n\ra \lambda$, if $x\lambda^n y\in A$ for all $n$, then $x\lambda y\in \overline A$.\fn{Note that linear closure of a set contains its ``algebraic closure."}


\prf[Proof of Proposition \ref{thm: ast}]
Theorem \ref{thm: mcarc} implies that $\succcurlyeq$ is mixture-continuous. It is clear from the argument in the proof of Proposition 1 of Galaabaatar-Khan-Uyanik \citeyearpar{gku19} that mixture-continuity and upper Archimedean properties imply upper strict-Archimedean property. Then,  Theorem \ref{thm: mcarc} implies that $\succcurlyeq$ is strict-Archimedean. Therefore, if we  show that $\succcurlyeq$ is reflexive and $\sim$ is transitive, then the proof follows from Galaabaatar-Khan-Uyanik \citeyearpar[Theorem 1]{gku19}. 

The following claim implies that the restriction of $\succ$ on the line segment connecting any two points is negatively transitive. 
\cl
For all $x, y\in X$ with  $x\succ y$,   $[x,y]\subseteq A_\succ(y)\cup A_\prec(x)$, where $[x,y]=\{x\lambda y| \lambda \in [0,1]\}$. 
\label{thm: ntransitive}
\cll

It is easy to see that transitivity of $\sim$ of an additive relation on a vector space is equivalent to the condition $A_\sim(0)+A_\sim(0)\subseteq A_\sim(0)$. Similarly, 
semi-transitivity of $\succcurlyeq$ is equivalent to the condition  
 $A_\succ(0)+A_\sim(0)\subseteq A_\succ(0)$.

\cl
$A_\succcurlyeq(0)=\overline{A_\succ(0)}$ and $\overline{A_\succ(0)}+A_\sim(0)\subseteq \overline{A_\succ(0)+A_\sim(0)}$.
\label{thm: lclosure}
\cll


\nt Semi-transitivity and Claim \ref{thm: lclosure} imply that $A_\succcurlyeq(0)+A_\sim(0)\subseteq \overline{A_\succ(0)}=A_\succcurlyeq(0)$. Therefore, $A_\sim(0)+A_\sim(0)\subseteq A_\succcurlyeq(0)$.   
 An analogous argument implies that $A_\sim(0)+A_\sim(0)\subseteq A_\preccurlyeq(0)$, hence $\sim$ is transitive. 
 
\cl
$\succcurlyeq$ is reflexive.
\label{thm: reflexive}
\cll 
\nt It follows from Galaabaatar-Khan-Uyanik \citeyearpar[Theorem 1]{gku19} that $\succcurlyeq$ is complete and transitive. It remains to prove Claims \ref{thm: ntransitive}, \ref{thm: lclosure} and \ref{thm: reflexive} in order to complete the proof.

\prf[Proof of Claim \ref{thm: ntransitive}]
Pick $x, y\in X$ with $x\succ y$ and $\lambda\in [0,1]$.    
 Assume $\lambda\in I_\succcurlyeq(l_{xy},y)$ and $\lambda\notin I_\succ(l_{xy},y)$. Then $x\lambda y\sim y$. Semi-transitivity implies that  $x\succ x\lambda y$, hence $\lambda\in I_\prec(l_{xy},x)$. Now assume $\lambda\in I_\preccurlyeq(l_{xy},x)$ and $\lambda\notin I_\prec(l_{xy},y)$.  Then $x\lambda y\sim x$. Semi-transitivity implies that  $x\lambda y\succ y$, hence $\lambda\in I_\succ(l_{xy},y)$. Therefore, it follows from $x\succ y$ and [0,1] is connected that 
 $$
[0,1]= I_\succcurlyeq(l_{xy},y)\cup I_\preccurlyeq(l_{xy},x) = I_\succ(l_{xy},y)\cup I_\prec(l_{xy},x).
 $$
 Since $A_\succ(y)\cap [x,y]=\{ x\lambda y| \lambda\in I_\succ(l_{xy},y)\}$, $A_\prec(x)\cap [x,y]=\{ x\lambda y| \lambda\in  I_\prec(l_{xy},x)\}$ and $l_{xy}([0,1])=[x,y]$, therefore  
$
 [x,y]\subseteq A_\succ(y)\cup A_\prec(x)
 $.
\prff

\prf[Proof of Claim \ref{thm: lclosure}]
We start by showing $A_\succcurlyeq(0)\subseteq \overline{A_\succ(0)}$. Note that non-triviality and additivity imply that $\bar x\succ 0$ for some $\bar x\in X$. Hence, $A_\succ(0)\neq\emptyset$.  Since $A_\succcurlyeq(0)=A_\succ(0)\cup A_\sim(0)$ and , therefore showing that $A_\sim(0)\subseteq  \overline{A_\succ(0)}$ is enough. To this end pick $x\sim 0$ and set $\lambda=0.5$.
 It follows from Claim \ref{thm: ntransitive}  and semi-transitivity that  $\bar x\lambda x\in A_\succ(0)\cup A_\prec(\bar x)$. We now show that $\bar x \lambda x\in A_\succ(0)$. If $\bar x\lambda x\succ 0$, then we are done. If $\bar x\succ \bar x \lambda 0$, then it follows form additivity that $\bar x \lambda x\succ x$.  And semi-transitivity implies that $\bar x \lambda x\succ 0$.  
 Repeating this argument by using Claim \ref{thm: ntransitive}, semi-transitivity and additivity imply that $\bar x \lambda^n x\in A_\succ (0)$ for all natural number $n$. Since  $\bar x \lambda^n x\ra x$, therefore  $x\in\overline{A_\succ(0)}$. 
 
For the inverse inclusion, pick $y\lambda^n z \in \overline{A_\succ(0)}$ such that $\lambda^n\ra \lambda$. Then mixture-continuity implies that $y\lambda z\in A_\succcurlyeq(0)$. Hence, $\overline{A_\succ(0)}\subseteq A_\succcurlyeq(0)$. 

We next prove $\overline{A_\succ(0)}+A_\sim(0)\subseteq \overline{A_\succ(0)+A_\sim(0)}$.  Pick $x^0\in \overline{A_\succ(0)}$ and $z\in A_\sim(0)$. Then there exists  $x\lambda^n y\in A_\succ(0)$ such that $\lambda^n\ra \lambda$ and $x^0=x\lambda y.$ Observing that $x\lambda^n y+z=(x+z)\lambda^n(y+z)\ra (x+z)\lambda (y+z)=x\lambda y+z\in \overline{A_\succ(0)+A_\sim(0)}$ finishes the proof. 
\prff

\prf[Proof of Claim \ref{thm: reflexive}]
 Reflexivity of an additive relation is equivalent to the condition $0\in A_\sim(0)$.  Claim \ref{thm: lclosure} implies that showing $0\in\overline{A_\succ(0)}$ is enough.   Set $\lambda=0.5$.
 It follows from Claim \ref{thm: ntransitive}  that  $\bar x\lambda 0\in A_\succ(0)\cup A_\prec(\bar x)$. We first show that $\bar x \lambda 0\in A_\succ(0)$. If $\bar x\lambda 0\succ 0$, then we are done. Then assume $\bar x\succ \bar x \lambda 0$. It follows form additivity that $\bar x \lambda 0\succ 0$.  
 Repeating this argument by using Claim \ref{thm: ntransitive} and additivity imply that $\bar x \lambda^n 0\in A_\succ (0)$ for all natural number $n$. Since  $\bar x \lambda^n 0\ra 0$, therefore  $0\in\overline{A_\succ(0)}$. 
 \prff

The proof of Proposition \ref{thm: ast} is complete. 
\prff

\subsection{Nine Technical Examples}\label{sec_examples} 


Section \ref{sec: results} established equivalence between different continuity postulates under convexity or monotonicity assumptions, and Section 4 applied the results to the antecedent economic literature. Through illustrative examples, this section shows that some of the assumptions in the various theorems are not redundant. 

The first example illustrates that linear continuity of a relation is not equivalent to its continuity on arbitrary convex subsets of a Euclidean space. Hence, the restrictive assumptions on the choice set we impose in this paper are not redundant. 

\exm
{\nf  Let $X=\{x\in \Re^2| x_1^2+x_2^2\leq 1\}$ be the unit sphere, $A=\{x\in X|x_1<0, x_1^2+x_2^2\neq 1\}\cup \{(-1,0)\}$ and $B=\{x\in X|x_1 > 0, x_1^2+x_2^2\neq 1\}\cup \{(1,0)\}$. Defene a binary relation $\succcurlyeq$ on $X$ as follows. For all $x\in A$ and all $y\in B$, $x\succ y$.  Assume there are no other comparable points. It is easy to see that $\succ$ has convex sections and satisfies strict-Archimedean  property. However, it does not have open sections since the sections of $\succ$ at any $x\in X$ is either $A$ or $B$ which are not open.  
}\label{exm: weakstrict}
\exmm


The following example illustrates that if we drop the order-denseness property from the (weak) Wold-continuity postulate, 
  then the binary relation can be discontinuous under weak monotonicity.\fn{This example is originally due to Mitra-Ozbek \citeyearpar{mo13}. They show that the relation in this example violates their scalar continuity property.} 

\exm
{\nf  Define a relation $\succcurlyeq$ on $\Re_+^2$ as follows: $0\sim 0$,  $x\succ 0$ for all $x\neq 0$ and $x\sim y$ for all $x,y>0$. It is clear that $\succcurlyeq$ is complete, transitive and weakly monotonic. Moreover, it  trivially satisfies the second part of Wold-continuity (hence of weak Wold-continuity) since there does not exist $x,y,z$ such that $x\succ y\succ z$. Since $A_\prec(x)= \{0\}$ is not open for all $x\neq 0$, $\succcurlyeq$ is not continuous. 
}\label{exm: wold}
\exmm
%
%
  Note that, under strong monotonicity, however, the second part of (weak) Wold-continuity implies the order-denseness property for a  transitive relation. In order to see this pick $x\succ y$. If $y=0$, then $x\succ x\lambda 0\succ 0$ for all $\lambda\in (0,1)$. Otherwise pick $z$ on the diagonal such that $z\gg x,y$. Then $z\succ x\succ y\succ 0$. Then there exists $0<\lambda< \delta<1$ such that $x\sim z\delta 0$ and $y\sim z\lambda 0$. For any $\mu\in (\lambda, \delta)$, $x\sim z\delta 0 \succ z\mu 0\succ  z\lambda 0\sim y$. Then transitivity completes the proof.

The following example shows that convexity and additivity properties of a relation do not imply each other. 

\exm{\nf 
It is clear that a relation with convex (upper) sections may not be additive.   
 In order to see that an additive relation may not have convex sections, consider the following preorder on $\Re$: $x\succcurlyeq y$ if and only if  $x-y\in \mathbb Z$. It is clear that $\succcurlyeq$ is additive. Since $A_\succcurlyeq(x)=A_\preccurlyeq(x)=\{x+k|k\in \mathbb Z|\}$ for all $x\in \Re$, therefore the sections of $\succcurlyeq$ is not convex. 
}
\label{exm: ac}
\exmm
\nt This implies that Theorem \ref{thm: mcarc} above is neither a corollary of the other results we present in this paper, nor of those presented in Neuefeind-Trockel \citeyearpar{nt95} or in \citet{ge15}, and that our results do not imply theirs.

The concept of local convexity of a set we define is slightly stronger than the ordinary local convexity concept in the literature which does not impose any restriction on those points that are not in the set; see 
\citet[p. 448]{kl51}. However, in terms of the main result in this literature, 
{\it for  closed, connected, and locally convex subset of $\Re^n$, the two properties are equivalent},\fn{Unlike Rockafellar's theorem, this main result, which is due to 1928 work of Tietze and Nakajima, is true for arbitrary topological vector spaces; see for example \citet{kl51}.} this definition is equivalent to the local convexity in the ordinary sense since the relevant set is closed in their result. The following example illustrates that our version of local convexity property is stronger. 

\exm{\nf 
Let $X=\bigcup_{n=1}^\infty \left( \frac{1}{n+1}, \frac{1}{n}\right)$. It is clear that $X$ is locally convex in the usual sense since it is open. However, $0\in \text{\nf cl}X$ and for all neighborhood $N_\varepsilon (0)$ of $0$, $N_\varepsilon (0)\cap X$ contains infinitely many disjoint intervals, hence the intersection is not convex. Therefore $X$ is not locally convex as we define.   
}\label{exm: localconvexity}
\exmm


The following example highlights the importance of the definition of the restricted relation we introduce in this paper for our results.   

\exm{\nf 
Let $X=[0,1]^2$ and $Y=\{y\in X~|~y_1+y_2=1, y_1>0, y_2>0\}$. Define $\succcurlyeq$ on $X$ as follows: for all $y\in Y$, $0\sim y$ and  there are no other comparable points. Then, $A_\succ(x)=A_{\prec}(x)=\emptyset$ for all $x\in X$, hence they are open and convex.  Moreover,  $A_{\succcurlyeq}(x)$ and $A_{\preccurlyeq}(x)$  are   convex for all $x\in X$.   However, $A_{\succcurlyeq}(0)$ is not closed. Hence, $\succcurlyeq$ is convex but not continuous.  This example satisfies all assumptions of Theorem \ref{thm: linearcontinuity}(b): $X$ is a polyhedron,   $\succ$ and $\succcurlyeq$ have (locally) convex sections.  We next show that if we define the restricted relation on a set $S$ as $\succcurlyeq\!\cap (S\times S)$, as mentioned in in the text, then $\succcurlyeq$  satisfies linear continuity (when $S$ is taken as a straight line in $X$). 
Note that the restriction of  $A_{\succcurlyeq}(x)$ and  $A_{\preccurlyeq}(x)$ on any straight line in $X$ for all $x\in X$ is either a singleton or empty, hence closed. It is trivial that the restriction of  $A_\succ(x)$ and $A_{\prec}(x)$  on any straight line in $X$ for all $x\in X$ is empty, hence open. Therefore, $\succcurlyeq$ is restricted continuous.   Since $\succcurlyeq$ is not continuous,   Theorem \ref{thm: linearcontinuity}  fails under this modified version of restricted relation.  
}\label{exm_restriction}
\exmm


The following example illustrates a preference relation that is linearly continuous, Wold-continuous and (trivially) weakly monotone but not  continuous whose domain is  a convex  set that fails property  {\bf B}.

\exm{\nf 
 Let $X=\{x\in [-1,0]\times [0,1]~|~x_1=-x_2\}$ and $f(x)=\text{sin}(1/x_1)$ if $x_1>0$ and $f(0,0)=1$. Then, it is clear that $f(x)\in [-1,1]$ for all $x\in X$. Define a binary relation $\succcurlyeq$ on $X$ as  $x\succcurlyeq y$ is and only if $f(x)\geq f(y)$. Note that $X$ does not satisfy property {\bf B} since for all $x,y\in X$, there is no $a,b\in X$ such that $x,y\geqq a$ and $b\geqq x,y$. Therefore, $\succcurlyeq$ is trivially monotone. By the arguments in the proof of Proposition \ref{thm: continuity0},  $\succcurlyeq$  is Archimedean and Wold-continuous, but not mixture continuous or continuous. 
}\label{exm_bdd}
\exmm

The following example illustrates a function that is quasi-convex, linearly continuous, but not jointly continuous whose domain is  a convex and closed set that is not a polyhedron, hence property {\bf C} fails.    

\exm{\nf 
Let $X=\left\{x\in \Re_+^2~|~x_1^2\leq x_2\right\}, Y=\left\{x\in [0,1]^2~|~x_1^2\leq x_2\leq ax_1\right\}, a>1,$ and $f: X\ra \Re$ defined as 
\begin{align*}
f(x)&=\frac{2x_1^2x_2}{x_1^4+x_2^2} \text{ for } x\neq 0 \text{ and } f(0)=0. 
\end{align*}
 It is clear that $X, Y$ are non-empty, convex and closed, and $Y\subseteq X$.    Note that, as $f$ is the function in the Genocchi-Peano example, both $f$  and  its restriction on $Y$  are linearly continuous but they are not jointly continuous. 
 
 We now show that  $f$ is  quasi-convex on $Y$.   
First, we show that $f$ is quasi-convex on int$X$ (the interior of $X$). Note that $f$ is twice differentiable on  int$X$. Hence, if the determinant of the first and second order bordered Hessian of $f$ is negative for all $x$ in int$X$, then $f$ is quasi-convex on int$X$. Pick $x\in \text{int}X$. Note that $x_1,x_2>0$. The first order bordered Hessian of $f$ at $x$ is 
$$
BH_1=\begin{bmatrix}
0 & \frac{4x_1x_2}{x_1^4+x_2^2} - \frac{8x_1^5x_2}{(x_1^4+x_2^2)^2}\\ 
& \\
\frac{4x_1x_2}{x_1^4+x_2^2} - \frac{8x_1^5x_2}{(x_1^4+x_2^2)^2} &  ~~~- \frac{56 x_1^4 x_2}{(x_1^4+x_2^2)^2}+ \frac{4x_2}{x_1^4+x_2^2}+\frac{64 x_1^8 x_2}{(x_1^4+x_2^2)^3}   
\end{bmatrix}
$$
The determinant of $BH_1(x)$ is 
$$
\text{det}(BH_1)=-\frac{16x_1^2 x_2^2(x_1^2+x_2)^2(x_1^2-x_2)^2}{(x_1^4+x_2^2)^4}
$$
Note that $x_1, x_2> 0$ and $x_2> x_1^2$. Therefore, det$(BH_1)<0$.   
The second order bordered Hessian of $f$ at $x$ is 
$$
BH_2=\begin{bmatrix}
0 & \frac{4x_1x_2}{x_1^4+x_2^2} - \frac{8x_1^5x_2}{(x_1^4+x_2^2)^2} & \frac{2 x_1^2}{x_1^4+x_2^2} - \frac{4x_1^2x_2^2}{(x_1^4+x_2^2)^2} \\ 
& & \\
\frac{4x_1x_2}{x_1^4+x_2^2} - \frac{8x_1^5x_2}{(x_1^4+x_2^2)^2} &  ~~~ - \frac{56 x_1^4 x_2}{(x_1^4+x_2^2)^2}+ \frac{4x_2}{x_1^4+x_2^2}+\frac{64 x_1^8 x_2}{(x_1^4+x_2^2)^3}   & ~~~ \frac{4x_1}{x_1^4+x_2^2} - \frac{8x_1 x_2^2+8x_1^5}{(x_1^4+x_2^2)^2}+ \frac{32x_1^5*x_2^2}{(x_1^4+x_2^2)^3} \\
& & \\
 \frac{2 x_1^2}{x_1^4+x_2^2} - \frac{4x_1^2x_2^2}{(x_1^4+x_2^2)^2} &  ~~~  \frac{4x_1}{x_1^4+x_2^2} - \frac{8x_1 x_2^2+8x_1^5}{(x_1^4+x_2^2)^2}+ \frac{32x_1^5*x_2^2}{(x_1^4+x_2^2)^3}  &  \frac{16 x_1^2*x_2^3}{(x_1^4+x_2^2)^3}-\frac{12 x_1^2x_2}{(x_1^4+x_2^2)^2}
\end{bmatrix}
$$
The determinant of $BH_2(x)$ is 
$$
\text{det}(BH_2)= -\frac{16x_1^4 x_2 (x_2^2-x_1^4)^3}{(x_1^4+x_2^2)^6}
$$
It follows from $x_1, x_2> 0$ and $x_2> x_1^2$ that  det$(BH_2)<0$. Since $x$ is an arbitrary point in int$X$, $f$ is quasi-convex on  int$X$. 

We next show that $f$ is quasi-convex on $Y$. Recall that $f$ is quasi-convex  if for all $x,y\in X$, all $\lambda\in (0,1)$, $f(\lambda x +(1-\lambda)y)\leq \text{max}\{f(x), f(y)\}$. Pick $x,y\in Y$ and $\lambda\in (0,1)$. If $x,y\in \text{int}X$, then by the argument above, quasi-convexity holds. Hence,  assume at least one of $x$ and $y$ is on the boundary of $Y$. Note that the  part of the boundary of $Y$ that is not contained in int$X$ consists of $\{z\in Y~|~z=0\text{ or } z_2=z_1^2\}$. It is easy to observe that $f(x)\leq 1$ for all $x\in Y$. Hence, if $x_2=x_1^2$ or $y_2=y_1^2$, then $f(\lambda x +(1-\lambda)y)\leq \text{max}\{f(x), f(y)\}$ is trivially true. Finally, let $y=0$ and $x\in \text{int}X$. Then, 
$$
f(\lambda x +(1-\lambda)y)=\frac{2\lambda^3x_1^2x_2}{\lambda^2(\lambda^2x_1^4+x_2^2)}=\frac{2 \lambda x_1^2x_2}{ \lambda^2 x_1^4+x_2^2}\leq \frac{2 x_1^2x_2}{ x_1^4+x_2^2}=f(x)=\text{max} \{f(x), f(y)\}. 
$$
Therefore, $f$ is quasi-convex. 
 
}\label{exm_polyhed}
\exmm

\nt Note that  even though this example is presented for functions,  the binary relation $\succcurlyeq$ on $X$ induced by the function $-f$ is complete, transitive, convex and linearly continuous but not continuous.


\medskip

The following example illustrates a function that is linearly continuous and quasi-concave (or quasi-convex)  in all of its variables, but it is not jointly continuous.  

\exm{\nf 
 

Let $X=\Re_+^2$. It is clear that $X$ satisfies condition {\bf C}. Let $f: X\ra \Re$ defined as 
$$
f(x)=\frac{2x_1 x_2^2}{x_1^2+x_2^4}.  
$$
Note that for each $i=1,2$ and each $x_i\in \Re_+$, $f(\cdot, x_i)$ is quasi-concave in $x_j$, $j\neq i$ since it is a single-peaked function on a subset of $\Re$.   Recall that $f$ is the function provided in the Genocchi-Peano's example and the discontinuity-point 0 is contained in $X$, hence $f$ is linearly continuous but not jointly  continuous.

 }\label{exm_quasiconv}
\exmm

The example above is presented for a function. The binary relation induced  by this function has upper sections whose restriction to any straight line parallel to any coordinate axis is convex.  When a function of two variables is quasi-concave in one index and quasi-convex in the other, the restriction of the lower sections of the induced relation on  any straight line parallel to one of the coordinate axis is convex and the restriction of the upper sections of the induced  relation on  any straight line parallel to the other coordinate axis is convex. The following example  illustrates  a binary relation that  satisfies this convexity property,  is  linearly continuous, but fails section continuity.  

\exm{\nf 
 

Let $X=[0,1]^2$. It is clear that $X$ satisfies condition {\bf C}. Let $Y=\{x\in X | x_2=x_1^2, x_1\in (0,1)\}$ Let $\succcurlyeq$ be a binary relation  on $X$ defined as follows: $x\succcurlyeq y$ if and only if $x,y\in Y$. It is clear that for all $x\in X$ and all straight lines $L$ in $X$ parallel to the first coordinate, $A_\succcurlyeq (x)\cap L$ is either empty or singleton, hence convex. Similarly, for all   $x\in X$ and all straight lines $L$ in $X$ parallel to the second coordinate, $A_\preccurlyeq (x)\cap L$ is either empty or singleton, hence convex. Note that for all $x\in X$ and all straight lines $L$   in $X$,  $A_\succcurlyeq (x)\cap L$ and $A_\preccurlyeq (x)\cap L$ is empty, or singleton or consists of two points, hence closed. Since no points in $X$ are strictly comparable, $A_\succcurlyeq(x)=A_\preccurlyeq(x)=\emptyset$ for all $x\in X$. Therefore, $\succcurlyeq$ is linearly continuous. However, $Y$ is not closed in $X$, hence $\succcurlyeq$ is not continuous.

 }\label{exm_qconcave_qconvex}
\exmm


\newpage



\setlength{\bibsep}{5pt}
\setstretch{0.96}



\ifx\undefined\BySame
\newcommand{\BySame}{\leavevmode\rule[.5ex]{3em}{.5pt}\ }
\fi
\ifx\undefined\textsc
\newcommand{\textsc}[1]{{\sc #1}}
\newcommand{\emph}[1]{{\em #1\/}}
\let\tmpsmall\small
\renewcommand{\small}{\tmpsmall\sc}
\fi

\end{document}